Nec sine te, nec tecum vivere possum. (Ovid)*

# Spin Physics Experiments at NICA-SPD with polarized proton and deuteron beams.

Compiled by the Drafting Committee:
I.A. Savin, A.V. Efremov, D.V. Peshekhonov, A.D. Kovalenko, O.V.Teryaev,
O.Yu. Shevchenko, A.P. Nagajcev, A.V. Guskov, V.V. Kukhtin, N.D. Topilin.

(Letter of Intent presented at the meeting of the JINR Program Advisory Committee (PAC) for Particle Physics on 25–26 June 2014.)

ABSTRACT

We propose to perform measurements of asymmetries of the DY pair's production in collisions of non-polarized, longitudinally and transversally polarized protons and deuterons which provide an access to all leading twist collinear and TMD PDFs of quarks and anti-quarks in nucleons. The measurements of asymmetries in production of J/Ψ and direct photons will be performed as well simultaneously with DY using dedicated triggers. The set of these measurements will supply complete information for tests of the quark-parton model of nucleons at the QCD twist-two level with minimal systematic errors.

**PAC Recommendations:**

….The PAC heard with interest a report on the preparation of the Letter of Intent "Spin physics experiments at NICA-SPD with polarized proton and deuteron beams" presented by I. Savin. The PAC is pleased to see the first steps toward formation of an international collaboration around the SPD experiment. The PAC regards the SPD experiment as an essential part of the NICA research program and encourages the authors of the Letter of Intent to prepare a full proposal and present it at one of the forthcoming meetings of the PAC. …

---

*) Neither without you, nor with you one can live.

## Updating list of participants

R. Abramishvili[1], V.V. Abramov[6], F. Ahmadov[1], R.R. Akhunzyanov[1], N. Akopov[23], V.A. Anosov[1], N.V. Anfimov[1], S. Anishchanka[12], X. Artru[15], A.A. Baldin[1], V.G. Baryshevsky[12], A.S. Belov[5], D.A. Bliznyuk[14], M. Bodlak[8], A.V. Butenko[1], A.P. Cheplakov[1], I.E. Chirikov-Zorin[1], G. Domanski[10], S.V. Donskov[6], G. L. Dorofeev[1], V. M. Drobin[1], V.V. Drugakov[17], M. Dziewiecki[10], A.V. Efremov[1], Yu.N. Filatov[1,3], V.V. Fimushkin[1], M. Finger (jun.)[7,1], M. Finger[7,1], S.G. Gerassimov[13], I.A. Golutvin[1], A.L.Gongadze[1], I.B. Gongadze[1], M.I. Gostkin[1], B.V. Grinyov[14], A. Gurinovich[12], A.V. Guskov[1], A.N. Ilyichev[17], Yu.I. Ivanshin[1], A.V. Ivanov[1], V. Jary[8], A. Janata[7,1], N. Javadov[1], Jen-Chieh Peng[20], L.L. Jenkovszky[4], V.D. Kekelidze[1], D.V. Kharchenko[1], A.P. Kobushkin[4], B. Konarzewski[10], A.M. Kondratenko[2], M.A. Kondratenko[2], I. Konorov[13], A.D. Kovalenko[1], O.M. Kouznetsov[1], G.A. Kozlov[1], A. D. Krisch[16], U.G. Kruchonak[1], Z.V. Krumshtein[1], V.V. Kukhtin[1], K. Kurek[9], P.K. Kurilkin[1], R. Kurjata[10], L.V. Kutuzova[1], N.K. Kuznetsov[1], V.P. Ladygin[1], R. Lednicky[1], A. Lobko[12], A.I. Malakhov[1], B. Marinova[1], J. Marzec[10], J. Matousek[7], G.V. Meshcheryakov[1], V.A. Mikhaylov[1], Yu.V. Mikhaylov[6], P.V. Moissenz[1], V.V. Myalkovskiy[1], A.P. Nagaytsev[1], J. Novy[8], I.A. Orlov[1], Baatar Otgongerel[22], B. Parsamyan[21], M. Pesek[7], D.V. Peshekhonov[1], V.D. Peshekhonov[1], V.A. Polyakov[6], Yu.V. Prokofichev[1], A.V. Radyushkin[1], Togoo Ravdandorj[22], V.K. Rodionov[1], N.S. Rossiyskaya[1], A. Rouba[12], A. Rychter[10], V.D. Samoylenko[6], A. Sandacz[9], I.A. Savin[1], G.A. Shelkov[1], N.M. Shumeiko[17], O.Yu. Shevchenko[1], S.S. Shimanskiy[1], A.V. Sidorov[1], D. Sivers[18], M. Slunechka[7,1], V. Slunechkova[7,1], A.V. Smirnov[1], G.I. Smirnov[1], N.B. Skachkov[1], J. Soffer[11], A.A. Solin[17], A.V. Solin[17], E.A. Strokovsky[1], O.V.Teryaev[1], A.V. Tkachenko[1,4], M. Tomasek[8], N.D. Topilin[1], Baatar Tseepeldorj[22], A.V.Turbabin[5], Yu.N. Uzikov[1], M.Virius[8], V.Vrba[8], K. Zaremba[10], P. Zavada[19], M.V. Zavertyaev[13], E.V. Zemlyanichkina[1], P.N. Zhmurin[14], M. Ziembicki[10], A.I. Zinchenko[1], V.N. Zubets[5], I.P.Yudin[1]

## Affiliations

[23]Alikhanyan National Science Laboratory (YerPhI), Yerevan, Armenia
[22]Institute of Physics and Technology MAS, Ulaanbaator, Mongolia
[21]Presently at the INFN section of Turin and University of Turin, Italy
[20]University of Illinois at Urbana, Illinois, USA
[19]Institute of Physics ASCzR, Prague, Czech Republic
[18]Portland Physics Institute, Portland, USA
[17]National Center of Particle and High Energy Physics, Belarusian State University, Minsk
[16]University of Michigan, USA
[15]CNRS, Lyon, France
[14]Institute for Scintillation Materials, NAS, Kharkov, Ukraine
[13]Lebedev Physics Institute, Moscow, Russia
[12]Research Institute for Nuclear Problems, Minsk, Belarus
[11]Temple University, Philadelphia, USA
[10]Warsaw University of Technology, Institute of Radio electronics, Warsaw, Poland
[9]National Center for Nuclear Research, Warsaw, Poland
[8]Technical University, Faculty of Nuclear Science and Physics Engineering, Prague, Czech Rep.
[7]Charles University, Faculty of Mathematics and Physics, Prague, Czech Republic
[6]Institute for High Energy Physics, Protvino, Russia
[5]Institute for Nuclear Research of Russian Academy of Sciences, Moscow, Russia
[4]Bogolyubov Institute for Theoretical Physics, Kiev, Ukraine
[3]Moscow Institute of Physics and Technology, Dolgoprudny, Russia
[2]Science and Technique Laboratory Zaryad, Novosibirsk, Russia
[1]Joint Institute for Nuclear Research, Dubna, Russia

## TABLE OF CONTENTS

# 1. Introduction

Main parts of this Letter of Intent (LoI) are related to the studies of the nucleon structure. The beginning of the nucleon structure story refers to the early 50-ties of the 20th century when in the famous Hofstadter's experiments at SLAC the proton electromagnetic form factor was measured determining thus the proton radius of $<r_p> = (0.74\pm0.24)\cdot10^{-13}$ cm. It means that the proton is not an elementary particle but the object with an internal structure. Later on, again at SLAC, the point-like ***constituents*** have been discovered in the proton and called ***partons***. After some time, in 1970-ties, partons were identified with ***quarks*** suggested early by Gell-Mann as structureless constituents of all hadrons. Three families of quarks, each containing two quarks and anti-quarks, are now the basic elements of the Standard Model (SM) of elementary particle structure. All six quarks are discovered.

The naive quark-parton model (***QPM***) of nucleons, i.e. of the proton and neutron, has been born. According to this model, the protons (neutron) consist of three spin-1/2 ***valence quarks***: two (one) of the ***u***-type and one (two) of the ***d***-type with a charge of (+2/3) *e* and (-1/3) *e*, respectively, where *e* is the absolute value of the electron charge. Quarks interact between themselves by ***gluon*** exchange. Gluons are also the nucleon constituents. Gluons can produce a ***sea*** of any type (***flavor***) quark-anti-quark pairs. Partons share between themselves fractions, *x*, of the total nucleon momentum. Parton Distribution Functions (***PDFs***) are universal characteristics of the internal nucleon structure.

Now the quark-parton structure of nucleons and respectively the quark-parton model of nucleons are becoming more and more complicated. In Quantum Chromo Dynamics (QCD), PDFs depend not only on *x*, but also on $Q^2$, four-momentum transfer (see below). Partons can have an internal momentum, *k*, with possible transverse component, $k_T$. A number of PDFs depends on the order of the QCD approximations. Measurements of the collinear (integrated over $k_T$) and Transverse Momentum Dependent (TMD) PDFs, the most of which are not well measured or not discovered yet, are proposed in this LoI. Main ideas of this document have been discussed at the specialized International Workshops [1]. General organization of the text follows the Table of contents.

## 1.1. Basic (twist-2) PDFs of the nucleon.

There are three collinear PDFs characterizing the nucleon structure at the leading QCD order (twist-2). These PDFs are: the distribution of the parton Number in non-polarized (*U*) nucleon (***Density)***, $f_1 (x, Q^2)$; the distribution of longitudinal polarization of quarks in longitudinally polarized (*L*) nucleon (***Helicity***), $g_1 (x, Q^2) \equiv g_{1L} (x, Q^2)$; and the distribution of transverse polarization of quarks in transversely polarized (*T*) nucleon (***Transversity***), $h_1 (x, Q^2)$. They are shown as diagonal terms in Fig. 1.1 with the nucleon polarization (*U, L, T*) along the vertical direction and the quark polarization along the horizontal direction. The PDF $h_1 (x, Q^2)$ is poorly studied. It is a chiral-odd function which can be measured in combination with another chiral-odd function.

If one takes into account the possible transverse momentum of quarks, $k_T$, there will be five additional Transverse Momentum Dependent (TMD) PDFs which are functions of three variables: $x, k_T, Q^2$. These TMD PDFs are: correlation between the transverse polarization of nucleon (transverse spin) and the transverse momentum of non-polarized quarks (***Sivers***), $f^{\perp}_{1T}$; correlation between the transverse spin and the longitudinal quark polarization (***Worm-gear-T***), $g^{\perp}_{1T}$; distribution of the quark transverse momentum in the non-polarized nucleon (***Boer-Mulders***), $h^{\perp}_{1}$; correlation between the longitudinal polarization of the nucleon (longitudinal spin) and the transverse momentum of quarks (***Worm-gear-L***), $h^{\perp}_{1L}$; distribution of the transverse momentum of quarks in the transversely polarized nucleon (***Pretzelosity***), $h^{\perp}_{1T}$. All new PDFs, except $f^{\perp}_{1T}$, are chiral-odd. The Sivers and Boer-Mulders PDFs are T-odd ones. At the sub-leading twist (twist-3), there are still 16 TMD PDFs containing the information on the nucleon structure. They have no definite physics interpretation yet.

The PDFs $\boldsymbol{f_1}$ and $\boldsymbol{g_1}$ are measured rather well (Section 1.2). The $\boldsymbol{h_1}$ has been measured recently but is still poorly investigated. All TMD PDFs are currently studied (Section 1.3).

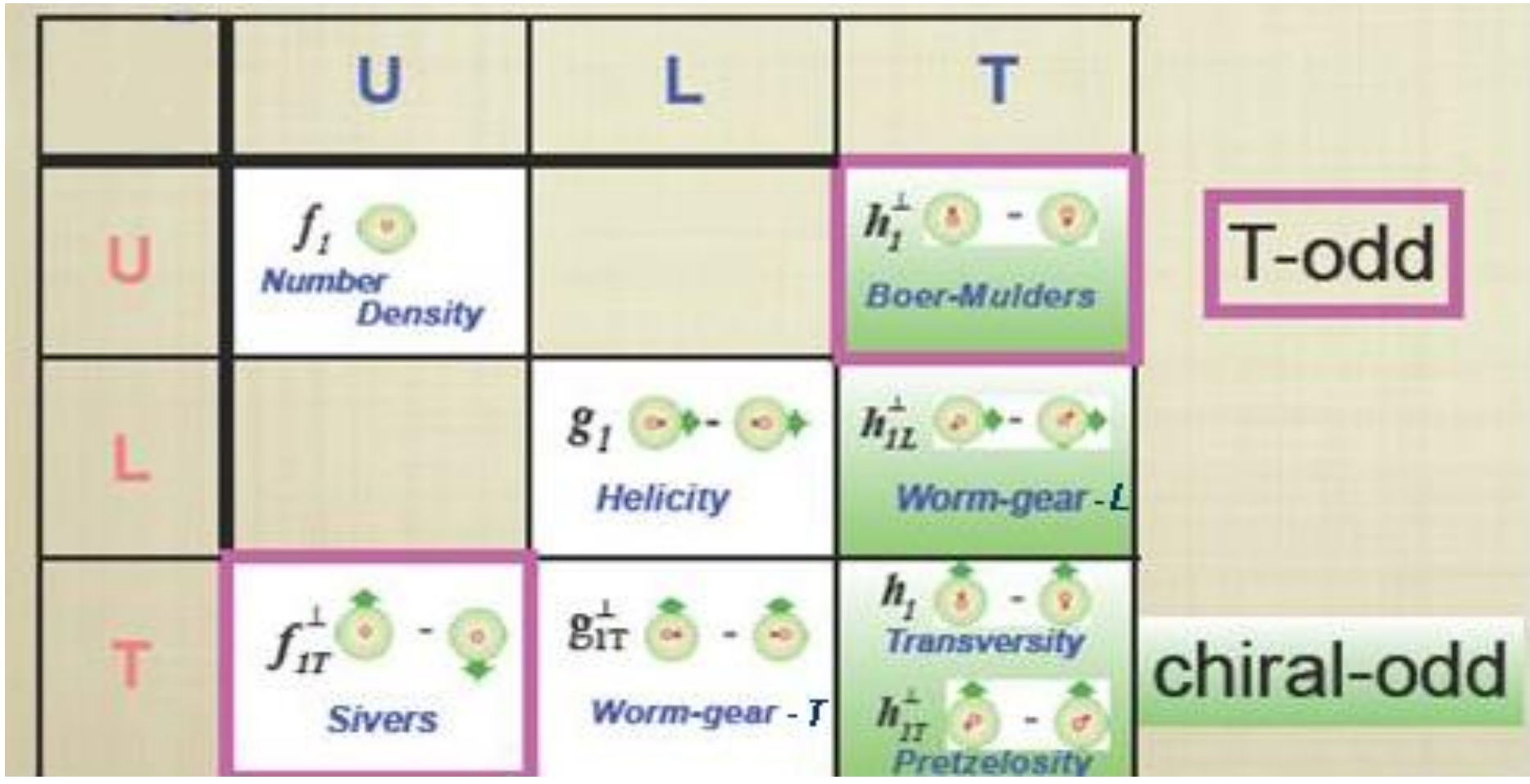

***Fig. 1.1***: *The twist-2 PDFs characterizing the nucleon structure.*

### 1.2. Deep Inelastic Scattering as the first microscope for the nucleon structure study. Status of the PDFs $\boldsymbol{f_1}$ and $\boldsymbol{g_1}$ studies.

The powerful methods to study the quark-parton structure of nucleons give reactions of the Deep Inelastic lepton-nucleon Scattering (DIS) illustrated by the Feynman diagrams in Fig. 1.2 where basic kinematic variables are introduced.

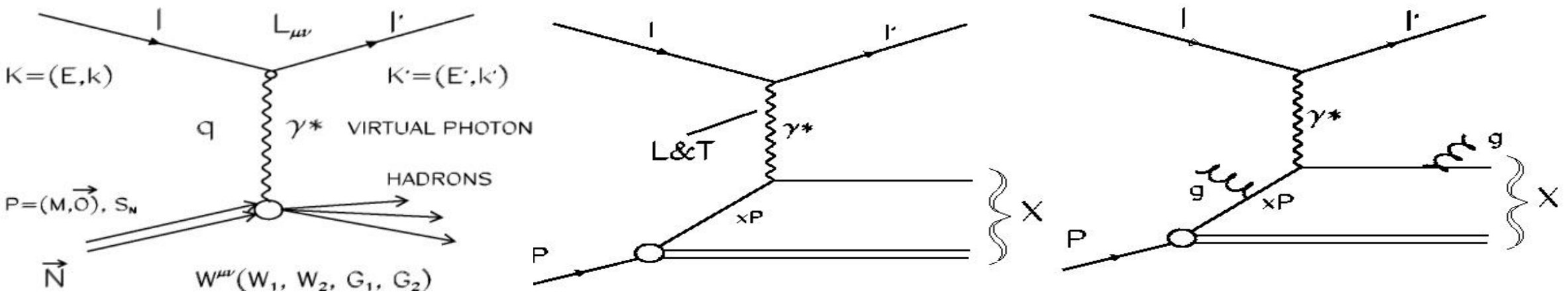

(a) (b) (c)

***Fig. 1.2***: *Feynman diagrams of DIS in one-photon exchange approximation:*
*(a) IDIS. The virtual photon transfers a four momentum squared, $Q^2$, and energy, $\nu$, from the incident lepton to the nucleon. Variables:* $-q^2 \equiv Q^2 = -(k-k')^2 = 4EE'\sin^2(\theta/2)$;
$\nu = P\cdot q/M$; $x = Q^2/2M\nu$; $y = \nu/E$.
*(b) IDIS in QPM. The constituent quark with a fraction **x** of the nucleon momentum, **q(x)**, absorbs the virtual photon $\gamma^*$ and then fragments into the final state **X**.*
*(c) IDIS in QCD improved QPM. The quark with a fraction **x** of the nucleon momentum absorbing the virtual photon with $Q^2$, **q(x, $Q^2$)**, can emit gluons before or after absorption.*

There are three types of DIS reactions:

- Inclusive (IDIS), when characteristics of incident (*l*), polarized or non-polarized, and scattered lepton (*l'*) are known (measured): $l + N \rightarrow l' + X$, nucleon (*N*) can be polarized;
- Semi-inclusive (SIDIS), when, additionally to the above mentioned, characteristics of the final state hadron (*h*) are known: $l + N \rightarrow l' + nh + X$, n $\geq$1, and
- Exclusive (EDIS), not considered here, when final states of the reaction are fully determined.

A quantitative characteristic of the IDIS reaction is a double differential cross section [2]. This cross section can be calculated theoretically assuming that the main contribution to it comes from the one-photon exchange process. It is known that the one-photon exchange IDIS cross section is defined as

$$\vec{\sigma}_{one-photon} \equiv \frac{d^2\vec{\sigma}^{S_\ell S_N}}{d\Omega dE'} = \left(\frac{4\alpha^2}{Q^4}\cdot\frac{E'}{E}\right)\cdot L_{\mu\nu}\cdot W^{\mu\nu} \qquad (1.1)$$

where the term in brackets characterizes the point-like interaction; $L_{\mu\nu}$ is the lepton current tensor representing the lepton vertex in Fig. 1.2 (a) and $W^{\mu\nu}$ is the hadronic tensor amplitude characterizing the hadrons' vertex structure. Each tensor has two parts, one of which (SIM) is independent of the spin orientations and the second one (ASIM) is spin-dependent:

$$L_{\mu\nu} = L_{\mu\nu}^{SIM} + iL_{\mu\nu}^{ASIM},$$
$$W^{\mu\nu} = W_{SIM}^{\mu\nu} + iW_{ASIM}^{\mu\nu}. \qquad (1.2)$$

The structure of $L_{\mu\nu}$ is exactly known from Quantum Electro Dynamics (QED). The hadronic tensor $W^{\mu\nu}$ is not calculated theoretically. It is a pure phenomenological quantity characterizing the nucleon structure. Theory tells us that, from the most general considerations, for electromagnetic interactions $W^{\mu\nu}$ should have the structure:

$$W_{\mathrm{SIM}}^{\mu\nu} = A_1^{\mu\nu}(q,q')\cdot W_1(Q^2,\nu) + A_2^{\mu\nu}(q,q')\cdot W_2(Q^2,\nu),$$
$$W_{\mathrm{ASIM}}^{\mu\nu} = B_1^{\mu\nu}(q,q')\cdot G_1(Q^2,\nu) + B_2^{\mu\nu}(q,q')\cdot G_2(Q^2,\nu), \qquad (1.3)$$

where $A_1$, $A_2$, $B_1$ and $B_2$ are known kinematic expressions, $W_1(Q^2,\nu)$ and $W_2(Q^2,\nu)$ are spin independent and $G_1(Q^2,\nu)$ and $G_2(Q^2,\nu)$ are spin dependent structure functions representing the nucleon structure. In general, these structure functions should be functions of two independent variables - either $(Q^2,\nu)$; or $(Q^2, x)$; or $(x, y)$, etc. Bjorken has assumed that in the DIS (scaling) limit, $(Q^2,\nu\to\infty$, $x$ fixed), the structure functions became the functions of the only one (Bjorken) scaling variable $x$:

$$M\cdot W_1(Q^2,\nu)\to F_1(x),$$
$$\nu\cdot W_2(Q^2,\nu)\to F_2(x),$$
$$\nu M^2\cdot G_1(Q^2,\nu)\to g_1(x),$$
$$\nu^2 M\cdot G_2(Q^2,\nu)\to g_2(x). \qquad (1.4)$$

But at the $Q^2$ of current experiments, this hypothesis is true only in the limited range of $x$.

Performing the calculations as prescribed above and summing over the spin orientations of scattered leptons, $S_e$, which are usually not known, one can get the cross section

$$\frac{d^2\vec{\sigma}^{S_eS_N}}{d\Omega dE'} = \frac{d^2\sigma^{unp}}{d\Omega dE'} + S_N S_e\frac{d^2\sigma^{pol}}{d\Omega dE'}, \qquad (1.5)$$

where $\sigma^{unp}$ ($\sigma^{pol}$) is the non-polarized (polarized) part of the cross section and $S_N = \pm 1$ is the orientation (helicity) of the nucleon spin. In the most commonly used notations the spin-independent part of the cross section, $\sigma^{unp}$, is expressed via two spin-independent structure functions $F_1$ and $F_2$:

$$\frac{d^2\sigma^{upn}}{dxdQ^2} = \frac{4\pi\alpha^2}{Q^2x}\left[xy^2(1-\frac{2m_e^2}{Q^2})F_1(x,Q^2) + (1-y-\frac{\gamma^2y^2}{4})F_2(x,Q^2)\right]. \qquad (1.6)$$

Here $m_e$ is the lepton mass and $\gamma = 2Mx/\sqrt{Q^2} = \sqrt{Q^2}/\nu$. There is a theoretical relationship between the structure functions $F_1$ and $F_2$ known under the name of Callan-Gross:

$$F_2(x, Q^2) = 2xF_1(x, Q^2). \qquad (1.7)$$

The $\sigma^{unp}$ is often expressed via $F_2(x, Q^2)$ and $R(x, Q^2) = \sigma_L/\sigma_T$ where $\sigma_L(\sigma_T)$ is the nucleon absorption cross section of the virtual photon with longitudinal (transverse) polarization:

$$\sigma^{unp} \equiv \frac{d^2\sigma^{unp}}{dxdQ^2} = \frac{4\pi\alpha^2}{Q^4 x} F_2(x,Q^2)\left[1 - y - \frac{y^2\gamma^2}{4} + \frac{y^2(1+\gamma^2)}{2(1+R(x,Q^2))}\right]. \quad (1.8)$$

The structure functions $R(x, Q^2)$ and $F_2(x, Q^2)$ have been measured by the well-known collaborations SLAC-MIT, EMC, BCDMS, NMC, ZEUS, $H_1$ and others.

By definition, the structure functions $F_1$ and $F_2$ are pure phenomenological. Their physics interpretations can be given only within certain models. In QPM of nucleons, IDIS is represented by the diagram in Fig. 1.2 (b) in which the virtual photon is absorbed by the nucleon's constituent quark carrying fraction $x$ of the nucleon momentum with a probability $q(x)$. In the QCD improved QPM (Fig. 1.2 (c), the quark with a probability $q(x, Q^2)$ can emit a gluon before or after absorption. Then the structure function $F_2$ is defined as:

$$F_2(x, Q^2) = x\sum_q e^2_q [q(x, Q^2) + \bar{q}(x, Q^2)],\ q=u, d, s, \quad (1.9)$$

where $e_q$ is the charge of the quark. From the global QCD analysis of all DIS data one can find the non-polarized nucleon PDFs $f^a{}_1$ (the superscript $a$ is usually omitted) for each parton, Fig.1.3.

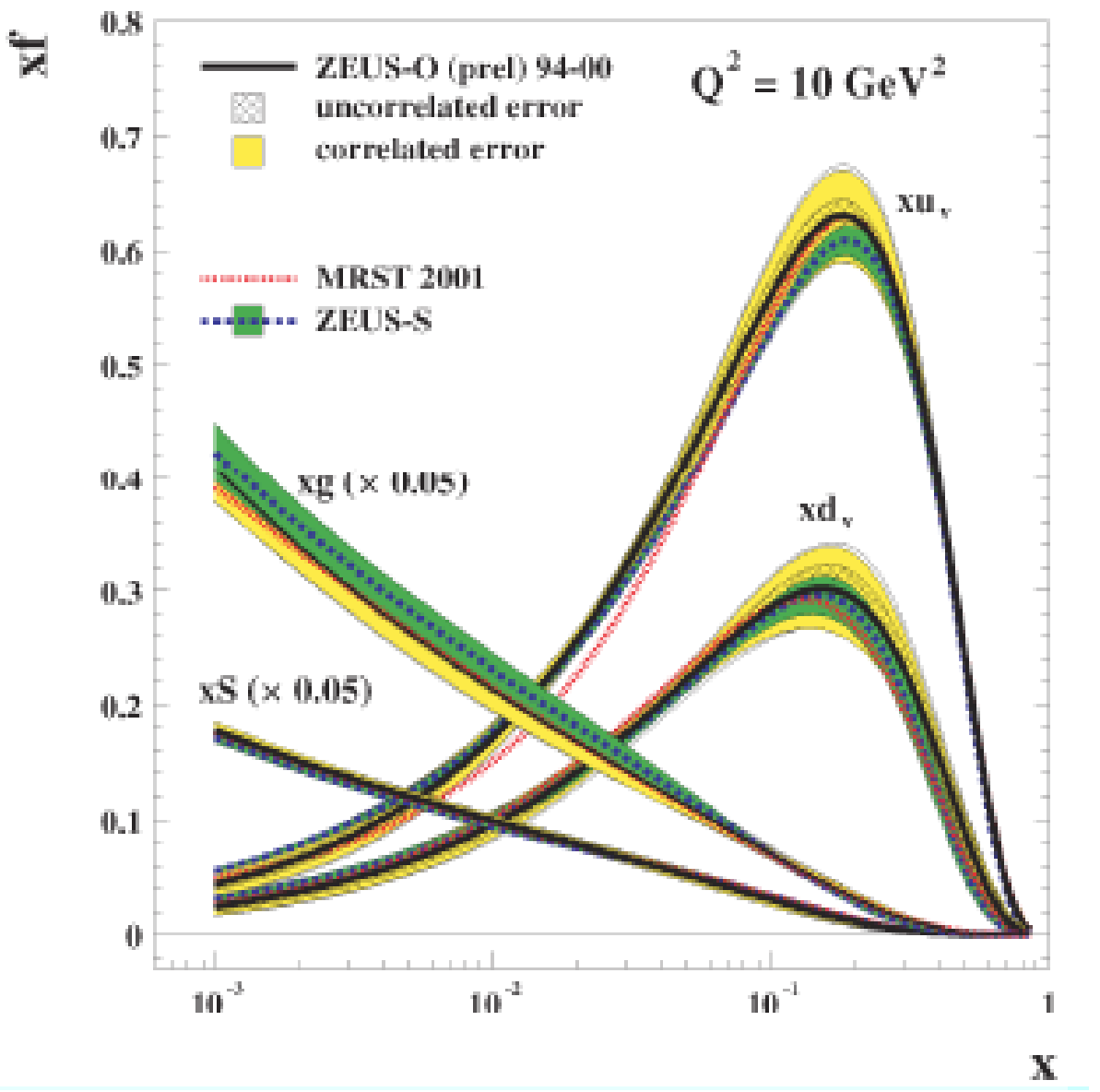

***Fig. 1.3:*** *Parton (Density) distributions in non-polarized nucleons at* $Q^2$*=10 GeV*$^2$ *vs. x.*

The spin-dependent part of the cross section, $\sigma^{pol}$, can be studied measuring so-called asymmetry which is proportional to the difference of cross sections (1.5) for two opposite target polarizations. In case of longitudinal target polarization, the longitudinal asymmetry, $\boldsymbol{A}_{||}$, is :

$$A_{||} = \frac{\Delta\sigma_{||}}{2\sigma^{unp}} = \frac{\sigma^{\rightarrow\Rightarrow} - \sigma^{\rightarrow\Leftarrow}}{\sigma^{\rightarrow\Leftarrow} + \sigma^{\rightarrow\Rightarrow}} \quad (1.10)$$

where arrows $\rightarrow$and $\Rightarrow$ indicate the directions of the incident lepton and of the target polarization, respectively. The difference between the cross sections, $\Delta\sigma_{||}$, is found to be:

$$\Delta\sigma_{||} \equiv \Delta\left(\frac{d^2\sigma_{||}^{pol}}{dxdQ^2}\right) = \frac{16\pi\alpha^2 y}{Q^4}\left[\left(1 - \frac{y}{2} - \frac{y^2\gamma^2}{4}\right)g_1 - \frac{y\gamma^2}{2} g_2\right], \quad (1.11)$$

and the experimental asymmetry $\boldsymbol{A}_{||}$ is related to the theoretical virtual photon asymmetries $A_1$ and $A_2$:

$$\boldsymbol{A}_{||} = D\,(A_1 + \eta A_2). \quad (1.12)$$

Here $D$ and $\eta$ are the known kinematic factors, $A_2 = \gamma\,(g_1 + g_2)/F_1$ is estimated to be small and

$$A_1 = (g_1 - \gamma^2 g_2)/F_1 \approx g_1/F_1$$

because the term $\gamma^2 g_2$ is also small or, alternatively,

$$A_1 = (\sigma_{1/2} - \sigma_{3/2})/(\sigma_{1/2} + \sigma_{3/2}) \quad (1.13)$$

where $\sigma_{1/2}$ and $\sigma_{3/2}$ are absorption cross sections of the virtual photon ($\gamma^*$) by the nucleon with the total photon-nucleon angular momentum along the $\gamma^*$ axis equal to 1/2 or 3/2, respectively. So, $\boldsymbol{A}_{||} \approx D \cdot A_1$ and, in the first approximation one can obtain a relation connecting experimentally measured asymmetry $\boldsymbol{A}_{||}$ and spin dependent structure function $g_1$:

$$\boldsymbol{A}_{||}/D \approx A_1 \approx g_1/F_1, \qquad (1.14)$$

The physics interpretation of $g_1$ can be obtained in QPM using the alternative expression (1.13) for the virtual photon asymmetry $A_1$. In QPM the IDIS is represented by the diagram in Fig. 1.2 (b, c): the virtual photon is absorbed by the constituent quark carrying the fraction $x$ of the nucleon momentum. Due to conservation of the total angular momentum, this photon can be absorbed only by a quark having the spin oriented in the opposite direction to the photon angular momentum. Taking this into account, one can obtain the QPM expressions for $\sigma_{1/2}$ and $\sigma_{3/2}$ :

$$\sigma_{1/2} = \Sigma_i\, e_i^2 \cdot q_i^{\uparrow}(x) \text{ and } \sigma_{3/2} = \Sigma_i\, e_i^2 \cdot q_i^{\downarrow}(x) \qquad (1.15)$$

where arrow ↑(↓) indicates that the spin of quark $i$ have the same (opposite) orientation as the spin of nucleon. Then the QPM expression for the proton asymmetry $A_1^p$ will be:

$$A_1^p = \frac{\sigma_{1/2}^p - \sigma_{3/2}^p}{\sigma_{1/2}^p + \sigma_{3/2}^p} = \frac{\sum e_i^2\left[q_i^{\uparrow}(x) - q_i^{\downarrow}(x)\right]}{\sum e_i^2\left[q_i^{\uparrow}(x) + q_i^{\downarrow}(x)\right]}. \qquad (1.16)$$

Comparing this expression with (1.14) one can associate its numerator with the structure function $g_1$, characterising the quark spin orientations (Helicity) with respect to orientation of the nucleon spin in the longitudinally polarized nucleon:

$$g_1(x) = \sum_i e_i^2\left[q_i^{\uparrow}(x) - q_i^{\downarrow}(x)\right] \equiv \sum_i e_i^2 g_{1L}^i(x) \equiv \sum_q e_q^2 \Delta q(x) \qquad (1.17)$$

Similarly, the denominator can be associated with the non-polarized structure function $F_1$.

The structure functions $g^p{}_1(x, Q^2)$ and $g^d{}_1(x, Q^2)$ for protons and deuterons have been determined from inclusive asymmetries $A_1$ measured by various collaborations at SLAC, CERN, DESY, JLAB. The summary of present $g_1$ data is shown in Fig. 1.4 [2]. The data are in very good agreement between themselves and with the QCD NLO predictions.

Inclusive and semi-inclusive asymmetries for proton and deuteron shown in Fig. 1.5, left permit to determine quark helicity distributions $\Delta q$, Fig. 1.5, right by using the following expression:

$$A_1^{h(p/d)}(x,z,Q^2) \approx \frac{\sum_q e_q^2 \Delta q(x,Q^2) D_q^h(z,Q^2)}{\sum_q e_q^2 q(x,Q^2) D_q^h(z,Q^2)} \qquad (1.18)$$

containing additionally to $\Delta q$ the parameterizations of non-polarized quark distributions $q(x, Q^2)$ and quark fragmentation functions (***FF***) $D^h{}_q(z, Q^2)$ measured in other experiments are used. The precision of this determination depends very much on the precision of the FFs. This is especially important for the strange quarks. Data shown in Fig. 1.5 give only values for $x\Delta S$, where $S$ is the sum of strange quarks and anti-quarks.

One can estimate the quark contributions to the nucleon spin integrating the helicity distributions over the covered $x$-range. As it is known, the longitudinal projection of the nucleon spin ($S_N$ ) is equal to ½ in units of the Plank constant. In QPM the $S_N$ is defined as a sum of contributions of quarks ($\Delta\Sigma$) , gluons ($\Delta G$ ) and their orbital momenta ($L$):

$$S_N = ½ = ½\, \Delta\Sigma + \Delta G + L^q{}_z + L^g{}_z. \qquad (1.19)$$

The present value of the quark contributions determined from the Helicity distributions amounts to about 33% of the $S_N$. This result confirms with high precision the original EMC observation that the quarks contribute little to the total nucleon spin (spin crisis). The COMPASS collaboration in the separate measurements, Fig. 1.6, has shown that the gluons contribute to the nucleon spin even smaller than that of quarks, almost zero. This is confirmed by the RHIC experiments. At the present knowledge, the nucleon spin crisis can be resolved by future measurements of Generalized Parton Distributions (GPD) accounting also for orbital momenta of nucleon constituents.

Similarly to the non-polarized PDF, the latest QCD analysis [3] of the $g^p{}_1(x, Q^2)$ and $g^d{}_1(x, Q^2)$ data has produced the Helicity distribution PDF $g^a{}_1$ (Fig. 1.7).

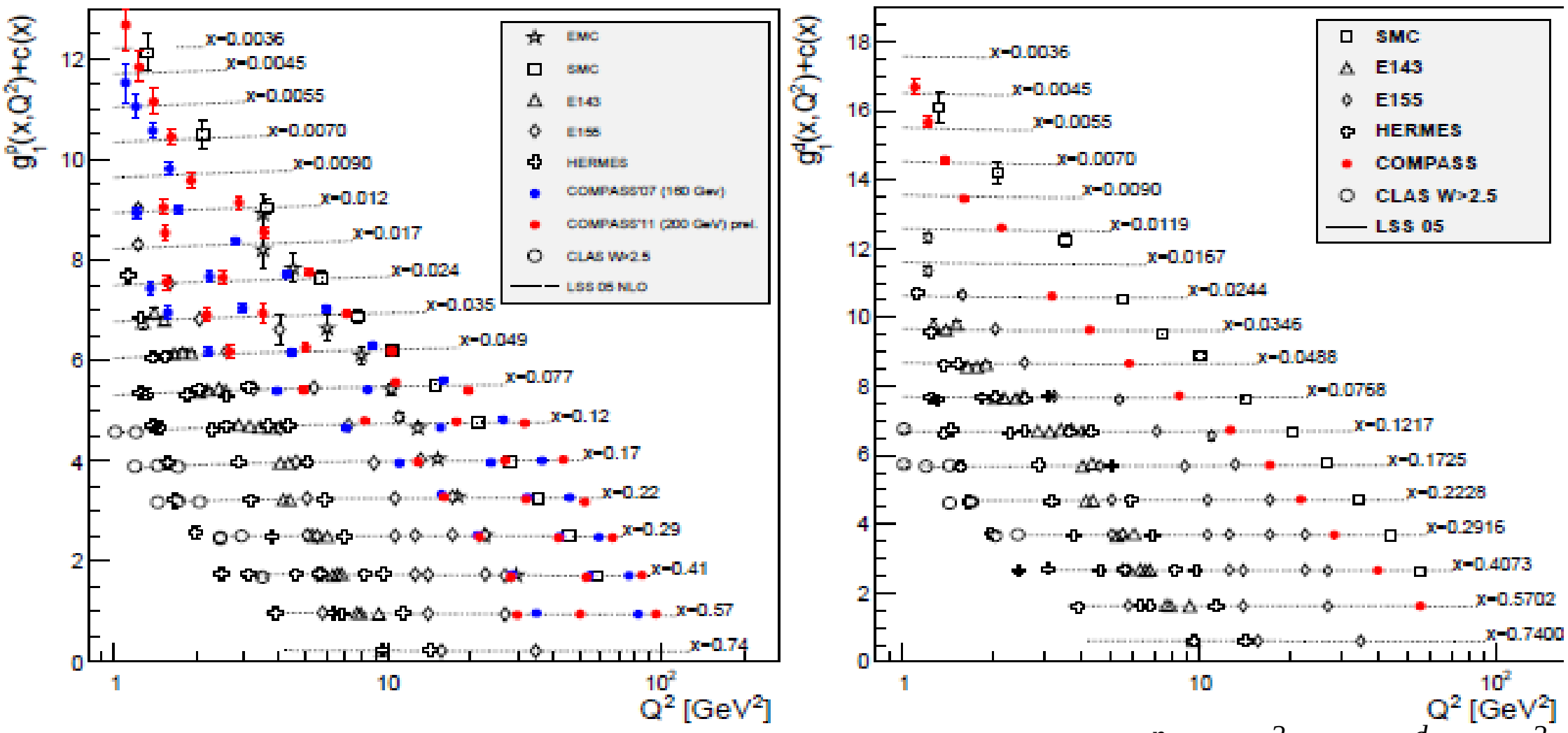

***Fig. 1.4:*** *Summary of the world data on the structure functions $g^p{}_1(x, Q^2)$ and $g^d{}_1(x, Q^2)$.*

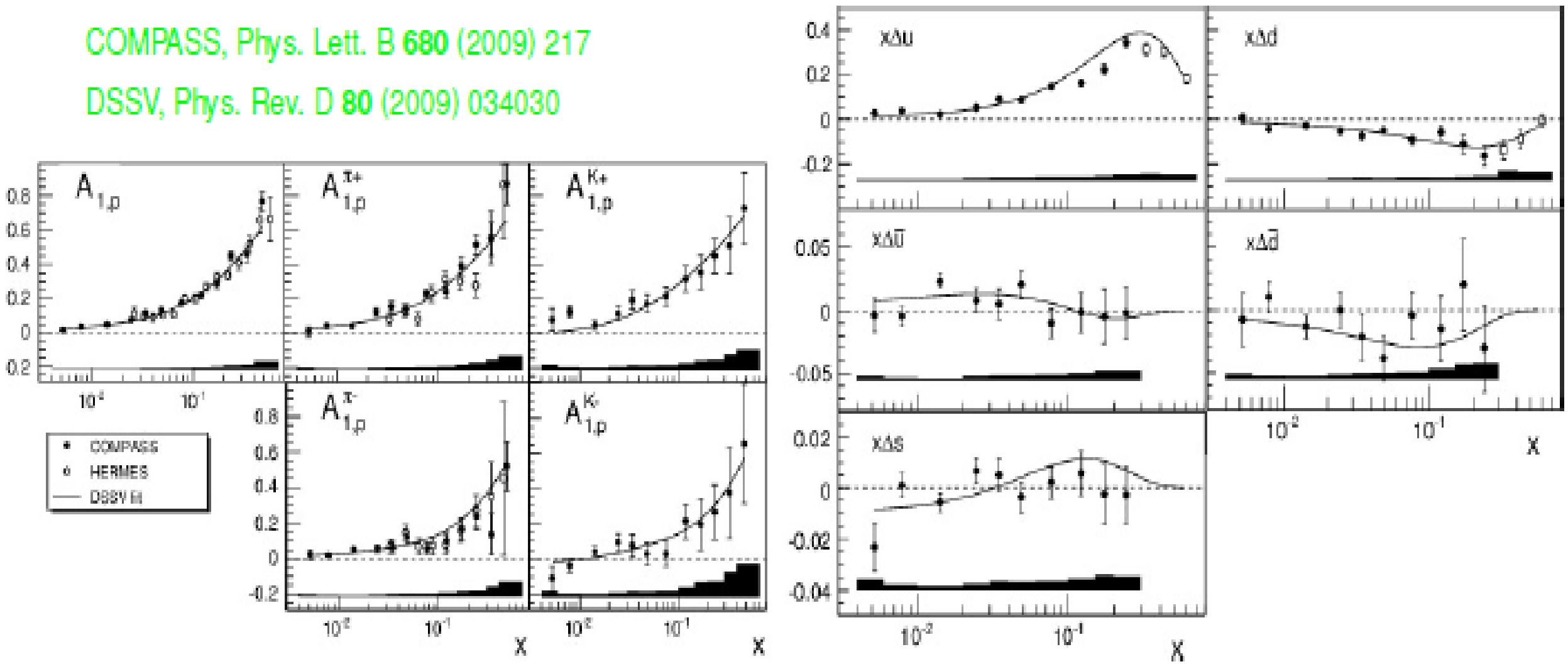

***Fig. 1.5:*** *Left: the proton inclusive and semi-inclusive asymmetries. Right: quark Helicity PDFs.*

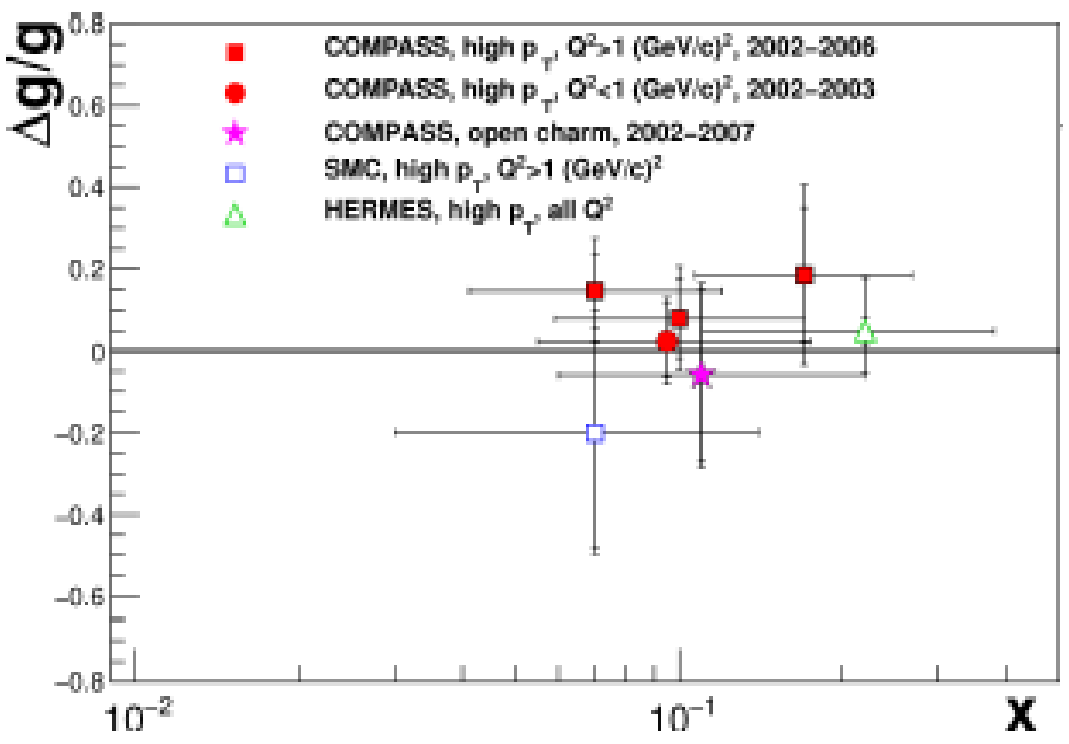

***Fig. 1.6:*** *Direct measurements of the gluon polarization in the nucleon.*

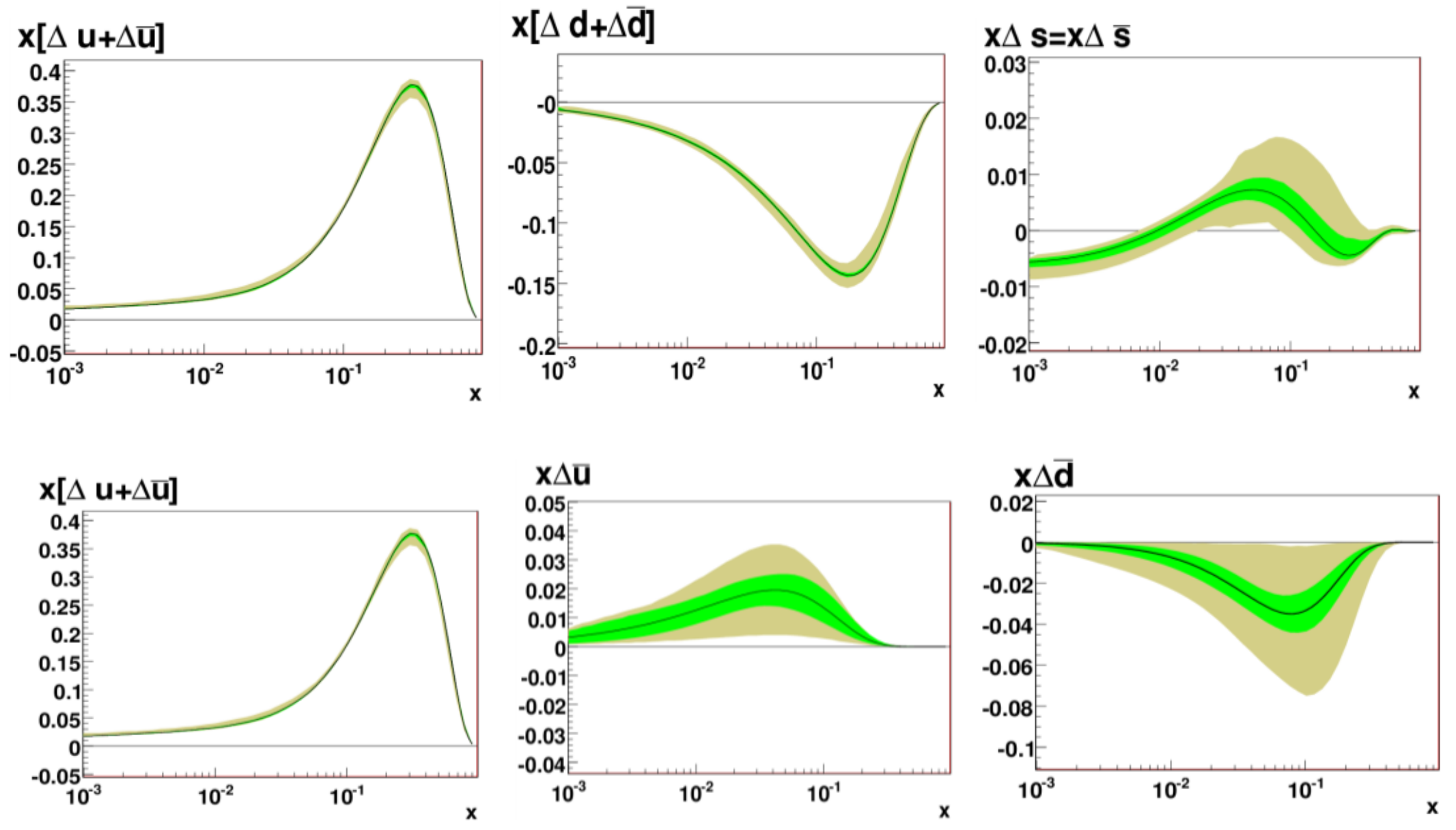

***Fig. 1.7:*** *Parton Helicity distributions in the longitudinally polarized nucleon at $Q^2$=3GeV$^2$ as a function of x.*

### 1.3. The TMD PDFs.

The new TMD PDFs are chiral odd and can be measured only in the SIDIS or DY processes, Fig. 1.8. So far data have been obtained for the polarized nucleon only from SIDIS by the HERMES and COMPASS collaborations. Polarized TMD PDFs from the DY processes in πp interactions are to be measured at COMPASS-II. There is a real opportunity and challenge to study TMD PDFs at NICA in polarized *pp*- and *pd*- collisions (see Section 2.1).

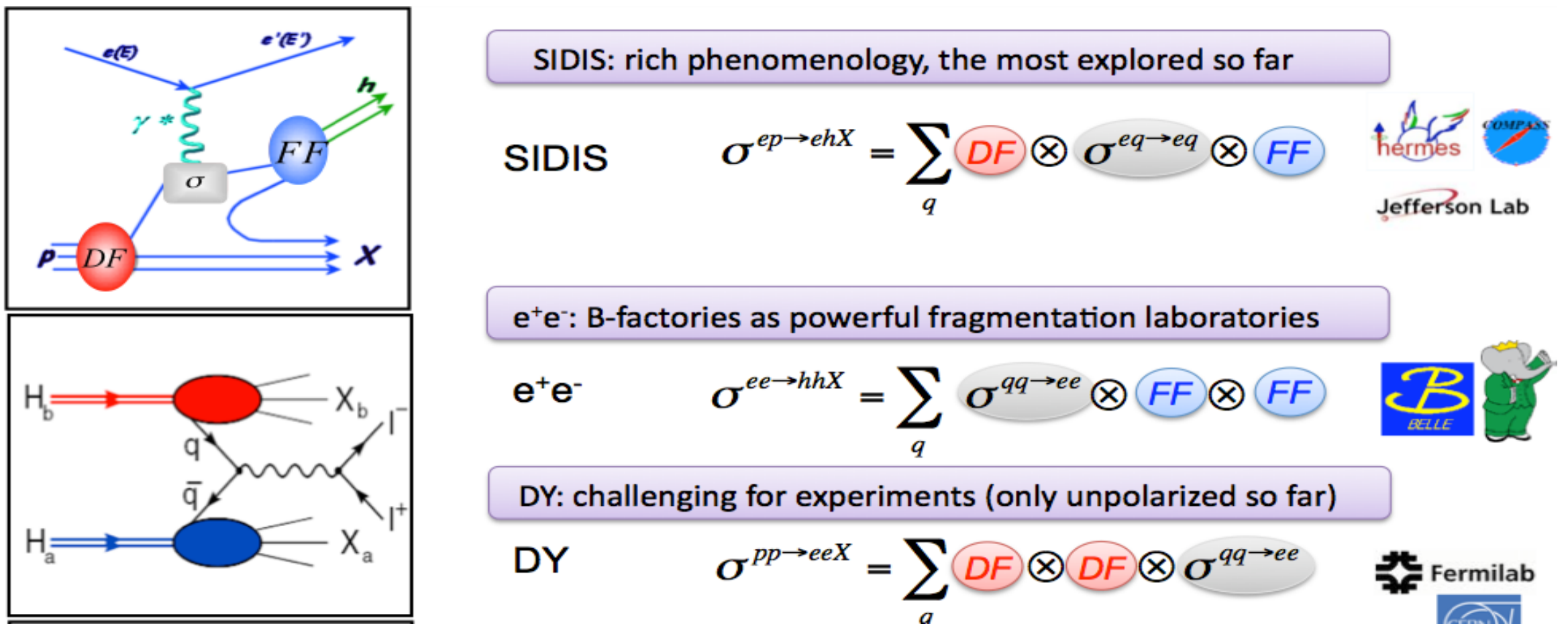

**DY, polarized πp** COMPASS

**DY, polarized pp, pd, dd** NICA

***Fig. 1.8:*** *Reactions for TMD PDF studies.*

In SIDIS, the chiral TMD PDFs can be obtained studying the azimuthal modulations of hadrons which are sensitive to convolution of PDF with the corresponding FF:

- Transversity: $A_{UT}^{sin(\phi_h+\phi_S)} \propto h_1 \otimes H_1^{\perp}$
- Sivers: $A_{UT}^{sin(\phi_h-\phi_S)} \propto f_{1T}^{\perp} \otimes D_1$
- Pretzelosity: $A_{UT}^{sin(3\phi_h-\phi_S)} \propto h_{1T}^{\perp} \otimes H_1^{\perp}$
- Boer-Mulders: $A_{UU}^{cos(2\phi_h)} \propto h_1^{\perp} \otimes H_1^{\perp}$
- Worm-Gears: $A_{UL}^{sin(2\phi_h)} \propto h_{1L}^{\perp} \otimes H_1^{\perp}$; $A_{LT}^{cos(\phi_h-\phi_S)} \propto g_{1T}^{\perp} \otimes D_1$ .

The first and second subscript labeling azimuthal modulations indicates beam and target polarizations, respectively; $\phi_h$ and $\phi_S$ are the azimuthal angles of the produced hadron and initial nucleon spin, defined with respect to the direction of the virtual photon in the lepton scattering plane; $H^{\perp}_1$ is the Collins FF which describes the distribution of non-polarized hadrons in the fragmentation of the transversely polarized quark and $D_1$ is the non-polarized $k_T$ -dependent FF. The Collins FF is chiral-odd; it is a partner of transversity. The status of these PDFs measurement is summarized in [4] and updated in [5].

#### *1.3.1. Transversity PDF $h_1$.*

The azimuthal modulations of hadrons' production in the SIDIS process $l+p(d)\to l+h+X$ with polarized protons and deuterons have been observed by the HERMES and COMPASS collaborations. The proton data are shown in Fig. 1.9. The COMPASS deuteron data on asymmetries are compatible with zero due to cancelations between the *u* and *d* quarks contributions. The Collins FF has been measured recently by the BELLE collaboration at KEK. The global analysis of the HERMES, COMPASS and BELLE data allowed obtaining the transversity distributions for *u* and *d* quarks (Fig. 1.9, right) although still with rather large uncertainties.

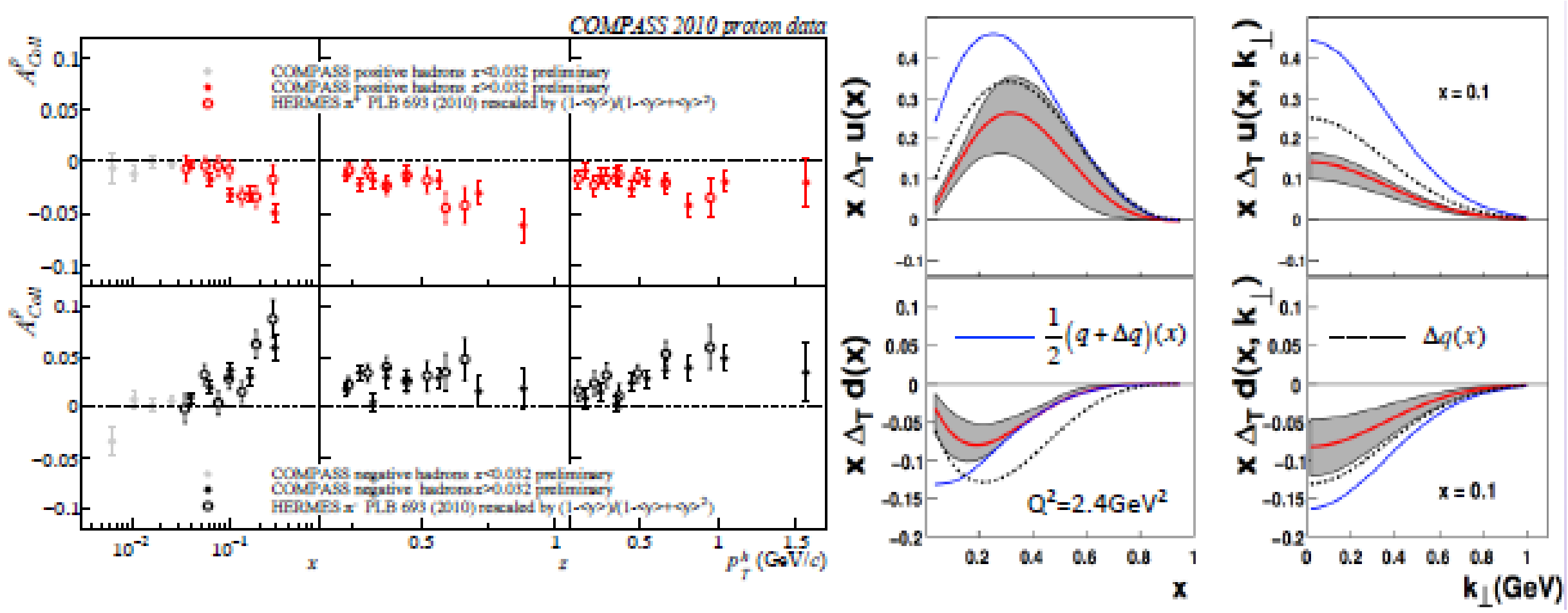

***Fig. 1.9: Left**: Collins asymmetry from COMPASS & HERMES.*
***Right**: Transversity PDFs, $\Delta_T q \equiv h_1 q$, extracted from the global analysis.*

#### *1.3.2. Sivers PDF $f^{\perp}_{1T}$.*

The Sivers correlation between the transverse nucleon spin and transverse momentum of its partons was originally proposed to explain large single-spin asymmetries observed in the hadron productions at Protvino and Fermilab. Later on, possibility of the Sivers effect existence has been confirmed for the Wilson-line TMD PDFs to enforce gauge invariance of QCD. The final state interactions in SIDIS (or initial state interactions in DY) are allowed for the non-zero T-odd Sivers PDFs but they must have opposite signs in SIDIS and DY.

Sivers asymmetries have been measured by the HERMES, COMPASS and JLAB collaborations on proton, deuteron, and $^3$He targets, respectively. Definite signals are observed for protons (Fig. 1.10, left). Because of cancelations between *u* and *d* quark contributions, Sivers asymmetries for the isoscalar targets are compatible with zero. From the global analysis of the HERMES and COMPASS data, the Sivers TMD PDFs for *u* and *d* quarks were determined (Fig. 1.10, right).

### ***1.3.3. Boer-Mulders ($h_1^{\perp}$), Worm-gear-T ($g^{\perp}_{1T}$) and Worm-gear-L ($h^{\perp}_{1L}$) PDFs.***

The Boer-Mulders TMD PDF, like the Sivers one, is T-odd and must have opposite signs once measured in SIDIS or DY. It can be observed (in convolution with the Collins FF) from the $cos(2\phi)$ azimuthal modulation of hadrons produced in the non-polarized SIDIS. Signals of this modulation have been seen by HERMES and COMPASS.

The Worm-gear-T PDF characterizing correlation between longitudinally polarized quarks inside a transversely polarized nucleon is very interesting. It is chiral-even and can be observed in SIDIS convoluted with non-polarized FF studying the $cos(\phi_h - \phi_S)$ modulation in hadron production by longitudinally polarized leptons on the transversely polarized target. Preliminary results were obtained by COMPASS and HERMES (Fig. 1.11).

Attempts to see the Worm-gear-L PDF were made by COMPASS. No signal is observed within the available statistical accuracy.

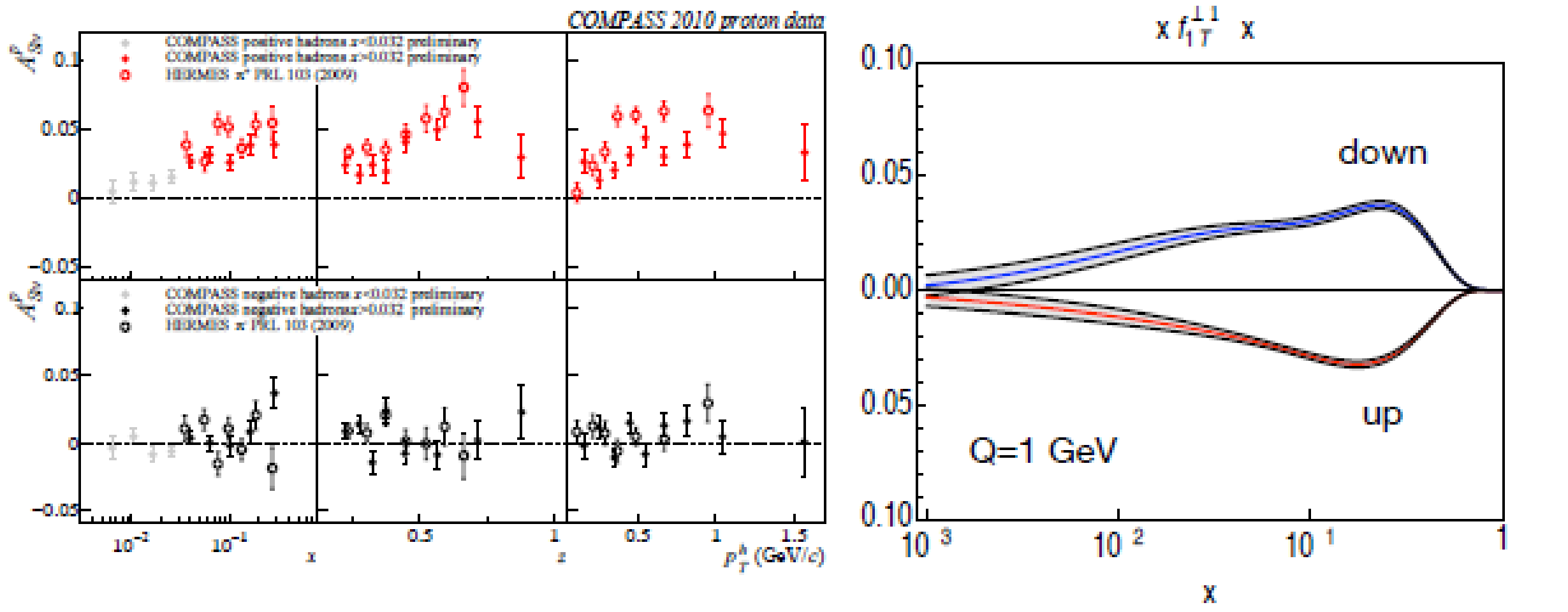

***Fig. 1.10: Left**: Sivers asymmetry from COMPASS and HERMES. **Right**: Sivers PDFs for the u and d quarks determined from the global analysis.*

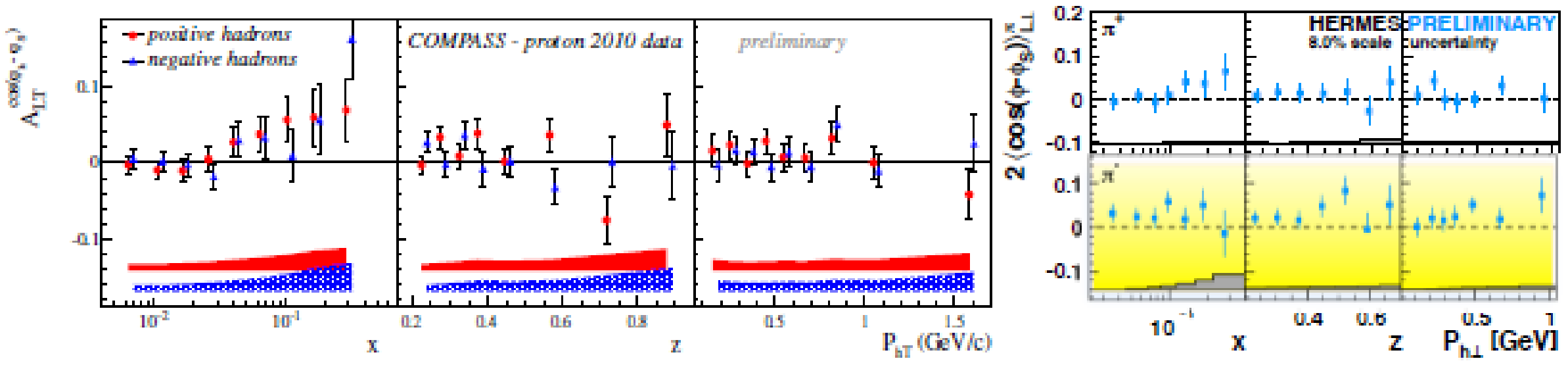

***Fig. 1.11:** Preliminary data on modulations characterizing the Worm-gear-T TMD PDF. **Left**: COMPASS, **right**: HERMES.*

### ***1.3.4. Pretzelosity PDF $h^{\perp}_{1T}$.***

Pretzelosity has been looked for by COMPASS. The $sin(3\phi_h - \phi_S)$ asymmetry modulations in hadrons' production are found to be compatible with zero within the available statistical accuracy. So, no signal of the pretzelosity is observed yet.

**Concluding the Section 1.3, one can summarize that the collinear and TMD PDFs are necessary for complete description of the nucleon structure at the level of twist-2**

**approximation. Its precision measurement at NICA can be the main subject of the NICA SPD spin program.**

### 1.4. Other actual problems of high energy physics.

There are actual problems in high energy physics which are partially solved or not solved at all. Among them one can mention the high-$p_T$ behavior of elastic cross sections (Fig. 1.12), the high-$p_T$ behavior of the asymmetry $A_n$ in elastic $pp$ scattering and inclusive hyperons polarization (Fig. 1.13), the deuteron wave function behavior as a function of $k$, (Fig. 1.14), and some others.

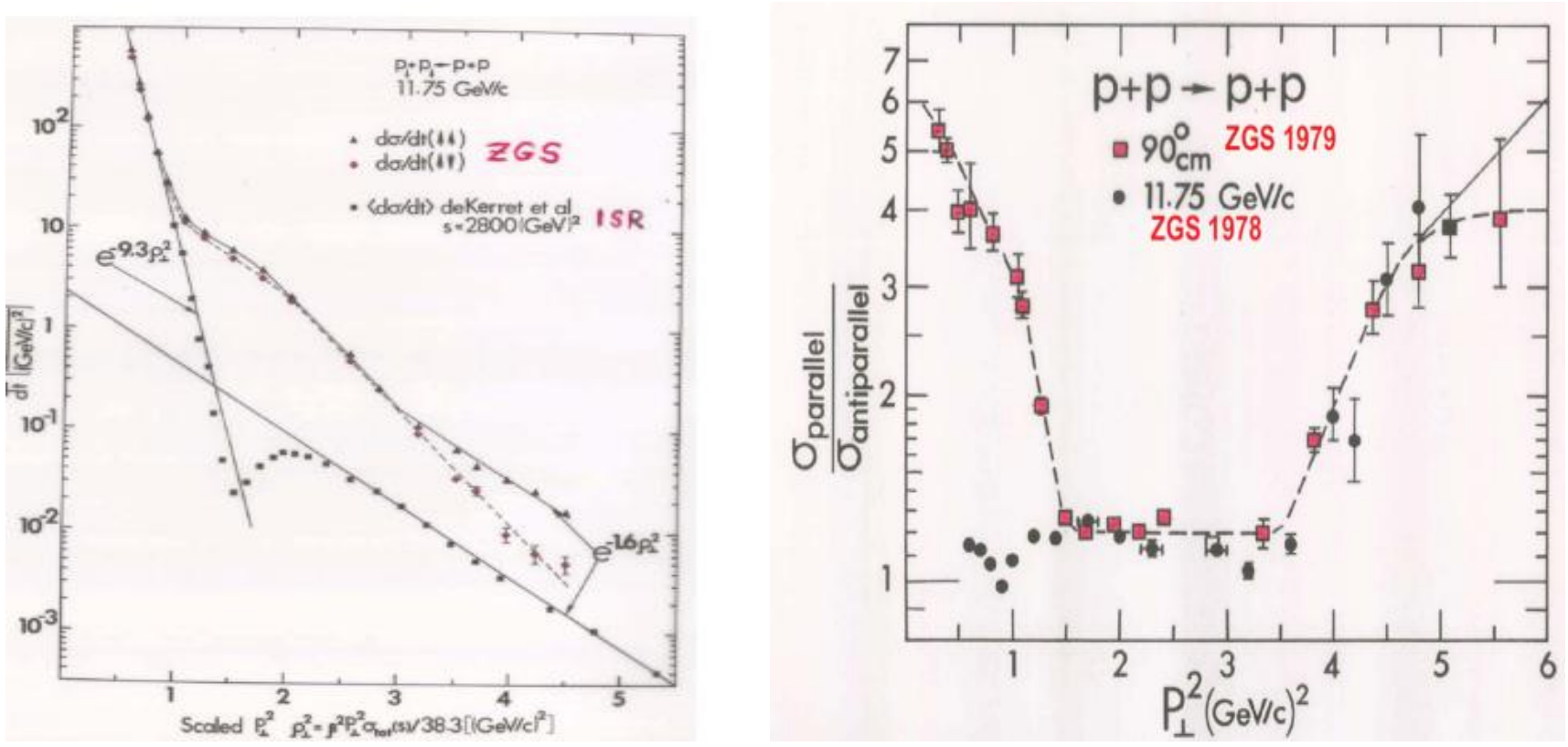

***Fig. 1.12:*** *The famous pp elastic scattering data at large $p_T$.*

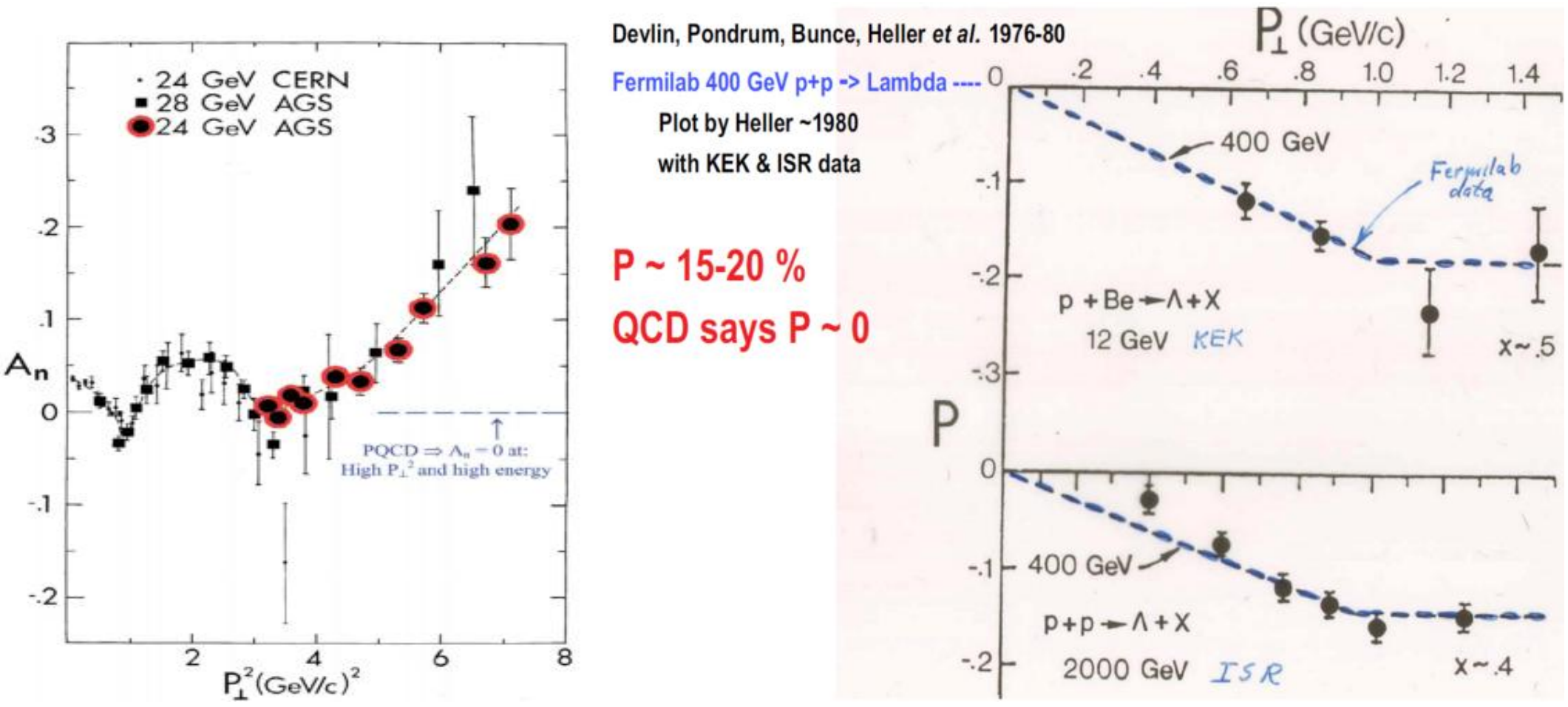

***Fig. 1.13: Right*** *- the Λ hyperons polarization in inclusive pp reactions;*
***left****- asymmetry $A_n$ in pp elastic scattering at high $p_T$.*

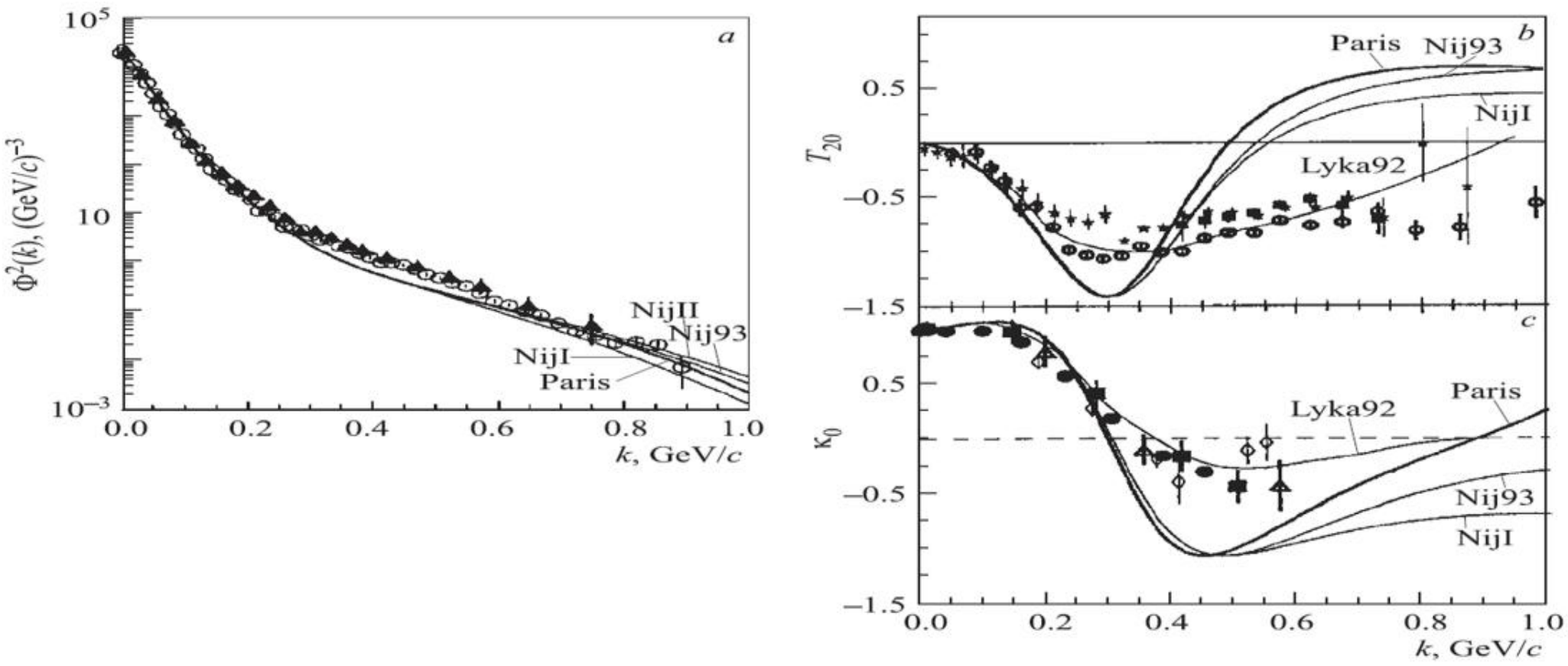

***Fig. 1.14:*** *World data on the deuteron wave function.*

# 2. Physics motivations.

## 2.1. The Drell-Yan mechanism as the second microscope of the nucleon structure studies.

### *2.1.1. The PDFs studies via asymmetry of the DY pairs production cross sections.*

The Drell-Yan (DY) process of the di-lepton production in high-energy hadron-hadron collisions (Fig. 2.1) is playing an important role in the hadron structure studies:

$$H_a(P_a, S_a) + H_b(P_b, S_b) \to l^-(l, \lambda) + l^+(l', \lambda') + X \,, \qquad (2.1.1)$$

where $\boldsymbol{P_a}$ ($\boldsymbol{P_b}$) and $\boldsymbol{S_a}$ ($\boldsymbol{S_b}$) are the momentum and spin of the hadron $H_a$ ($H_b$), respectively, while $\boldsymbol{l}$ ($\boldsymbol{l'}$) and $\boldsymbol{\lambda}$ ($\boldsymbol{\lambda'}$) are the momentum and spin of the lepton, respectively.

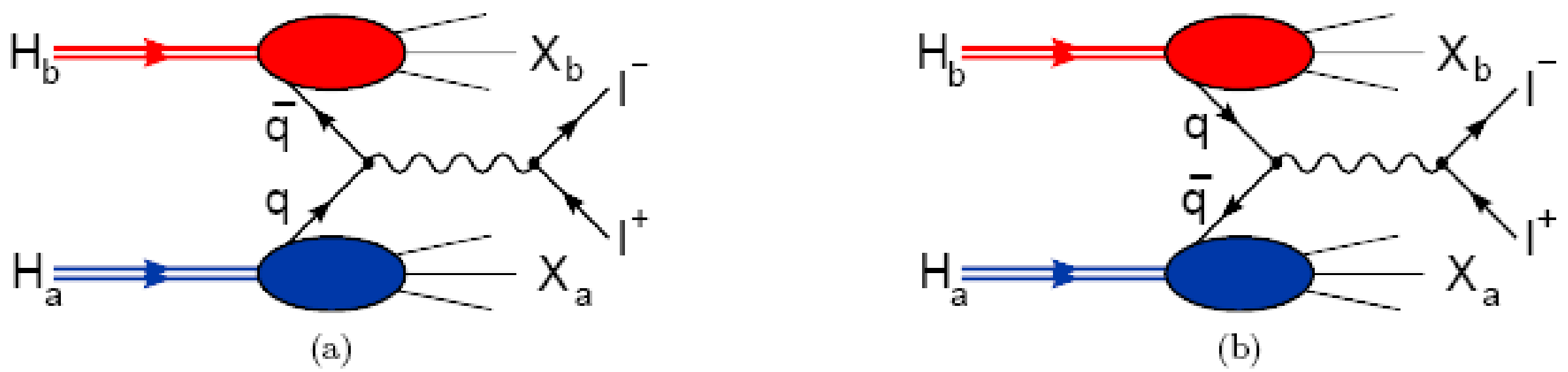

***Fig. 2.1****: The parton model diagrams of the di-lepton production in collisions of hadrons $H_a$ ($P_a$, $S_a$) with hadrons $H_b$ ($P_b$, $S_b$). The constituent quark (anti-quark) of the hadron $H_a$ annihilates with constituent anti-quark (quark) of the hadron $H_b$ producing the virtual photon which decays into a pair of leptons $l^\pm$ (electron-positron or $\mu^\pm$). The hadron spectator systems $X_a$ and $X_b$ are usually not detected. Both diagrams have to be taken into account.*

The kinematics of the Drell-Yan process can be most conveniently considered in the Collins-Soper (CS) reference frame [1-4], Fig. 2.2.

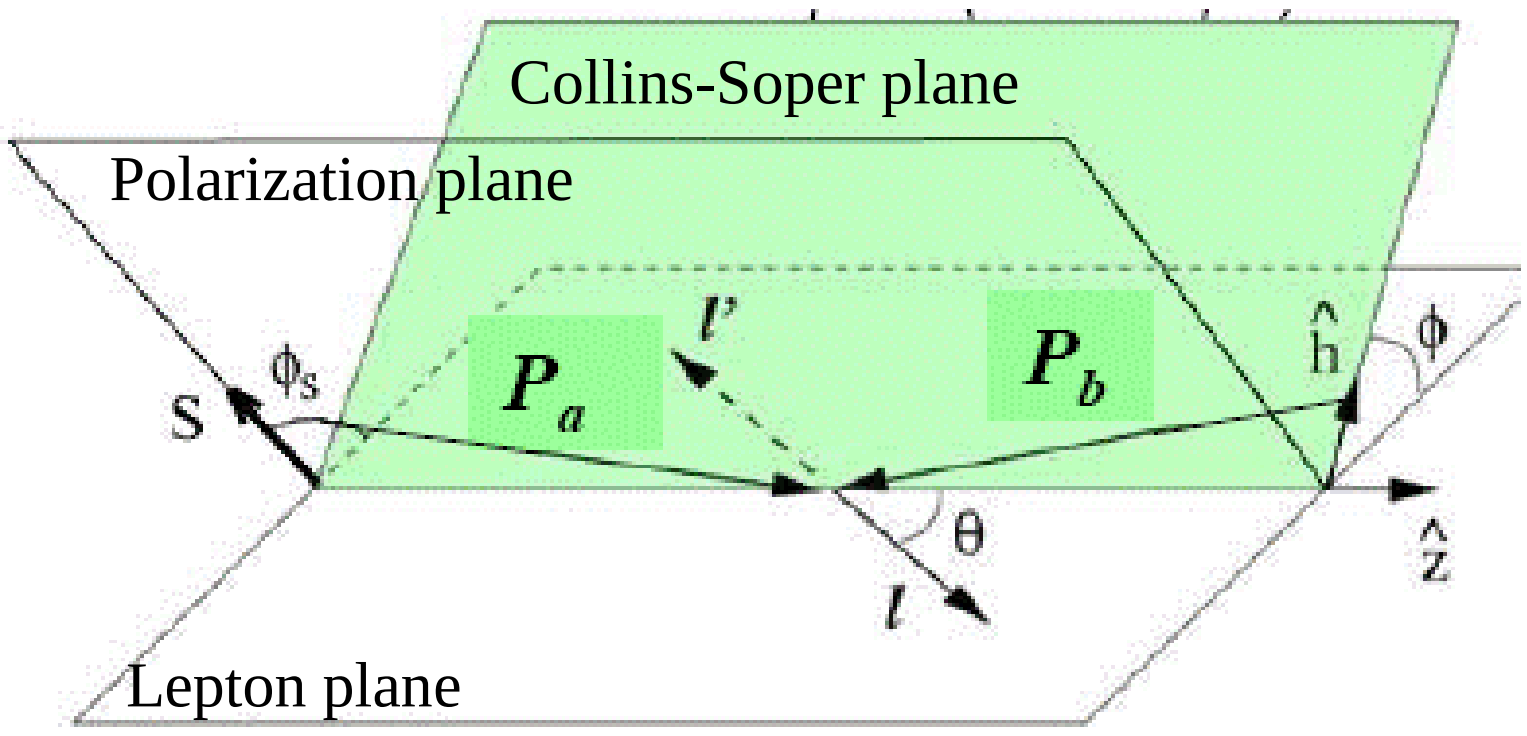

***Fig. 2.2:*** *Kinematics of the Drell-Yan process in the Collins-Soper reference frame.*

The transition from the hadrons-center-of-mass frame (CM-frame) to the CS-frame is described in [1]. The CS-frame includes three intersecting planes. The first one is the *Lepton plane* containing vectors of the lepton momenta, ***l, l'*** (in the lepton rest frame), and the unit vector in the z-direction, $\hat{z} \equiv \hat{e}_{z,CS}$ :

$$\hat{e}_{z,CS} = (\vec{P}_{a,CS} / |\vec{P}_{a,CS}| - \vec{P}_{b,CS} / |\vec{P}_{b,CS}|) / 2\cos\alpha, \quad (\hat{e}_{x,CS} = -(\vec{P}_{a,CS} / |\vec{P}_{a,CS}| + \vec{P}_{b,CS} / |\vec{P}_{b,CS}|) / 2\sin\alpha\,),$$

where tg$\alpha = \boldsymbol{q_T}/q$, $\boldsymbol{q_T}$ ($q = l + l'$, $q \equiv Q$) is the transverse momentum (momentum) of the virtual photon in the CM-frame. The second plane, the *Hadron or Collins-Soper* plane, contains the momentum of colliding hadrons, $P_a$, $P_b$, and vector $\hat{h}$ – is the unit vector in the direction of the photon transverse momentum, $\hat{h} \equiv \vec{h} = \vec{q}_T / q_T$, and the third plane – *Polarization plane* – contains the polarization vector $S \equiv \boldsymbol{S_T}$ **($\boldsymbol{S_{aT}}$, $\boldsymbol{S_{bT}}$)** and the unit vector $\hat{e}_{z,CS}$. The $\phi$ is the azimuthal angle between the *Lepton* and *Hadron* planes; $\phi_S$ (i.e. $\phi_{Sa}$ or $\phi_{Sb}$) is the angle between the *Lepton* and *Polarization* planes and $\theta$ is the polar angle of $\boldsymbol{l}$ in the CS-frame.

The most complete theoretical analysis of this process, for cases when both hadrons $H_a$ and $H_b$, in our case protons or deuterons, are polarized or non-polarized, was performed in [5] which we will follow below. Let us consider the regime where $q_T << q$. In this region the TMD PDFs enter the description of the DY process in a natural way. Our treatment is restricted to the leading twist, i.e. to the leading order of TMDs expansion in powers of *1/q*. Because of the potential problems of the sub-leading-twist -TMD PDFs- factorization pointed out in Refs. [6, 7], we refrain from including in considerations the twist-3 case. Moreover, we neither take into account higher order hard scattering corrections nor effects associated with soft gluon radiation. In this approximation the Eq. (57) of Ref. [5] for the differential cross section of the DY pair's production in the quark-parton model via PDFs is rewritten by us in the more convenient variables with a change of notations of the azimuthal angle polarization vectors as in Fig. 2.2:

$$\frac{d\sigma}{dx_a dx_b d^2 q_T d\Omega} = \frac{\alpha^2}{4Q^2} \times$$

$$\Big\{ \Big( (1+\cos^2\theta) F_{UU}^1 + \sin^2\theta \cos 2\phi F_{UU}^{\cos 2\phi} \Big) + S_{aL} \sin^2\theta \sin 2\phi F_{LU}^{\sin 2\phi} + S_{bL} \sin^2\theta \sin 2\phi F_{UL}^{\sin 2\phi}$$

$$+ \left|\vec{S}_{aT}\right| \Big[ \sin(\phi - \phi_{S_a}) \big(1+\cos^2\theta\big) F_{TU}^{\sin(\phi-\phi_{S_a})} + \sin^2\theta \Big( \sin(3\phi - \phi_{S_a}) F_{TU}^{\sin(3\phi-\phi_{S_a})} + \sin(\phi+\phi_{S_a}) F_{TU}^{\sin(\phi+\phi_{S_a})} \Big) \Big]$$

Lepton plane

$$+ \left|\vec{S}_{bT}\right| \Big[ \sin(\phi - \phi_{S_b}) \big(1+\cos^2\theta\big) F_{UT}^{\sin(\phi-\phi_{S_b})} + \sin^2\theta \Big( \sin(3\phi - \phi_{S_b}) F_{UT}^{\sin(3\phi-\phi_{S_b})} + \sin(\phi+\phi_{S_b}) F_{UT}^{\sin(\phi+\phi_{S_b})} \Big) \Big]$$

$$+ S_{aL} S_{bL} \Big[ \big(1+\cos^2\theta\big) F_{LL}^1 + \sin^2\theta \cos 2\phi F_{LL}^{\cos 2\phi} \Big] \qquad (2.1.2)$$

$$+ S_{aL} \left|\vec{S}_{bT}\right| \Big[ \cos(\phi - \phi_{S_b}) \big(1+\cos^2\theta\big) F_{LT}^{\cos(\phi-\phi_{S_b})} + \sin^2\theta \Big( \cos(3\phi - \phi_{S_b}) F_{LT}^{\cos(3\phi-\phi_{S_b})} + \cos(\phi+\phi_{S_b}) F_{LT}^{\cos(\phi+\phi_{S_b})} \Big) \Big]$$

$$+ \left|\vec{S}_{aT}\right| S_{bL} \Big[ \cos(\phi - \phi_{S_a}) \big(1+\cos^2\theta\big) F_{TL}^{\cos(\phi-\phi_{S_a})} + \sin^2\theta \Big( \cos(3\phi - \phi_{S_a}) F_{TL}^{\cos(3\phi-\phi_{S_a})} + \cos(\phi+\phi_{S_a}) F_{TL}^{\cos(\phi+\phi_{S_a})} \Big) \Big]$$

$$+ \left|\vec{S}_{aT}\right| \left|\vec{S}_{bT}\right| \Big[ \big(1+\cos^2\theta\big) \Big( \cos(2\phi - \phi_{S_a} - \phi_{S_b}) F_{TT}^{\cos(2\phi-\phi_{S_a}-\phi_{S_b})} + \cos(\phi_{S_b} - \phi_{S_a}) F_{TT}^{\cos(\phi_{S_b}-\phi_{S_a})} \Big) \Big]$$

$$+ \left|\vec{S}_{aT}\right| \left|\vec{S}_{bT}\right| \Big[ \sin^2\theta \Big( \cos(\phi_{S_a} + \phi_{S_b}) F_{TT}^{\cos(\phi_{S_a}+\phi_{S_b})} + \cos(4\phi - \phi_{S_a} - \phi_{S_b}) F_{TT}^{\cos(4\phi-\phi_{Sa}-\phi_{S_b})} \Big) \Big]$$

$$+ \left|\vec{S}_{aT}\right| \left|\vec{S}_{bT}\right| \Big[ \sin^2\theta \Big( \cos(2\phi - \phi_{S_a} + \phi_{S_b}) F_{TT}^{\cos(2\phi-\phi_{S_a}+\phi_{S_b})} + \cos(2\phi + \phi_{S_a} - \phi_{S_b}) F_{TT}^{\cos(2\phi+\phi_{S_a}-\phi_{S_b})} \Big) \Big] \Big\}$$

where $F_{jk}^i$ are the Structure Functions (SFs) connected to the corresponding PDFs. The SFs depend on four variables $P_a \cdot q$, $P_b \cdot q$, $\boldsymbol{q_T}$ and $q^2$ or on $\boldsymbol{q_T}$, $q^2$ and the Bjorken variables of colliding hadrons, $x_a$, $x_b$,

$$x_a = \frac{q^2}{2P_a \cdot q} = \sqrt{\frac{q^2}{s}} e^{y}, \quad x_b = \frac{q^2}{2P_b \cdot q} = \sqrt{\frac{q^2}{s}} e^{-y}, \quad y \text{ is the CM rapidity and} \qquad (2.1.3)$$

$s$ is the hadrons-CM-total-energy squared. The SFs $F_{jk}^i$ introduced here give more detailed information on the nucleon structure than usual structure functions depending on two variables $x_{Bj}$ and $Q^2$. Equation (2.1.2) includes 24 leading twist SFs. Each of them is expressed through a weighted convolution, $C$, of corresponding leading twist TMD PDFs :

$$C\Big[ w(\vec{k}_{aT}, \vec{k}_{bT}) f_1 \bar{f}_2 \Big] \equiv \frac{1}{N_c} \sum_q e_q^2 \int d^2\vec{k}_{aT} d^2\vec{k}_{bT} \delta^2(\vec{q}_T - \vec{k}_{aT} - \vec{k}_{bT}) w(\vec{k}_{aT}, \vec{k}_{bT}) \times \qquad (2.1.4)$$

$$\Big[ f_{1q}(x_a, \vec{k}_{aT}^2) \bar{f}_{2q}(x_b, \vec{k}_{bT}^2) + \bar{f}_{1q}(x_a, \vec{k}_{aT}^2) f_{2q}(x_b, \vec{k}_{bT}^2) \Big],$$

where $w$ is a weight, $k_{aT}\,(k_{bT})$ is the transverse momentum of quark (anti-quark) in the hadron $H_a$ ($H_b$) and $f_1$ ($f_2$) is a TMD PDF of the corresponding hadron. The particular SF can include a linear combination of several PDFs. Eventually; one can find expressions for all leading twist SFs of **quarks** and **anti-quarks** entering Eq. (2.1.2). When both hadrons are non-polarized, they are:

$$F_{UU}^{1} = C\left[ f_1 \bar{f}_1 \right], \quad F_{UU}^{\cos 2\phi} = C\left[ \frac{2(\vec{h}\cdot\vec{k}_{aT})(\vec{h}\cdot\vec{k}_{bT}) - \vec{k}_{aT}\cdot\vec{k}_{bT}}{M_a M_b} h_1^{\perp}\bar{h}_1^{\perp} \right], \tag{2.1.5}$$

when only one hadron is polarized (proton or deuteron):

$$F_{LU}^{\sin 2\phi} = C\left[ \frac{2(\vec{h}\cdot\vec{k}_{aT})(\vec{h}\cdot\vec{k}_{bT}) - \vec{k}_{aT}\cdot\vec{k}_{bT}}{M_a M_b} h_{1L}^{\perp}\bar{h}_1^{\perp} \right], \quad F_{UL}^{\sin 2\phi} = -C\left[ \frac{2(\vec{h}\cdot\vec{k}_{aT})(\vec{h}\cdot\vec{k}_{bT}) - \vec{k}_{aT}\cdot\vec{k}_{bT}}{M_a M_b} h_1^{\perp}\bar{h}_{1L}^{\perp} \right],$$

$$F_{UT}^{\sin(\phi-\phi_{S_b})} = C\left[ \frac{\vec{h}\cdot\vec{k}_{bT}}{M_b} f_1 \bar{f}_{1T}^{\perp} \right], \qquad F_{TU}^{\sin(\phi-\phi_{S_a})} = -C\left[ \frac{\vec{h}\cdot\vec{k}_{aT}}{M_a} f_{1T}^{\perp} \bar{f}_1 \right],$$

$$F_{TU}^{\sin(3\phi-\phi_{S_a})} = C\left[ \frac{2(\vec{h}\cdot\vec{k}_{aT})[2(\vec{h}\cdot\vec{k}_{aT})(\vec{h}\cdot\vec{k}_{bT}) - \vec{k}_{aT}\cdot\vec{k}_{bT}] - \vec{k}_{aT}^2(\vec{h}\cdot\vec{k}_{bT})}{2M_a^2 M_b} h_{1T}^{\perp}\bar{h}_1^{\perp} \right], \tag{2.1.6}$$

$$F_{UT}^{\sin(3\phi-\phi_{S_b})} = -C\left[ \frac{2(\vec{h}\cdot\vec{k}_{bT})[2(\vec{h}\cdot\vec{k}_{aT})(\vec{h}\cdot\vec{k}_{bT}) - \vec{k}_{aT}\cdot\vec{k}_{bT}] - \vec{k}_{bT}^2(\vec{h}\cdot\vec{k}_{aT})}{2M_a M_b^2} h_1^{\perp}\bar{h}_{1T}^{\perp} \right],$$

$$F_{TU}^{\sin(\phi+\phi_{S_a})} = C\left[ \frac{\vec{h}\cdot\vec{k}_{bT}}{M_b} h_1 \bar{h}_1^{\perp} \right], \quad F_{UT}^{\sin(\phi+\phi_{S_b})} = -C\left[ \frac{\vec{h}\cdot\vec{k}_{aT}}{M_a} h_1^{\perp} \bar{h}_1 \right],$$

when both hadrons are polarized:

$$F_{LL}^{1} = -C\left[ g_{1L}\bar{g}_{1L} \right], \quad F_{LL}^{\cos 2\phi} = C\left[ \frac{2(\vec{h}\cdot\vec{k}_{aT})(\vec{h}\cdot\vec{k}_{bT}) - \vec{k}_{aT}\cdot\vec{k}_{bT}}{M_a M_b} h_{1L}^{\perp}\bar{h}_{1L}^{\perp} \right],$$

$$F_{LT}^{\cos(\phi-\phi_{S_b})} = -C\left[ \frac{\vec{h}\cdot\vec{k}_{bT}}{M_b} g_{1L}\bar{g}_{1T} \right], \qquad F_{TL}^{\cos(\phi-\phi_{S_a})} = -C\left[ \frac{\vec{h}\cdot\vec{k}_{aT}}{M_a} g_{1T}\bar{g}_{1L} \right],$$

$$F_{TL}^{\cos(\phi+\phi_{S_a})} = C\left[ \frac{\vec{h}\cdot\vec{k}_{bT}}{M_b} h_1 \bar{h}_{1L}^{\perp} \right], \quad F_{LT}^{\cos(\phi+\phi_{S_b})} = C\left[ \frac{\vec{h}\cdot\vec{k}_{aT}}{M_a} h_{1L}^{\perp} \bar{h}_1 \right],$$

$$F_{LT}^{\cos(3\phi-\phi_{S_b})} = C\left[ \frac{2(\vec{h}\cdot\vec{k}_{bT})[2(\vec{h}\cdot\vec{k}_{aT})(\vec{h}\cdot\vec{k}_{bT}) - \vec{k}_{aT}\cdot\vec{k}_{bT}] - \vec{k}_{bT}^2(\vec{h}\cdot\vec{k}_{aT})}{2M_a M_b^2} h_{1L}^{\perp}\bar{h}_{1T}^{\perp} \right],$$

$$F_{TL}^{\cos(3\phi-\phi_{S_a})} = C\left[ \frac{2(\vec{h}\cdot\vec{k}_{aT})[2(\vec{h}\cdot\vec{k}_{aT})(\vec{h}\cdot\vec{k}_{bT}) - \vec{k}_{aT}\cdot\vec{k}_{bT}] - \vec{k}_{aT}^2(\vec{h}\cdot\vec{k}_{bT})}{2M_a^2 M_b} h_{1T}^{\perp}\bar{h}_{1L}^{\perp} \right],$$

$$F_{TT}^{\cos(2\phi-\phi_{S_a}-\phi_{S_b})} = C\left[ \frac{2(\vec{h}\cdot\vec{k}_{aT})(\vec{h}\cdot\vec{k}_{bT}) - \vec{k}_{aT}\cdot\vec{k}_{bT}}{2M_a M_b} (f_{1T}^{\perp}\bar{f}_{1T}^{\perp} - g_{1T}\bar{g}_{1T}) \right], \tag{2.1.7}$$

$$F_{TT}^{\cos(\phi_{S_b}-\phi_{S_a})} = -C\left[ \frac{\vec{k}_{aT}\cdot\vec{k}_{bT}}{2M_a M_b} (f_{1T}^{\perp}\bar{f}_{1T}^{\perp} + g_{1T}\bar{g}_{1T}) \right], \qquad F_{TT}^{\cos(\phi_{S_b}+\phi_{S_a})} = C\left[ h_1 \bar{h}_1 \right],$$

$$F_{TT}^{\cos(2\phi-\phi_{S_a}+\phi_{S_b})} = C\left[ \frac{2(\vec{h}\cdot\vec{k}_{aT})^2 - \vec{k}_{aT}^2}{2M_a^2} h_{1T}^{\perp}\bar{h}_1 \right], \quad F_{TT}^{\cos(2\phi+\phi_{S_a}-\phi_{S_b})} = C\left[ \frac{2(\vec{h}\cdot\vec{k}_{bT})^2 - \vec{k}_{bT}^2}{2M_b^2} h_1 \bar{h}_{1T}^{\perp} \right],$$

$$F_{TT}^{\cos(4\phi-\phi_{Sa}-\phi_{Sb})} = C\left[\left(\frac{4(\vec{h}\cdot\vec{k}_{aT})(\vec{h}\cdot\vec{k}_{bT})[2(\vec{h}\cdot\vec{k}_{aT})(\vec{h}\cdot\vec{k}_{bT})-\vec{k}_{aT}\cdot\vec{k}_{bT}]}{4M_a^2M_b^2} + \frac{\vec{k}_{aT}^2\vec{k}_{bT}^2 - 2\vec{k}_{aT}^2(\vec{h}\cdot\vec{k}_{bT})^2 - 2\vec{k}_{bT}^2(\vec{h}\cdot\vec{k}_{aT})^2}{4M_a^2M_b^2}\right) h_{1T}^{\perp}\bar{h}_{1T}^{\perp}\right].$$

Note that the exchange $H_a \leftrightarrow H_b$ in these expressions leads to the reversal of the z-direction in Fig. 2.2 which, in particular, implies exchanges:

$$\phi_{Sa} \leftrightarrow -\phi_{Sb},\ \phi \to -\phi,\ \theta \to \pi - \theta. \tag{2.1.8}$$

The cross section (2.1.2) cannot be measured directly because there is no single beam containing particles with the *U*, *L* and *T* polarization. To measure SFs entering this equation one can use the following procedure: first, to integrate Eq. (2.1.2) over the azimuthal angle $\phi$, second, following the SIDIS practice, to measure azimuthal asymmetries of the DY pair's production cross sections.

The integration over the azimuthal angle $\phi$ gives:

$$\sigma_{\text{int}} \equiv \frac{d\sigma}{dx_a dx_b d^2q_T d\cos\theta} = \frac{\pi\alpha^2}{2q^2}\times\left(1+\cos^2\theta\right)\Big[F_{UU}^1 + S_{aL}S_{bL}F_{LL}^1 + \left|\vec{S}_{aT}\right|\left|\vec{S}_{bT}\right|\left(\cos(\phi_{S_b}-\phi_{S_a})F_{TT}^{\cos(\phi_{S_b}-\phi_{S_a})} + D\cos(\phi_{S_a}+\phi_{S_b})F_{TT}^{\cos(\phi_{S_a}+\phi_{S_b})}\right)\Big] \tag{2.1.9}$$

The azimuthal asymmetries can be calculated as ratios of cross sections differences to the sum of the integrated over $\phi$ cross sections. The numerator of the ratio is calculated as a difference of the DY pair's production cross sections in the collision of hadrons $H_a$ and $H_b$ with different polarizations. The denominator of the ratio is calculated as a sum of $\sigma_{\text{int}}$ 's calculated for the same hadron polarizations and same $\boldsymbol{x_a}$, $\boldsymbol{x_b}$ regions as in numerator. The azimuthal distribution of DY pair's produced in non-polarized hadron collisions, $A_{UU}$, and azimuthal asymmetries of the cross sections in polarized hadron collisions, $A_{jk}$, are given by Eqs. (2.1.10):

$$A_{UU} \equiv \frac{\sigma^{00}}{\sigma_{\text{int}}^{00}} = \frac{1}{2\pi}(1 + D\cos 2\phi A_{UU}^{\cos 2\phi})$$

$$A_{LU} \equiv \frac{\sigma^{\to 0} - \sigma^{\leftarrow 0}}{\sigma_{\text{int}}^{\to 0} + \sigma_{\text{int}}^{\leftarrow 0}} = \frac{|S_{aL}|}{2\pi} D\sin 2\phi A_{LU}^{\sin 2\phi}$$

$$A_{UL} \equiv \frac{\sigma^{0\to} - \sigma^{0\leftarrow}}{\sigma_{\text{int}}^{0\to} + \sigma_{\text{int}}^{0\leftarrow}} = \frac{|S_{bL}|}{2\pi} D\sin 2\phi A_{UL}^{\sin 2\phi}$$

$$A_{TU} \equiv \frac{\sigma^{\uparrow 0} - \sigma^{\downarrow 0}}{\sigma_{\text{int}}^{\uparrow 0} + \sigma_{\text{int}}^{\downarrow 0}} = \frac{|\vec{S}_{aT}|}{2\pi}\Big[A_{TU}^{\sin(\phi-\phi_{S_a})}\sin(\phi-\phi_{S_a}) + D\left(A_{TU}^{\sin(3\phi-\phi_{S_a})}\sin(3\phi-\phi_{S_a}) + A_{TU}^{\sin(\phi+\phi_{S_a})}\sin(\phi+\phi_{S_a})\right)\Big]$$

$$A_{UT} \equiv \frac{\sigma^{0\uparrow} - \sigma^{0\downarrow}}{\sigma_{\text{int}}^{0\uparrow} + \sigma_{\text{int}}^{0\downarrow}} = \frac{|\vec{S}_{bT}|}{2\pi}\Big[A_{UT}^{\sin(\phi-\phi_{S_b})}\sin(\phi-\phi_{S_b}) + D\left(A_{UT}^{\sin(3\phi-\phi_{S_b})}\sin(3\phi-\phi_{S_b}) + A_{UT}^{\sin(\phi+\phi_{S_b})}\sin(\phi+\phi_{S_b})\right)\Big]$$

$$A_{LL} \equiv \frac{\sigma^{\to\to} + \sigma^{\leftarrow\leftarrow} - \sigma^{\to\leftarrow} - \sigma^{\leftarrow\to}}{\sigma_{\text{int}}^{\to\to} + \sigma_{\text{int}}^{\leftarrow\leftarrow} + \sigma_{\text{int}}^{\to\leftarrow} + \sigma_{\text{int}}^{\leftarrow\to}} = \frac{|S_{aL}S_{bL}|}{2\pi}\left(A_{LL}^1 + DA_{LL}^{\cos 2\phi}\cos 2\phi\right)$$

$$A_{TL} \equiv \frac{\sigma^{\uparrow\to} + \sigma^{\downarrow\leftarrow} - \sigma^{\downarrow\to} - \sigma^{\uparrow\leftarrow}}{\sigma_{\text{int}}^{\uparrow\to} + \sigma_{\text{int}}^{\downarrow\leftarrow} + \sigma_{\text{int}}^{\downarrow\to} + \sigma_{\text{int}}^{\uparrow\leftarrow}} = \frac{|\vec{S}_{aT}\,|S_{bL}|}{2\pi}\left[A_{TL}^{\cos(\phi-\phi_{S_a})}\cos(\phi-\phi_{S_a}) + D\begin{pmatrix} A_{TL}^{\cos(3\phi-\phi_{S_a})}\cos(3\phi-\phi_{S_a}) \\ +A_{TL}^{\cos(\phi+\phi_{S_a})}\cos(\phi+\phi_{S_a}) \end{pmatrix}\right]$$

$$A_{LT} \equiv \frac{\sigma^{\rightarrow\uparrow} + \sigma^{\leftarrow\downarrow} - \sigma^{\rightarrow\downarrow} - \sigma^{\leftarrow\uparrow}}{\sigma_{\rm int}^{\rightarrow\uparrow} + \sigma_{\rm int}^{\leftarrow\downarrow} + \sigma_{\rm int}^{\rightarrow\downarrow} + \sigma_{\rm int}^{\leftarrow\uparrow}} = \frac{S_{aL}\,|\vec{S}_{bT}|}{2\pi}\left[ A_{LT}^{\cos(\phi-\phi_{S_b})}\cos(\phi-\phi_{S_b}) + D\begin{pmatrix} A_{LT}^{\cos(3\phi-\phi_{S_b})}\cos(3\phi-\phi_{S_b}) \\ +A_{LT}^{\cos(\phi+\phi_{S_b})}\cos(\phi+\phi_{S_b}) \end{pmatrix}\right]$$

$$A_{TT} \equiv \frac{\sigma^{\uparrow\uparrow} + \sigma^{\downarrow\downarrow} - \sigma^{\uparrow\downarrow} - \sigma^{\downarrow\uparrow}}{\sigma_{\rm int}^{\uparrow\uparrow} + \sigma_{\rm int}^{\downarrow\downarrow} + \sigma_{\rm int}^{\uparrow\downarrow} + \sigma_{\rm int}^{\downarrow\uparrow}} = \frac{|\vec{S}_{aT}\,\|\,\vec{S}_{bT}|}{2\pi}\Big[ A_{TT}^{\cos(2\phi-\phi_{S_a}-\phi_{S_b})}\cos(2\phi-\phi_{S_a}-\phi_{S_b}) + A_{TT}^{\cos(\phi_{S_b}-\phi_{S_a})}\cos(\phi_{S_b}-\phi_{S_a})$$

$$+D\Big( A_{TT}^{\cos(\phi_{S_b}+\phi_{S_a})}\cos(\phi_{S_a}+\phi_{S_b}) + A_{TT}^{\cos(4\phi-\phi_{S_a}-\phi_{S_b})}\cos(4\phi-\phi_{S_a}-\phi_{S_b})$$

$$+A_{TT}^{\cos(2\phi-\phi_{S_a}+\phi_{S_b})}\cos(2\phi-\phi_{S_a}+\phi_{S_b}) + A_{TT}^{\cos(2\phi+\phi_{S_a}-\phi_{S_b})}\cos(2\phi+\phi_{S_a}-\phi_{S_b})\Big)\Big] \quad . \qquad (2.1.10)$$

In these expressions $D = \sin^2\theta/(1+\cos^2\theta)$ is the depolarization factor and ratio $A_{jk}^{i} = F_{jk}^{i}/F_{UU}^{1}$ of the SFs defined in Eqs. (2.1.5-7). The superscripts of the $\sigma^{pq}$ in (2.1.10) mean:

$\rightarrow(\leftarrow)$ – positive (negative) longitudinal beam polarization in the direction of $\boldsymbol{P}_{a\,cm}$;

$\uparrow(\downarrow)$ – transverse beam polarization with the azimuthal angle $\phi_{Sa}$ or $\phi_{Sb}$ ($\phi_{Sa}+\pi$ or $\phi_{Sb}+\pi$);

0 – non-polarized hadron $H_a$ or $H_b$.

Applying the Fourier analysis to the measured asymmetries, one can separate each of all ratios $A_{jk}^{i} = F_{jk}^{i}/F_{UU}^{1}$ entering Eq. (2.1.10). This will be the ultimate task of the experiments proposed for SPD. The extraction of different TMD PDFs from those ratios is a task of the global theoretical analysis (a challenge for the theoretical community) since each of the SFs $F_{jk}^{i}$ is a result of convolutions of different TMD PDFs in the quark transverse momentum space. For this purpose one needs either to assume a factorization of the transverse momentum dependence for each TMD PDFs, having definite mathematic form (usually Gaussian) with some parameters to be fitted [8], or to transfer $F_{jk}^{i}$ to impact parameter representation space and to use the Bessel weighted TMD PDFs [9].

A number of conclusions can be drawn comparing some asymmetries to be measured. Let us compare the measured asymmetries $A_{LU}$ and $A_{UL}$ and assume that during these measurements the beam polarizations are equal, i.e. $|S_{aL}|=|S_{bL}|$ , and hadrons $a,b$ are identical. Then one can intuitively expect that the integrated over $x_a$ and $x_b$ asymmetries $A_{LU} = A_{UL}$. Similarly, comparing the asymmetries $A_{TU}$ and $A_{UT}$ or $A_{TL}$ and $A_{LT}$ one can expect that $A^1{}_{TU} = A^1{}_{UT}$ and $A^1{}_{TL} = A^1{}_{LT}$ . The tests of these expectations will be a good check of the parton model approximations.

One can close this section with following comments.

**1.** The Structure Functions $F_{jk}^{i}$ depend on the variables ($x_a$, $x_b$, $q_T$ , $q^2$). Instead of $q_T$ one may also work with the transverse momentum of one of the hadrons in the CS-frame.

**2.** Eqs. (2.1.5 - 2.1.7) define 24 SFs out of the 48 [5]. This means that in the considered kinematic region $q_T << q$ there is exactly half of the total number of leading twist SFs.

**3.** The Structure Functions in Eq. (2.1.2) are understood in the CS-frame. Exactly the same expressions for SFs can be obtained in the Gottfried-Jackson (GJ) frame, because difference between values of SFs in CS and GJ frames is of the order of $O$ ($q_T/q$).

**4.** In the $q_T$ -dependent cross section, all the chiral-odd parton distributions disappear after integrating over the azimuthal angle $\phi$. On the other hand, all the chiral-even effects survive this integration.

**5.** The large number of independent SFs to be determined from the polarized DY processes at NICA (24 for identical hadrons in the initial state) is sufficient to map out all eight leading twist TMD PDFs for quarks and anti-quarks. This fact indicates the high potential of the polarized DY process for studying new PDFs. This process has also a certain advantage over SIDIS [10, 11] which also capable of mapping out the leading twist TMD PDFs but requires knowledge of fragmentation functions.

**6.** The transverse single-spin asymmetries depending on the Structure Functions $F_{UT}^{1}$ or $F_{TU}^{1}$ are of the particular interests. The both SFs contain the Sivers PDF which was predicted to have the opposite sign in DY as compared to SIDIS [12, 13, 14]. As the sign reversal is at the core of our present understanding of transverse single spin asymmetries in hard scattering processes, the experimental check of this prediction is of the utmost importance.

**7.** The expected sign reversal of T-odd TMDs can also be investigated through the structure functions $F_{TU}^{\sin(2\phi-\phi_a)}$ or $F_{UT}^{\sin(2\phi-\phi_b)}$ in which the Boer-Mulders PDF enters (see [15, 16, 17] ).

**8.** It is very important to measure those new TMD PDFs which are still not measured or measured with large uncertainties. These are Worm-gear-T, L and Pretzelosity PDFs. The last one would give new information (at least within some models) on the possible role of constituent's orbital momenta in the resolution of the nucleon spin crisis.

**9.** For the complete success of the nucleon structure study program it is mandatory that NICA provides beams of all above mentioned configurations (see also Section 3). The expected effects are of the order of a few percent. So the high luminosity, $\geq 10^{32}$ cm$^{-2}$s$^{-1}$, is necessary to guaranty a corresponding statistical accuracy of measurements.

**10.** As usual, the new facility, i.e. NICA and SPD, prior to measurements of something unknown, should show its potentials measuring already known quantities. So, the program of the nucleon structure study at NICA should start with measurements of non-polarized SFs. Measuring $\sigma^{00}_{\text{int}}$ (Eq. 2.1.9) we could obtain the structure function $F^1_{UU}$ which is proportional to the Density PDF $f_1$ (Eq.2.1.5) – quite well measured in SIDIS experiments. Additionally from measurements of $A_{UU}$ (Eq. 2.1.10) we obtain $F^{\cos 2\phi}_{UU}$ which is proportional to the Boer-Mulders PDF and still poor measured.

**11.** Next step in the program should be measurements of the $A_{LL}$ asymmetry which provide the access to the SFs $F^1_{LL}$ and $F^{\cos 2\phi}_{LL}$. The first one is proportional to the Helicity PDF, well measured in SIDIS, while the second one is proportional to the still unknown Worm-gear-L PDF.

### *2.1.2. Studies of PDFs via integrated asymmetries.*

The set of asymmetries (2.1.10) gives the access to all eight leading twist TMD PDFs. However, sometimes one can work with integrated asymmetries. Integrated asymmetries are useful for the express analysis of data and checks of expected relations between asymmetries mentioned in Section 2.1. They are also useful for model estimations and determination of required statistics (see Section 6.2). Let us consider several examples starting from the case when only one of colliding hadrons (for instance, hadron "*b*") is transversely polarized. In this case the DY cross section Eq. (2.1.2) with SFs given by Eq. (2.1.6) is reduced to the expression (2.1.11):

$$\begin{aligned}
\frac{d\sigma}{dx_a dx_b d^2\mathbf{q}_T d\Omega} &= \frac{\alpha^2}{4Q^2}\Big\{(1+\cos^2\theta)\, C\big[f_1 \bar{f}_1\big] \\
&+\sin^2\theta\cos 2\phi\, C\left[\frac{2(\vec{h}\cdot\vec{k}_{aT})(\vec{h}\cdot\vec{k}_{bT})-\vec{k}_{aT}\cdot\vec{k}_{bT}}{M_a M_b} h_1^\perp \bar{h}_1^\perp\right] \\
&+|S_{bT}|\Big[(1+\cos^2\theta)\sin(\phi-\phi_{S_b})\, C\left[\frac{\vec{h}\cdot\vec{k}_{bT}}{M_b} f_1 \bar{f}_{1T}^\perp\right]-\sin^2\theta\sin(\phi+\phi_{S_b})\, C\left[\frac{\vec{h}\cdot\vec{k}_{aT}}{M_a} h_1^\perp \bar{h}_1\right] \\
&-\sin^2\theta\sin(3\phi-\phi_{S_b})\, C\left[\frac{2(\vec{h}\cdot\vec{k}_{bT})[2(\vec{h}\cdot\vec{k}_{aT})(\vec{h}\cdot\vec{k}_{bT})-\vec{k}_{aT}\cdot\vec{k}_{bT}]-\vec{k}_{bT}^2(\vec{h}\cdot\vec{k}_{aT})}{2M_a M_b^2} h_1^\perp \bar{h}_{1T}^\perp\right]\Big]\Big\},
\end{aligned} \tag{2.1.11}$$

which, being integrated over $\phi_{Sb}$, allows to construct the weighted asymmetries given by Eqs. (2.1.12) where $\phi_{S_b} \equiv \phi_S$ (the weight function ($w$) is shown in the superscript of the asymmetry). They provide access to the Boer-Mulders, Sivers, and Pretzelosity TMD PDFs.

The integrated and additionally $q_T$-weighted asymmetries $A_{UT}^{w\left[\sin(\phi+\phi_S)\frac{q_T}{M_N}\right]}$ and $A_{UT}^{w\left[\sin(\phi-\phi_S)\frac{q_T}{M_N}\right]}$ given by Eqs. (2.1.13-14) provide access to the first moments of the Boer-Mulders, $h_{1q}^{\perp}(x,k_T^2)$, and Sivers, $f_{q1T}^{\perp(1)}(x,k_T^2)$, PDFs given by Eqs. (2.1.15).

$$A_{UT}^{w[\sin(\phi+\phi_S)]} = \frac{\int d\Omega\, d\phi_S \sin(\phi+\phi_S)\left[d\sigma^{\uparrow}-d\sigma^{\downarrow}\right]}{\int d\Omega d\phi_S\left[d\sigma^{\uparrow}+d\sigma^{\downarrow}\right]/2} = -\frac{1}{2}\frac{C\left[\frac{\vec{h}\cdot\vec{k}_{aT}}{M_a}h_1^{\perp}\bar{h}_1\right]}{C\left[f_1\bar{f}_1\right]},$$

$$A_{UT}^{w[\sin(\phi-\phi_S)]} = \frac{\int d\Omega\, d\phi_S \sin(\phi-\phi_S)\left[d\sigma^{\uparrow}-d\sigma^{\downarrow}\right]}{\int d\Omega d\phi_S\left[d\sigma^{\uparrow}+d\sigma^{\downarrow}\right]/2} = \frac{1}{2}\frac{C\left[\frac{\vec{h}\cdot\vec{k}_{bT}}{M_b}f_1\bar{f}_{1T}^{\perp}\right]}{C\left[f_1\bar{f}_1\right]}, \qquad (2.1.12)$$

$$A_{UT}^{w[\sin(3\phi-\phi_S)]} = \frac{\int d\Omega\, d\phi_S \sin(3\phi-\phi_S)\left[d\sigma^{\uparrow}-d\sigma^{\downarrow}\right]}{\int d\Omega d\phi_S\left[d\sigma^{\uparrow}+d\sigma^{\downarrow}\right]/2} =$$

$$= -\frac{1}{2}\frac{C\left[\frac{2(\vec{h}\cdot\vec{k}_{bT})[2(\vec{h}\cdot\vec{k}_{aT})(\vec{h}\cdot\vec{k}_{bT})-\vec{k}_{aT}\cdot\vec{k}_{bT}]-\vec{k}_{bT}^{2}(\vec{h}\cdot\vec{k}_{aT})}{2M_aM_b^2}h_1^{\perp}\bar{h}_{1T}^{\perp}\right]}{C\left[f_1\bar{f}_1\right]}.$$

$$A_{UT}^{w\left[\sin(\phi+\phi_S)\frac{q_T}{M_N}\right]} = \frac{\int d\Omega\int d^2\mathbf{q}_T(|\mathbf{q}_T|/M_p)\sin(\phi+\phi_S)\left[d\sigma^{\uparrow}-d\sigma^{\downarrow}\right]}{\int d\Omega\int d^2\mathbf{q}_T\left[d\sigma^{\uparrow}+d\sigma^{\downarrow}\right]/2}$$

$$= -\frac{\sum_q e_q^2\left[\bar{h}_{1q}^{\perp(1)}(x_p)h_{1q}(x_{p\uparrow})+(q\leftrightarrow\bar{q})\right]}{\sum_q e_q^2\left[\bar{f}_{1q}(x_p)f_{1q}(x_{p\uparrow})+(q\leftrightarrow\bar{q})\right]}, \qquad (2.1.13)$$

$$A_{UT}^{w\left[\sin(\phi-\phi_S)\frac{q_T}{M_N}\right]} = \frac{\int d\Omega\int d^2\mathbf{q}_T(|\mathbf{q}_T|/M_p)\sin(\phi-\phi_S)\left[d\sigma^{\uparrow}-d\sigma^{\downarrow}\right]}{\int d\Omega\int d^2\mathbf{q}_T\left[d\sigma^{\uparrow}+d\sigma^{\downarrow}\right]/2}$$

$$= 2\frac{\sum_q e_q^2\left[f_{1T}^{\perp(1)q}(x_{p\uparrow})f_{1q}(x_p)+(q\leftrightarrow\bar{q})\right]}{\sum_q e_q^2\left[\bar{f}_{1q}(x_{p\uparrow})f_{1q}(x_p)+(q\leftrightarrow\bar{q})\right]}, \qquad (2.1.14)$$

where

$$h_{1q}^{\perp(1)}(x) = \int d^2k_T\left(\frac{k_T^2}{2M_p^2}\right)h_{1q}^{\perp}(x_p,k_T^2) \text{ and } f_{q1T}^{\perp(1)}(x) = \int d^2k_T\left(\frac{k_T^2}{2M_p^2}\right)f_{q1T}^{\perp(1)}(x,k_T^2). \qquad (2.1.15)$$

For the *pp* collisions, there are two limiting cases when one can neglect contributions to the asymmetries from sea part of PDFs either of polarized or non-polarized protons. The first case corresponds to the region of $x_{Bj}$ values where $x_{unpol} >> x_{pol}$ while the second one-- to the region $x_{unpol} << x_{pol}$ . In these cases one can obtain the approximate expressions for asymmetries (2.1.13-14) which are given by Eqs. (2.1.16-17)

$$A_{UT}^{w\left[\sin(\phi-\phi_S)\frac{q_T}{M_N}\right]}\Bigg|_{x_p>>x_{p\uparrow}} \approx 2\frac{\bar{f}_{1uT}^{\perp(1)}(x_{p\uparrow})}{\bar{f}_{1u}(x_{p\uparrow})} \quad ; \quad A_{UT}^{w\left[\sin(\phi+\phi_S)\frac{q_T}{M_N}\right]}\Bigg|_{x_p>>x_{p\uparrow}} \approx -\frac{h_{1u}^{\perp(1)}(x_p)\bar{h}_{1u}(x_{p\uparrow})}{f_{1u}(x_p)\bar{f}_{1u}(x_{p\uparrow})} \quad ; (2.1.16)$$

$$A_{UT}^{w\left[\sin(\phi-\phi_S)\frac{q_T}{M_N}\right]}\Bigg|_{x_p<<x_{p\uparrow}} \approx 2\frac{f_{1uT}^{\perp(1)}(x_{p\uparrow})}{f_{1u}^{\perp(1)}(x_{p\uparrow})} \quad ; \quad A_{UT}^{w\left[\sin(\phi+\phi_S)\frac{q_T}{M_N}\right]}\Bigg|_{x_p<<x_{p\uparrow}} \approx -\frac{\bar{h}_{1u}^{\perp(1)}(x_p)h_{1u}(x_{p\uparrow})}{\bar{f}_{1u}(x_p)f_{1u}(x_{p\uparrow})}. \qquad (2.1.17)$$

So far the *pp*-collisions have been considered. At NICA the *pd*- and *dd*-collisions will be investigated as well. As it is known from COMPASS experiment, the SIDIS asymmetries on polarized deuterons are consisted with zero. At NICA one can expect that asymmetries $A_{UT}^{w\left[\sin(\phi\pm\phi_S)\frac{q_T}{M_N}\right]}\Big|_{pD^{\uparrow}}$ , $A_{UT}^{w\left[\sin(\phi\pm\phi_S)\frac{q_T}{M_N}\right]}\Big|_{DD^{\uparrow}}$ also will be consistent with zero (subject of tests).

But asymmetries in $Dp\uparrow$ collisions are expected to be non-zero. In the limiting cases $x_D >> x_{p\uparrow}$ and $x_D << x_{p\uparrow}$ these asymmetries (**accessible only at NICA** ) are given by expressions (2.1.18).

$$A_{UT}^{w\left[\sin(\phi-\phi_S)\frac{q_T}{M_N}\right]}(x_D >> x_{p\uparrow})\Bigg|_{Dp\uparrow\to l^+l^-X} \approx \frac{4\bar{f}_{1uT}^{\perp(1)}(x_{p\uparrow})+\bar{f}_{1dT}^{\perp(1)}(x_{p\uparrow})}{4\bar{f}_{1u}^{\perp(1)}(x_{p\uparrow})+\bar{f}_{1d}^{\perp(1)}(x_{p\uparrow})},$$

$$A_{UT}^{w\left[\sin(\phi-\phi_S)\frac{q_T}{M_N}\right]}(x_D << x_{p\uparrow})\Bigg|_{Dp\uparrow\to l^+l^-X} \approx 2\frac{4f_{1uT}^{\perp(1)}(x_{p\uparrow})+f_{1dT}^{\perp(1)}(x_{p\uparrow})}{4f_{1u}^{\perp(1)}(x_{p\uparrow})+f_{1d}^{\perp(1)}(x_{p\uparrow})}, \qquad (2.1.18)$$

$$A_{UT}^{w\left[\sin(\phi+\phi_S)\frac{q_T}{M_N}\right]}(x_D >> x_{p\uparrow})\Bigg|_{Dp\uparrow\to l^+l^-X} \approx -\frac{[h_{1u}^{\perp(1)}(x_D)+h_{1d}^{\perp(1)}(x_D)][4\bar{h}_{1u}(x_{p\uparrow})+\bar{h}_{1d}(x_{p\uparrow})]}{[f_{1u}(x_D)+f_{1d}(x_D)][4\bar{f}_{1u}(x_{p\uparrow})+\bar{f}_{1d}(x_{p\uparrow})]},$$

$$A_{UT}^{w\left[\sin(\phi+\phi_S)\frac{q_T}{M_N}\right]}(x_D << x_{p\uparrow})\Bigg|_{Dp\uparrow\to l^+l^-X} \approx -\frac{[\bar{h}_{1u}^{\perp(1)}(x_D)+\bar{h}_{1d}^{\perp(1)}(x_D)][4h_{1u}(x_{p\uparrow})+h_{1d}(x_{p\uparrow})]}{[\bar{f}_{1u}(x_D)+\bar{f}_{1d}(x_D)][4f_{1u}(x_{p\uparrow})+f_{1d}(x_{p\uparrow})]}.$$

In collisions of transversely polarized hadrons, instead of complicated analysis of the $A_{TT}$ asymmetry given by Eq. (2.1.10), the direct access to the transversity PDF $h_1$ one can have via the weighted asymmetry, $A^{w[\cos(\phi_{Sb}+\phi_{Sa})q_T/M]}$, integrated over the angles $\phi_{Sb}$ and $\phi_{Sa}$:

$$A_{TT}^{w[\cos(\phi_{Sb}+\phi_{Sa})q_T/M]} \equiv A_{TT}^{\text{int}} = \frac{\sum_q e_q^2\left(\bar{h}_{1q}(x_1)h_{1q}(x_2)+(x_1\leftrightarrow x_2)\right)}{\sum_q e_q^2\left(\bar{f}_{1q}(x_1)f_{1q}(x_2)+(x_1\leftrightarrow x_2)\right)}. \qquad (2.1.19)$$

The method of integrated asymmetries requires calculations of corresponding cross sections prior to their integration. It means that the detector acceptance and collider luminosity should be under perfect control.

## 2.2. New nucleon PDFs and *J/Ψ* production mechanisms.

The *J/Ψ*-meson, a bound state of charm and anti-charm quarks, was discovered in 1974 at BNL [18] and SLAC [19]. The production and binding mechanisms of these two quarks are still not completely known. It is important to note that some of *J/Ψ* mesons observed so far are not directly produced from collisions but are the result of decays of other charmonium states. Recently it has been estimated that 30 ± 10 % of *J/Ψ* mesons come from $\chi_c$ decays, and 59 ±10 % of them are produced directly [20]. The *J/Ψ* production mechanism, included in the PYTHIA simulation code and intended for collider applications, considers two approaches: "color singlet" and "color octet" ones. The "color singlet" approach considers *gg* fusion processes, while "color octet" considers $gg$, $gq$, $q\bar{q}$ and $g\bar{q}$ processes. According to PYTHIA [21], the cross section of the $J/\Psi$ production in $pp$-collisions at $\sqrt{s}$=24 GeV via $gg$ processes (singlet and octet) and $gq$ plus $qq$ processes are about equal (~53 and ~50 nb, respectively). Some of the $g\bar{q}$ and $q\bar{q}$ processes can proceed via various charmonium states subsequently decaying into $J/\Psi$. So, these processes could be sensitive to the TMD PDFs. It is interesting to note that the $g\bar{q}$ processes have the largest cross sections (see the Table 1 in Appendix 1).

The production of lepton pairs in the $q\bar{q}$ processes, via $J/\Psi$ with it subsequent decay into leptons, $H_a + H_b \rightarrow J/\Psi + X \rightarrow l^+ + l^- + X$, is analogous to the DY production mechanism of Eq. (2.1.1). The analogy is correct if the $J/\Psi$ interactions with quarks and leptons are of the vector type. This analogy is known under the name "duality model" [22, 23]. For the TMD PDF studies, the "duality model" can predict [24] a similar behavior of asymmetries $A^i_{jk} = F^i_{jk} / F^1_{UU}$ in the lepton pair's production calculated via DY (Eq. (2.1.10)) and via $J/\Psi$ events. This similarity follows from the duality model idea to replace the coupling $e_q^2$ in the convolutions for $F^i_{jk}$ (Eq.2.1.4) by $J/\Psi$ vector coupling with $q\bar{q}$ $(g_q^V)^2$. The vector couplings are expected to be the same for $u$ and $d$ quarks [22] and cancel in the ratios $A^i_{jk} = F^i_{jk} / F^1_{UU}$ for large $x_a$ or $x_b$. For instance, we can compare the Sivers asymmetry $A_{UT}^{w\left[\sin(\phi-\phi_S)\frac{q_T}{M_N}\right]}$ given in the DY case by Eq. (2.1.14) with the same asymmetry given in $J/\Psi$ case by Eq. (2.1.14) with omitted quark charges. At NICA such a comparison can be performed at various colliding beam energies.

### 2.3. Direct photons.

Direct photon productions in the non-polarized and polarized $pp$ $(pd)$ reactions provide information on the gluon distributions in nucleons (Fig. 2.3). There are two main hard processes where direct photons can be produced: gluon Compton scattering, $g + q \rightarrow \gamma + X$, and quark-anti-quark annihilation, $q + \bar{q} \rightarrow \gamma + X$. As it has been pointed out in [25], "the direct photon production in non polarized $pp$ collisions can provide a clear test of short-distance dynamics as predicted by the perturbative QCD, because the photon originates in the hard scattering sub-process and does not fragment. This immediately means that Collins effect is not present. The process is very sensitive to the non polarized gluon structure function, since it is dominated by quark-gluon Compton sub process in a large photon transverse momentum range".

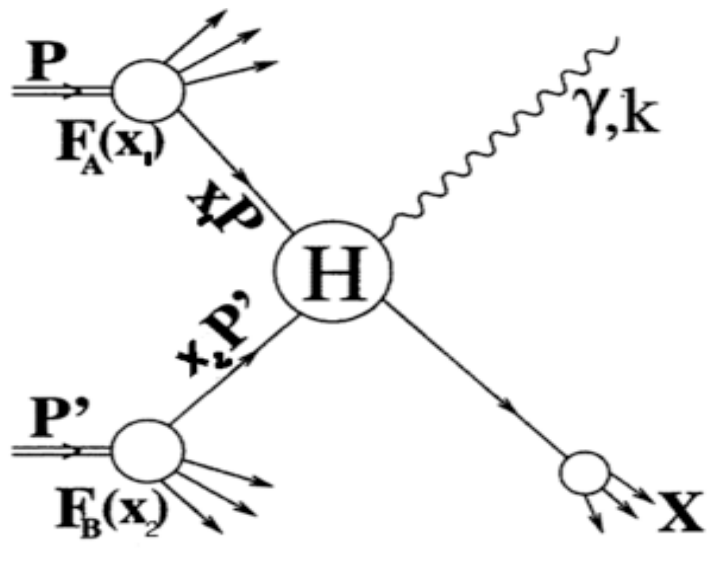

***Fig. 2.3***: *Diagram of the direct photon production. Vertex H corresponds to* $q + \bar{q} \rightarrow \gamma + g$ *or* $g + q \rightarrow \gamma + q$ *hard processes.*

The non- polarized cross section for production of a photon with the transverse momentum $p_T$ and rapidity $y$ in the reaction $p + p \rightarrow \gamma + X$ is written [25] as follows:

$$d\sigma = \sum_i \int_{x_{min}}^{1} dx_a \int d^2\mathbf{k}_{Ta} d^2\mathbf{k}_{Tb} \frac{x_a x_b}{x_a - (p_T/\sqrt{s})e^y} \left[ q_i(x_a, \mathbf{k}_{Ta}) G(x_b, \mathbf{k}_{Tb}) \times \frac{d\hat{\sigma}}{d\hat{t}}(q_i G \to q_i \gamma) + G(x_a, \mathbf{k}_{Ta})\ q_i(x_b, \mathbf{k}_{Tb}) \frac{d\hat{\sigma}}{d\hat{t}}(G q_i \to q_i \gamma) \right], \qquad (2.3.1)$$

where $k_{Ta}$ ( $k_{Tb}$ ) is the transverse momentum of the interacting quark (gluon), $x_a$ ($x_b$) is the fraction of the proton momentum carried by them and $q_i(x, k_T)$ , [$G(x, k_T)$] is the quark (gluon) distribution function with the specified $k_T$ [25]. The total cross section of the direct photon production in the *pp*-collision at √s=24 GeV via the Compton scattering (according to PYTHIA 6.4) is equal to1100 nb, while the cross section of the $q\bar{q}$ annihilation is about 200 nb. So, the gluon Compton scattering is the main mechanism of the direct photon production. One can show [25], that the above expression can be used also for extraction of the polarized gluon distribution (Sivers gluon function) from measurement of the transverse single spin asymmetry $A_N$ defined as follows:

$$A_N = \frac{\sigma^{\uparrow} - \sigma^{\downarrow}}{\sigma^{\uparrow} + \sigma^{\downarrow}} \qquad (2.3.2)$$

Here $\sigma\uparrow$ and $\sigma\downarrow$ are the cross sections of the direct photon production for the opposite transverse polarizations of one of the colliding protons. In [26] it has been pointed out that the asymmetry $A_N$ at large positive $x_F$ is dominated by quark-gluon correlations while at large negative $x_F$ [27] it is dominated by pure gluon-gluon correlations. The further development of the corresponding formalism can be found in [28], [29].

Predictions for the value of $A_N$ at $\sqrt{s}$ = 30 GeV, $p_T$ = 4 GeV/c can be found in [28] for negative $x_F$ (Fig. 2.4 (left)) and in [26] for positive $x_F$ (Fig. 2.4 (right)). In both cases the $A_N$ values remain sizable.

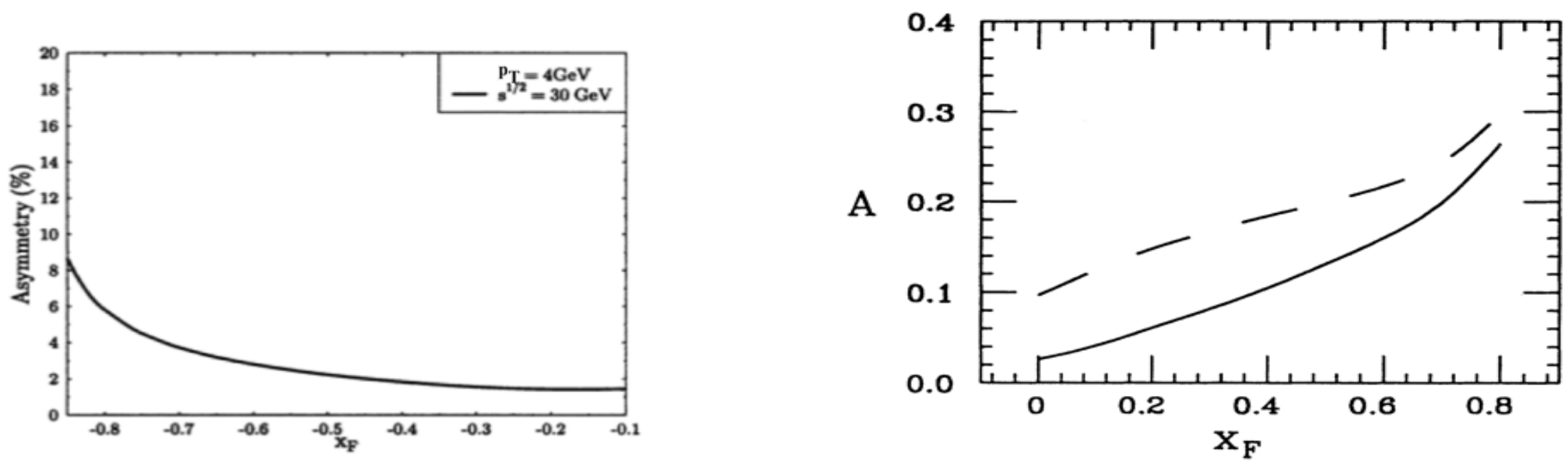

***Fig. 2.4***: *Predictions for $A_N$ at √s=30 GeV, $p_T$=4 GeV/c: from [28](left), from [26] (right).*

The first attempt to measure $A_N$ at $\sqrt{s}$=19.4 GeV was performed in the fixed target experiment E704 at Fermilab [30] in the kinematic range -0.15<$x_F$<0.15 and 2.5<$p_T$<3.1 GeV/c. Results are consistent with zero within large statistical and systematic uncertainties (Fig. 2.5).

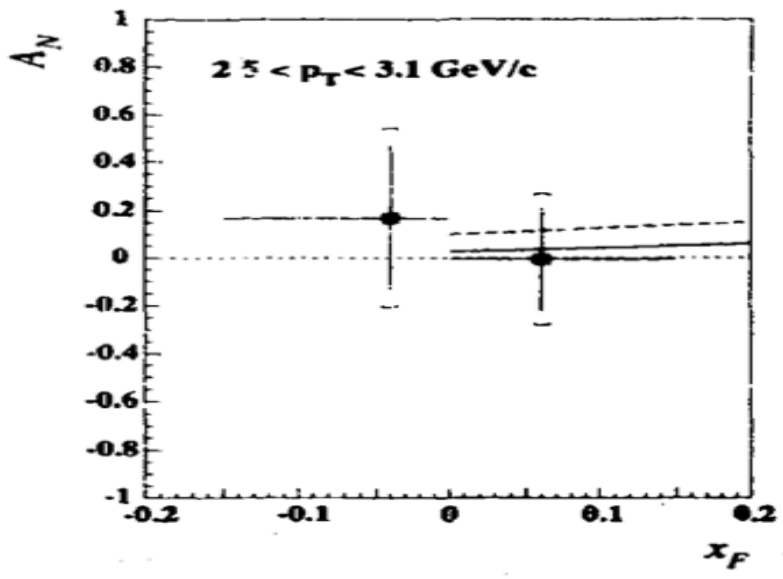

***Fig. 2.5:*** *The single transverse spin asymmetry $A_N$ measured in the E704 experiment. Curves are predictions of [26].*

The single spin asymmetries in the direct photon production will be measured also by PHENIX [31] and STAR [32] at RHIC.

Production of direct photons at large transverse momentum with longitudinally polarized proton beams is a very promising method to measure gluon polarization $\Delta g$ [33]. Longitudinal double spin asymmetry $A_{LL}$, defined as:

$$A_{LL} = \frac{(\sigma_{++} + \sigma_{--}) - (\sigma_{+-} + \sigma_{-+})}{(\sigma_{++} + \sigma_{--}) + (\sigma_{+-} + \sigma_{-+})} \qquad (2.3.3)$$

where $\sigma_{\pm\pm}$ are cross sections for all four helicity combinations can be expressed (assuming dominance of the Compton process) as [34]:

$$A_{LL} \approx \frac{\Delta g(x_1)}{g(x_1)} \cdot \left[\frac{\sum_q e_q^2\,[\Delta q(x_2) + \Delta\bar{q}(x_2)]}{\sum_q e_q^2\,[q(x_2) + \bar{q}(x_2)]}\right] \cdot \hat{a}_{LL}(gq \to \gamma q) + (1 \leftrightarrow 2), \qquad (2.3.4)$$

where the second factor is known as $A_1^p$ asymmetry (Section 1.1) from polarized SIDIS and $a_{LL}(gq \to \gamma q)$ is spin asymmetry for sub-process $gq \to \gamma q$.

Measurement of $A_{LL}$ at $\sqrt{s}$>100 GeV is included in the long range program of RHIC [34].

## 2.5. Spin-dependent effects in elastic *pp, dp* and *dd* scattering.

There are several spin-dependent effects in elastic and quasi-elastic scattering reactions which could be further studied at NICA.

### *2.5.1. The charge-exchange $dp \to (pp)_S\, n$ reaction.*

For the *dp* spin correlation experiments discussed in Sections 2.5.1-3 we adopt the kinematics shown in Fig. 2.6. In this figure the *Z* axis is directed along the deuteron beam momentum *k*; *k'* is the momentum of scattered *d* or $(pp)_s$ ; the *Y* axis is along the vector ***n*** normal to the scattering plane, $\boldsymbol{n}=[\boldsymbol{k} \times \boldsymbol{k'}]$; and the *X* axis is chosen such as to form a right-handed coordinate system; θ is the scattering angle. For the so called collinear kinematics, $\theta = 0$.

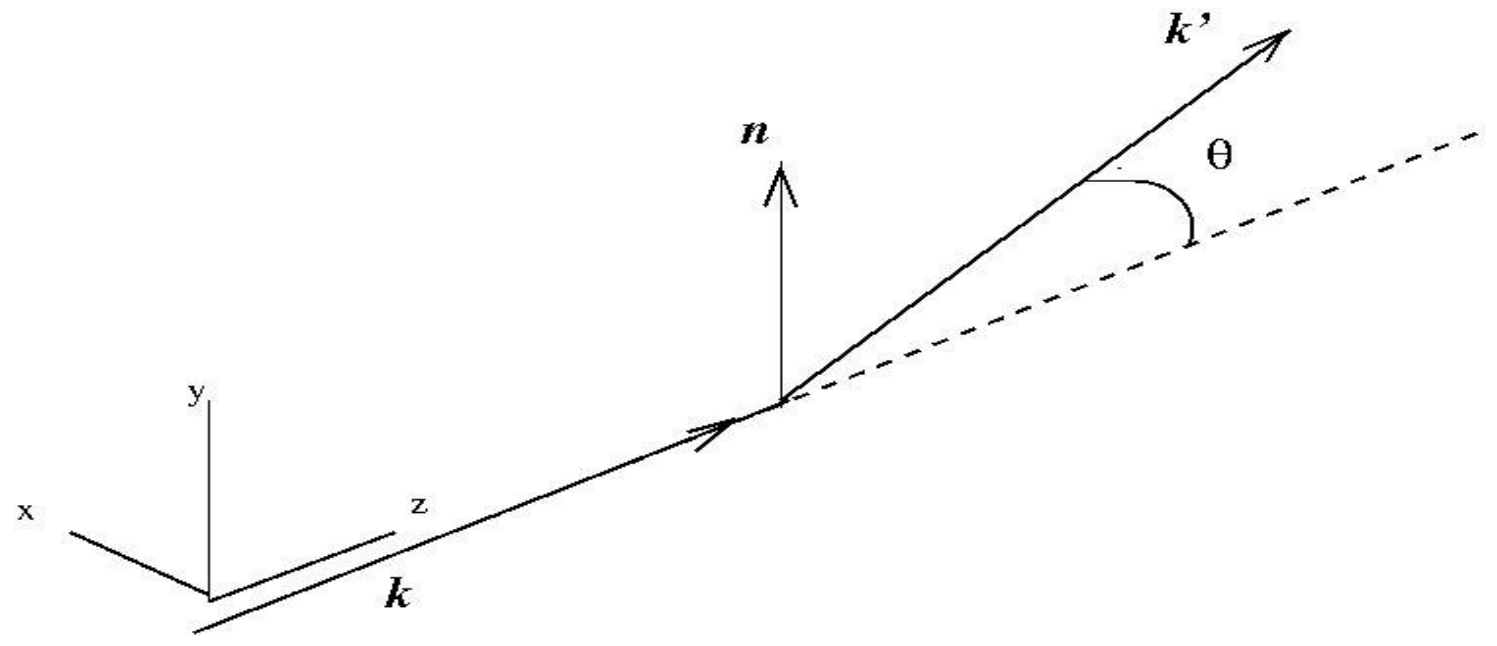

***Fig. 2.6:*** *Kinematics of the dp reactions. The proton beam momentum (dotted line) assumed to be opposite to the deuteron beam momentum.*

We use standard notations $C_{a,b}$ for spin correlation coefficients, where the first subscript index refers to the deuteron and the second one to the proton beam. The vector analyzing power is denoted as $A_y$ and the tensor analyzing powers are denoted as $A_{i,j}$ $(i,j = x,y,z)$. In this notations index *z* corresponds to the longitudinal (*L*) and *x*, *y* – to transverse (*T*) beam polarization wile *xz* denotes the mixed (*LT*) alignment (polarization) of the deuteron. The differential cross section of the reaction $dp \to (pp)_S\, n$ with polarized deuterons and protons, in units of the non-polarized differential cross section $\sigma_{UU}$, is given by Eq. (6.8) in [54]. For our purpose this formula in the collinear kinematics can be reduced to:

$$\sigma/\sigma_{UU} = 1 + (1/3)\, P^T_{yy} \cdot A_{yy} + (3/2)\, P^T_y \cdot Q^T_y \cdot C_{y,y} + (2/3)\, P^{TL}_{xz} \cdot Q^T_y \cdot C_{xz} \qquad (2.5.1)$$

where $P^T_y$ and $Q^T_y$ are transverse (vector) polarizations of the deuteron and proton, respectively, and $P^T_{yy}$ and $P^{TL}_{xz}$ are the transverse (tensor) polarizations (alignments) of the deuteron.

So, to study the spin-dependent effects in elastic or quasi-elastic *dp* scattering and to test Eq. (2.5.1) one needs to perform measurements with the non-polarized *d* and *p* beams (*UU*), with the transversally (tensor) polarized *d* and non-polarized *p* beams (*TU*) and with transversally (vector) polarized *d* & *p* beams (*TT*).

The reaction $dp \to (pp)_S\, n$ at low momentum transfer from the incident deuteron to the final di-proton $(pp)_S$ allows to measure a spin-flip part of the nucleon-nucleon charge-exchange amplitudes [44, 45]. Selecting two final protons with low energy, typically $E_{pp} < 3$ MeV, one can have the emerging di-proton $(pp)_S$ dominantly in the $^1S_0$ state. So, the reaction then involves a transition from the initial spin triplet state of the deuteron to the spin singlet state of di-proton. The transition amplitude in impulse approximation is proportional to that for the $np \to pn$, times a form factor reflecting overlap of the initial deuteron and final di-proton wave functions. This approach assumes measurements of the differential cross section, tensor and vector analyzing powers and spin correlation coefficients in $dp \to (pp)_S\, n$ reaction provided the contamination of *P*- and higher partial waves in the final *pp* system is taken into account [46]. A systematic study of this reaction has been started by ANKE@COSY in both single [46] and double polarized [47] experiments. Such kind of measurements at NICA would allow studying the elementary spin amplitudes of the $np \to pn$ transitions at higher energies. In collinear kinematics (defined above) the non-polarized cross section $\sigma^{UU}$ and spin observables (asymmetries) $C_{y,y}$ , $A_{yy}$ and $C_{xz,y}$ (Eq.2.5.1) measured with transversally polarized protons and deuterons (vector & tensor) constitute a complete polarization experiment.

### *2.5.2. Forward elastic pd-scattering and pN → pN amplitudes.*

As was shown recently [48, 49], the modified Glauber theory of multistep scattering, accounting the full spin dependence of the elementary *pN*-elastic scattering amplitudes and the deuteron spin structure, allows to explain quantitatively the non-polarized differential cross section, vector and tensor analyzing powers and spin correlation parameters of the elastic *pd* scattering in the forward hemisphere in the GeV- region. In this approach, the elementary *pN* scattering amplitudes are used as input from the SAID data base [50] available for *pp* scattering up to 3 GeV and for *pn* scattering up to 1.3 GeV.

At higher energies the Glauber theory of the diffraction scattering is a solid theoretical basis for description of the hadron-nuclei scattering data. Therefore, elastic *pd* scattering with longitudinally and transversally polarized proton and/or deuteron can be used at NICA energies as a test for spin amplitudes of the elastic *pN* scattering, at least in the region where the single *pN* scattering mechanism dominates and, therefore, inelastic shadowing corrections are negligible.

### *2.5.3. Backward elastic pd-scattering and the hard deuteron breakup $pd \to (pp)_S\, n$.*

The reaction $pd \to (pp)_S\, n$ with formation of the $^1S_0$-di-proton ($E_{pp}$= 0 - 3 MeV) in the backward elastic scattering $pd \to dp$ was studied at COSY in the GeV region [51]. Due to different quantum numbers of the deuteron ($J$ = 1, $I$ = 0) and di-proton ($J$ = 0, $I$ = 1), the dynamics of the reaction in these two channels is essentially different, providing suppression of the isovector meson exchange in the di-proton channel as compared to the deuteron channel. A combined analysis of these measurements provide a definite conclusion with respect to properties of the short-range *NN* interaction and deuteron wave function at large internal momenta [52], $k$ = 0.5 - 0.6 GeV/c. According to expectations [53], measurement of tensor analyzing power in the breakup reaction $pd \to (pp)_S\, n$ would shed light on the old $T_{20}$ puzzle observed in the inclusive deuteron breakup and in the $pd \to dp$. Measurements of the cross sections and spin observables of these reactions at NICA can be extended to higher internal momentum $k$ = 1-2 GeV/c, i.e. to very short distances inside the deuteron and di-proton, where quark-gluon degrees of freedom are expected to be a natural language for the deuteron structure.

This study would be complementary to the program of experiments planned at Jefferson Laboratory (Virginia, USA) with electron-deuteron scattering.

## 2.6. Spin-dependent reactions in heavy ion collisions.

### *2.6.1. Investigation of the birefringence phenomenon at NICA facility.*

One of the most interesting quasi-optical effects – the birefringence phenomenon for deuterons (or other particles with spin $S \geq 1$) passing through matter – has recently become the area of research [35]. Birefringence occurs when spin $S \geq 1$ particles pass through isotropic non-polarized matter and is due to the inherent anisotropy of particles with spin $S \geq 1$ (as distinct from spin- ½ particles). The birefringence effect leads to the rotation of the beam polarization vector when a non-polarized deuteron beam passes through a non-polarized target. Moreover, the appearing spin dichroism effect (the different absorption of deuterons in states with $m = \pm 1$ and $m = 0$) gives rise to a tensor polarization of the initially non-polarized deuteron beam that has passed through the non-polarized target [35].

The experimental investigation of the birefringence effect began with the observation of the spin dichroism effect for low- and high-energy deuterons. The experiments with 5 - 20 MeV deuterons were performed at the electrostatic accelerator of Cologne University (Germany) [36]. Tensor polarization acquired by the beam was obtained by varying the thickness of carbon targets and the initial energy of the beam. The experiments using carbon targets and deuterons with a momentum of 5 GeV/c were performed at Nuclotron-M accelerator. The measured values of tensor polarization acquired by the beam passing through a set of variable-thickness targets are given in Fig. 2.7 [37].

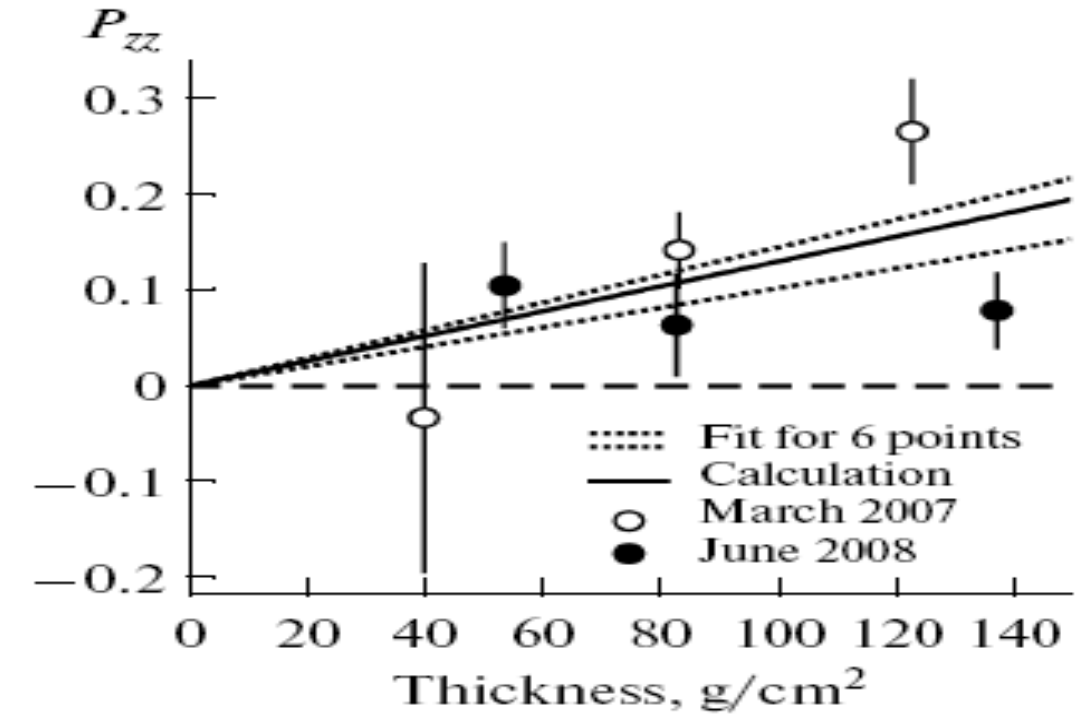

***Fig. 2.7:*** *Tensor polarization value acquired by deuterons of 5 GeV/c crossing the carbon target of various thicknesses.*

The birefringence phenomena can be further studied at NICA:
- in few-nucleon systems involving protons and deuterons;
- appearing through the interaction of protons or deuterons with heavy nuclei;
- for heavy nuclei with spin $S \geq 1$.
- with vector particles produced in inelastic collisions.

### *2.6.2. Inclusive particle polarizations in heavy-ion collisions.*

It is well known that the hyperon and other particle polarization observed in proton-nuclear ($p+A$) reactions and in heavy ions ($A_1+A_2$) collisions have significant value in a wide range of energies. Examples of the data on the hyperon transverse polarization at medium energy [55] and the data obtained at the higher collider energies on the so called "global polarization", i.e. measured along the normal to the reaction plane [50-58], are shown below. From a theoretical point of view the origin of sizable hyperon polarization in $p+A$ and $A_1+A_2$ collisions represents a significant problem since in the perturbative QCD it is expected to be small [59]. Several phenomenological models have been proposed to explain observed polarizations [60-64].

Recently the more general model, so called Chromo-Magnetic Polarization of Quarks (CMPQ), has been developed. It explains the origin of single-spin asymmetry, hyperon and vector meson polarization at the phenomenological level [65-69]. The mechanism of the CMPQ model is based on the Standard Model but requires a number of additional assumptions. The model assumes that:

-an effective transverse circular chromo-magnetic field is created during the interaction process;

-the spin-dependent Stern-Gerlach type forces, appearing due to the interaction of chromo-magnetic moments of the probe quarks (which fragments into the observed hadrons) with the inhomogeneous effective color magnetic field, are acting on these quarks;

-the spin of the probe quark precess in the effective color field, resulting in oscillations of hyperons (anti-hyperons) polarization as a function of Feynman $x_F$ and other variables;

-the strength of the effective color field is a linear function of the number of spectator quarks and antiquarks with weights determined by the color factors for $qq$ and $q\bar{q}$ interactions.

These assumptions are characterized by free parameters which can be determined from the experimental data. The CMPQ model permits also to make predictions for new data.

Comparison of the transverse polarization, $P_N$ , of $\Lambda$ produced in the reaction $Au+Au\rightarrow \Lambda^{\uparrow} +X$ with corresponding expectations from CMPQ model is shown in Fig. 2.7. A resonance like behavior of $P_N$ vs. $\sqrt{s}$ is seen. In CMPQ model it is expected due to strong interaction of color charges of the spectator quarks and the probe $s$-quark in the $\Lambda$ hyperon. The oscillation of $P_N$ vs. $x_F$ is a consequence of the $s$-quark spin precession in a color field [67].

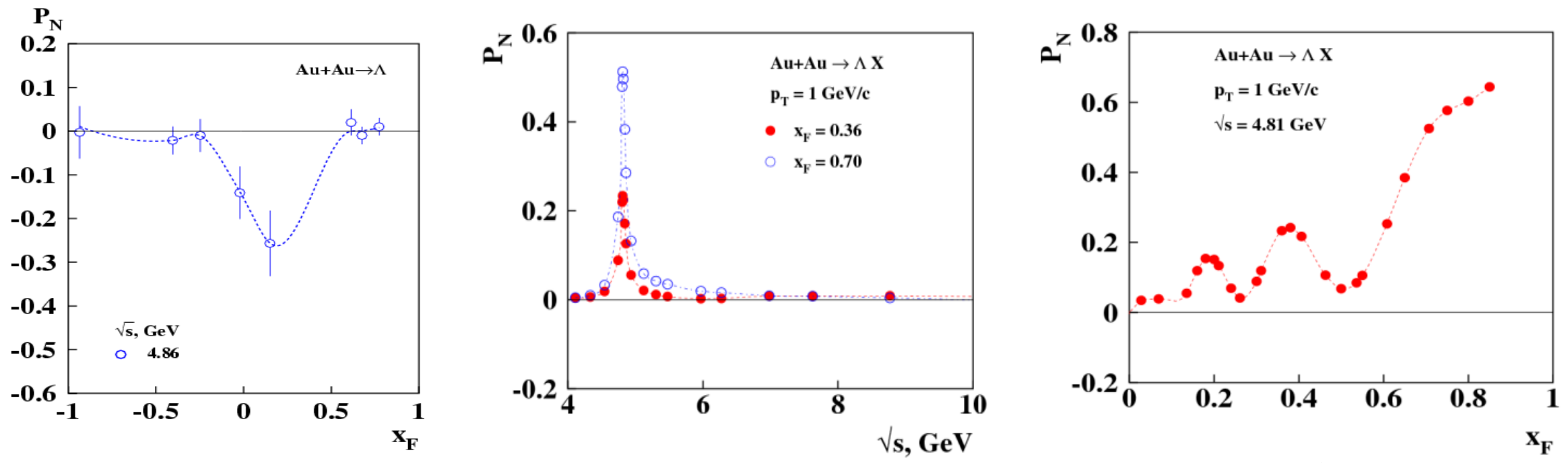

***Fig. 2.7:*** *Transverse polarization $P_N$ vs. $x_F$ of $\Lambda$ from the reaction $Au+Au\rightarrow \Lambda^{\uparrow} +X$ at $\sqrt{s}$=4.86 GeV [1] in comparison with the CMPQ model expectations (dashed curve) [66] (left panel); $P_N$ vs. $\sqrt{s}$ for two values of $x_F$ (central panel) and $P_N$ vs. $x_F$ (right panel) [67].*

The global polarization in the reaction $Au+Au\rightarrow \Lambda^{\uparrow} +X$ at $\sqrt{s}$ = 62 and 200 GeV is shown in Fig. 2.8 [50] as a function of the $\Lambda$ transverse momentum $p_T$. It is sizable (unfortunately with large errors) only for $p_T$ above 2.7 GeV/c. The oscillation of $P_N(p_T)$ seen in this Figure is due to $s$-quark spin precession in a color field [69]. It would be interesting to compare the global and transverse hyperons' polarizations within the same experiment.

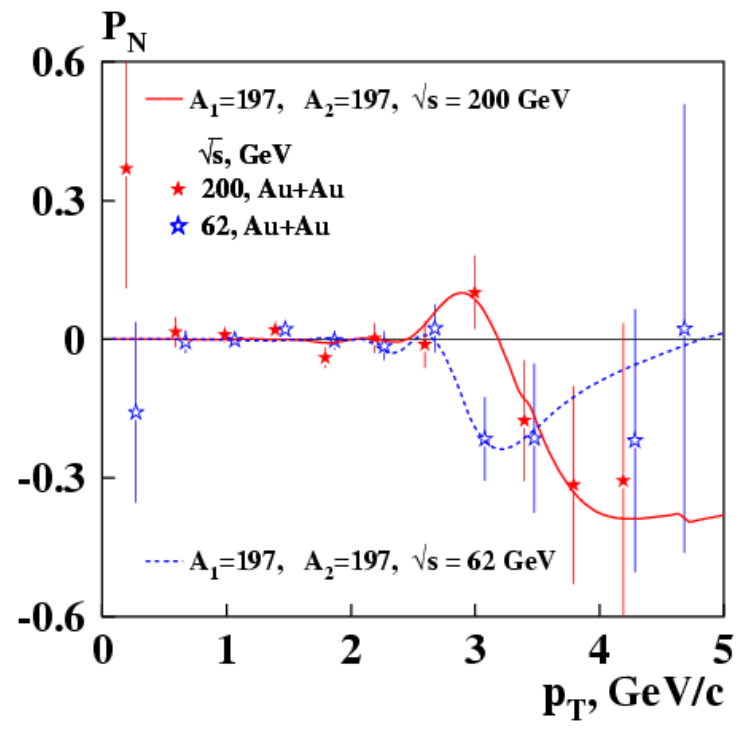

***Fig. 2.8:*** *Global polarization $P_N$ as a function of $p_T$ in the reaction $Au+Au\rightarrow \Lambda^{\uparrow} +X$ at $\sqrt{s}$=62 and 200 GeV [50] compared with the CMPQ model expectations [69].*

Predictions for $P_N$ vs. the pseudo rapidity $\eta$ ($\eta$ = -ln tan $(\theta/2)$) for the reaction $A_1+A_2 \rightarrow \Lambda^{\uparrow} + X$ are shown in Fig. 2.9 for the energies $\sqrt{s}$ = 7 and 9 GeV [52]. The oscillations of $P_N(\eta)$ are expected due to the *s*-quark spin precession in the color field. The frequency of $P_N(\eta)$ oscillation increases with the atomic weight *A* of colliding ions. This is due to higher number of spectator quarks creating the color field of heavier ions. The spin precession frequency is also a rising function of *A*, because the effective chromo-magnetic field strength is increasing with the atomic weight. These predictions are also valid for heavier hyperons and other inclusive particles like $\tilde{K}$*(892), $\varphi$(1020), $\rho$, $\omega$ , $J/\psi$.

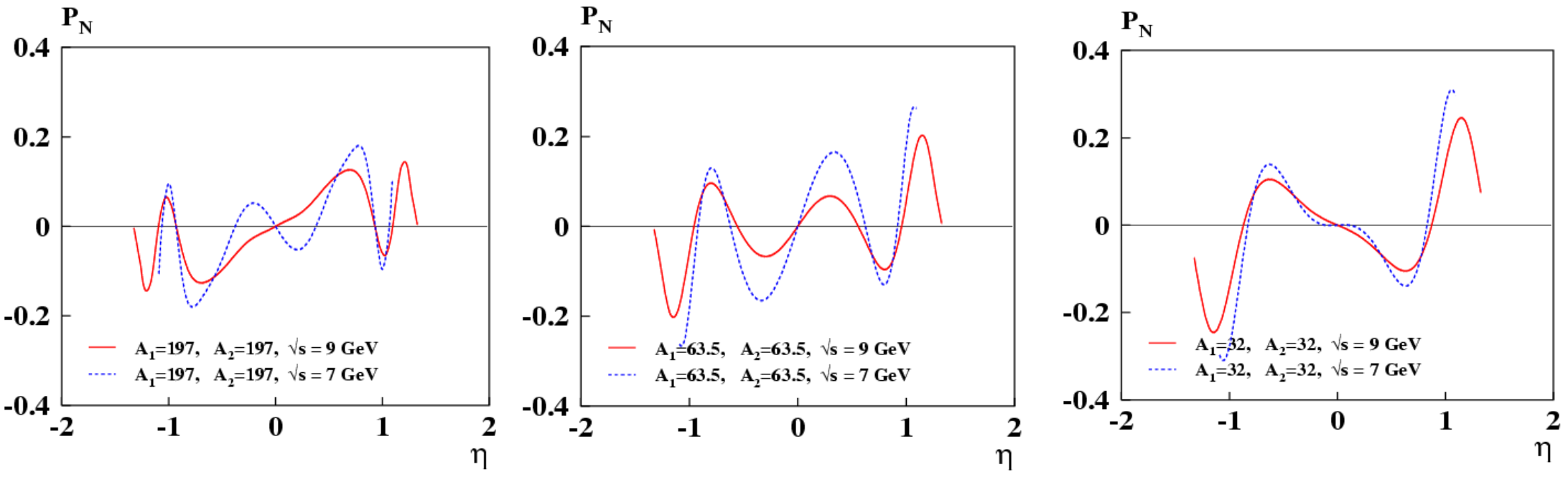

***Fig. 2.9:*** *Predictions of the CMPQ model for $P_N(\eta)$ of $\Lambda$ from the reaction Au+Au$\rightarrow \Lambda^{\uparrow}$ +X (left panel), from the reaction Cu+Cu$\rightarrow \Lambda^{\uparrow}$ +X (central panel) and from the reaction S+S$\rightarrow \Lambda^{\uparrow}$ +X (right panel). The value of $p_T$ for $\Lambda$ is 2.35 GeV/c.*

Concluding this subsection it is worthwhile to note that systematic studies of inclusive transverse and global polarizations of hyperons, anti-hyperons and vector mesons vs. kinematic variables and as function of the energy and atomic weight of colliding ions can be subjects of systematic studies at SPD.

### 2.7. Future DY experiments on nucleon structure in the world.

The measurements of DY processes using various beams and targets have started in 1970 with the non-polarized proton beam of AGS accelerator in Brookhaven. Since that time series of DY experiments were performed at FNAL and CERN but only two of them directly connected with studies of the nucleon structure. Those are experiments NA51 [38] and E866 [39]. Both of them have measured the ratio of the anti-d and anti-u quarks in the nucleons.

Present list of the DY experiment in the world (Table 1 below) includes fixed target and collider experiments aimed to study spin-dependent and spin-independent processes in a wide range of energies. Physics goals of the experiments include studies of one or several TMD PDFs.

The first fixed target polarized DY measurements will be performed at CERN by the COMPASS-II experiment [40]. It will start to take data in 2014 using the 190 GeV (or $\sqrt{s}$ ~ 19 GeV) $\pi^-$ beam and polarized hydrogen target. The FNAL E-906 [41] non-polarized experiment has started already. Recently FNAL has initiated the workshops on polarized DY experiments. The PANDA [42] at FAIR will start somewhat later. The SPASCHARM experiment at Protvino is under preparations.

Future collider DY experiments are included in the long range programs of the PHENIX and STAR at RHIC [43]. They are planning to carry out DY measurements with 500 GeV longitudinally polarized as well as with 200 GeV transversely polarized protons.

The SPD experiments, proposed at the second intersection point of the NICA collider, will have a number of advantages for DY measurements related to nucleon structure studies. These advantages include:

- operations with *pp*, *pd* and *dd* beams,
- scan of effects on beam energies,

- measurement of effects via muon and electron-positron pairs simultaneously,
- operations with non-polarized, transverse and longitudinally polarized beams or their combinations. Such possibilities permit for **the first time** to perform comprehensive studies of **all leading twist PDFs** of nucleons in a single experiment with minimum systematic errors.

Table 1: List of the present and future DY experiments in the world.

| Experiment | CERN, COMP.-II | FAIR, PANDA | FNAL, E-906 | SPAS-CHARM | RHIC, STAR | RHIC, PHENIX | NICA, SPD |
|---|---|---|---|---|---|---|---|
| ***mode*** | ***FixTar*** | ***FixTar*** | ***FixTar*** | ***FixTar*** | ***collider*** | ***collider*** | ***collider*** |
| ***Beam/target*** | ***π- , p*** | ***anti-p, p*** | ***π- , p*** | ***π±, pol.p*** | ***pp*** | ***pp*** | ***pp, pd,dd*** |
| ***Polarization:b/t*** | ***0; 0.8*** | ***0; 0*** | ***0; 0*** | ***0; 0.5*** | ***0.5*** | ***0.5*** | ***0.9*** |
| ***Luminosity*** | $2{\cdot}10^{33}$ | $2{\cdot}10^{32}$ | $3.5{\cdot}10^{35}$ | | $5{\cdot}10^{32}$ | $5{\cdot}10^{32}$ | $10^{32}$ |
| ***√s , GeV*** | ***19*** | ***6*** | ***16*** | ***8*** | ***200, 500*** | ***200, 500*** | ***10-26*** |
| $x_{1(beam)}$ ***range*** | ***0.1-0.9*** | ***0.1-0.6*** | ***0.1-0.9*** | ***0.1-0.3*** | ***0.03-1.0*** | ***0.03-1.0*** | ***0.1-0.8*** |
| $q_T$***, GeV*** | ***0.5 -4.0*** | ***0.5 -1.5*** | ***0.5 -3.0*** | | ***1.0 -10.0*** | ***1.0 -10.0*** | ***0.5 -6.0*** |
| ***Lepton pairs,*** | ***μ-μ+*** | ***μ-μ+*** | ***μ-μ+*** | | ***μ-μ+*** | ***μ-μ+*** | ***μ-μ+, e+e-*** |
| ***Data taking*** | ***2014*** | ***>2018*** | ***2013*** | | ***>2016*** | ***>2016*** | ***>2018*** |
| Transversity | NO | NO | NO | | YES | YES | YES |
| Boer-Mulders | YES | YES | YES | | YES | YES | YES |
| Sivers | YES | YES | YES | | YES | YES | YES |
| Pretzelosity | YES (?) | NO | NO | | NO | YES | YES |
| Worm Gear | YES (?) | NO | NO | | NO | NO | YES |
| J/Ψ | YES | YES | NO | | NO | NO | YES |
| Flavour separ. | NO | NO | YES | | NO | NO | YES |
| Direct γ | NO | NO | NO | | YES | YES | YES |

## 3. Requirements to the NUCLOTRON-NICA complex.

The research program outlined in Section 2 requires definite characteristics of beams and technical infrastructure.

***Beams.*** The following beams will be needed, polarized and non-polarized:

$$pp,\ pd,\ dd,\ pp\uparrow,\ pd\uparrow,\ p\uparrow p\uparrow,\ p\uparrow d\uparrow,\ d\uparrow d\uparrow.$$

Beam polarizations both at MPD and SPD: longitudinal and transversal. Absolute values of polarizations during the data taking should be 90-50%.The life time of the beam polarization should be long enough. Measurements of Single Spin and Double Spin asymmetries in DY require running in different beam polarization modes: *UU, LU, UL,TU, UT,LL ,LT and TL* (spin flipping for every bunch or group of bunches should be considered).

Beam energies: $p\uparrow p\uparrow(\sqrt{s_{pp}})$ = 12 ÷ ≥ 27 GeV (5 ÷ ≥12.6 GeV kinetic energy),

$d\uparrow d\uparrow(\sqrt{s_{NN}})$ = 4 ÷ ≥13.8 GeV (2 ÷ ≥5.9 GeV/u ion kinetic energy).

Asymmetric beam energies should be considered also.

Beam luminosities: in the *pp* mode: $L_{average} \geq 1{\cdot}10^{32}$ cm$^{-2}$s$^{-1}$ (at $\sqrt{s_{pp}}$ = 27 GeV),

in the *dd* mode: $L_{average} \geq 1{\cdot}10^{30}$ cm$^{-2}$s$^{-1}$ (at $\sqrt{s_{NN}}$ = 14 GeV).

For estimations of the expected statistics of events, we assume that total efficiency of the NICA complex will be ≥ 80%.

***Infrastructure.*** The infrastructure of the Nuclotron-NICA complex should include:

- source(s) of polarized (non-polarized) protons and deuterons,
- system of the beams polarization control and absolute measurements (3-5%),
- system of luminosity control and absolute measurements, including a space inside the NICA ring for the beam polarimeters,

- system(s) for transferring data on the beam(s) polarization and luminosity to the experiments. The infrastructure tasks should be subjects of the separate project(s).
Local polarization and luminosity monitors at SPD are discussed in Section 5.4.
***Beams intersection area.*** The area of ± 3m along and across of the beams second intersection point, where the detector for the spin physics experiment will be situated, must be free of any collider elements and equipment. The beam pipe diameter in this region should be less than 10 cm. In this case the angular acceptance of SPD will be close to 4π. The walls of the beam pipe in the region ± 1m of the beams intersection should have a minimal thickness and made of the low-*Z* material, e.g. beryllium.

# 4. Polarized beams at NICA.

The NICA complex at JINR has been approved in 2008 assuming two phases of the construction. The first phase, is being realized now, includes construction of facilities for heavy ion physics program [1] while the second phase should include facilities for the program of spin physics studies with accelerated polarized protons and deuterons. In this document we communicate briefly the status of the NICA project in relation to experiments with polarized beams.

## 4.1. Layout of the complex.

The main elements of NICA complex are shown in Figure 4.1. They include: the heavy ion source and source of polarized ions (proton and deuteron), SPI, with corresponding Linacs, existing superconducting accelerator Nuclotron upgraded to Nuclotron M, new superconducting Booster synchrotron , new collider NICA with two detectors – MPD (Multi-Purpose Detector) for heavy ion interaction studies and SPD (Spin Physics Detector), as well as experimental hall for fixed target experiments with beams extracted from Nuclotron M.

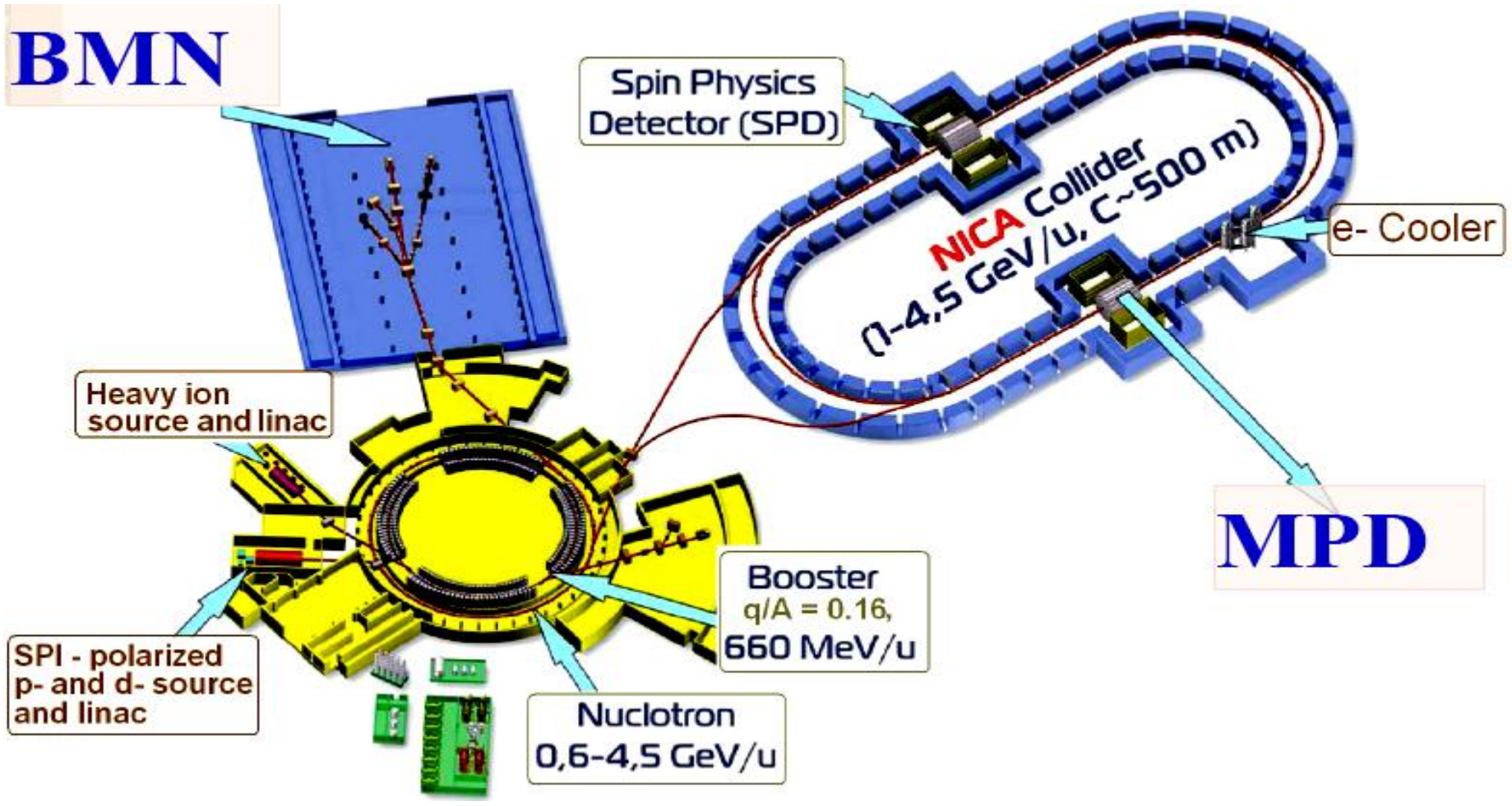

*Fig. 4.1: The NICA complex of JINR.*

The functional scheme of facility approved for the first phase of construction scenario is presented in Fig. 4.2. The chain of beams injection to the collider rings in the case of polarized protons and deuterons includes: SPI, the modernized injection Linac LU-20 equipped with the new pre-injector (PI), Booster, Nuclotron and NICA. The main goals of the Booster in polarized case are the following: 1) formation of the required beam emittance with electron cooling and 2) fast extraction of the accelerated beam. The chain bypassing Booster is also considered [2]. Feasibilities to fulfill requirements to the NICA complex formulated in previous Section are considered below moving along the chain: SPI – LU-20 – Nuclotron (Booster) – NICA.

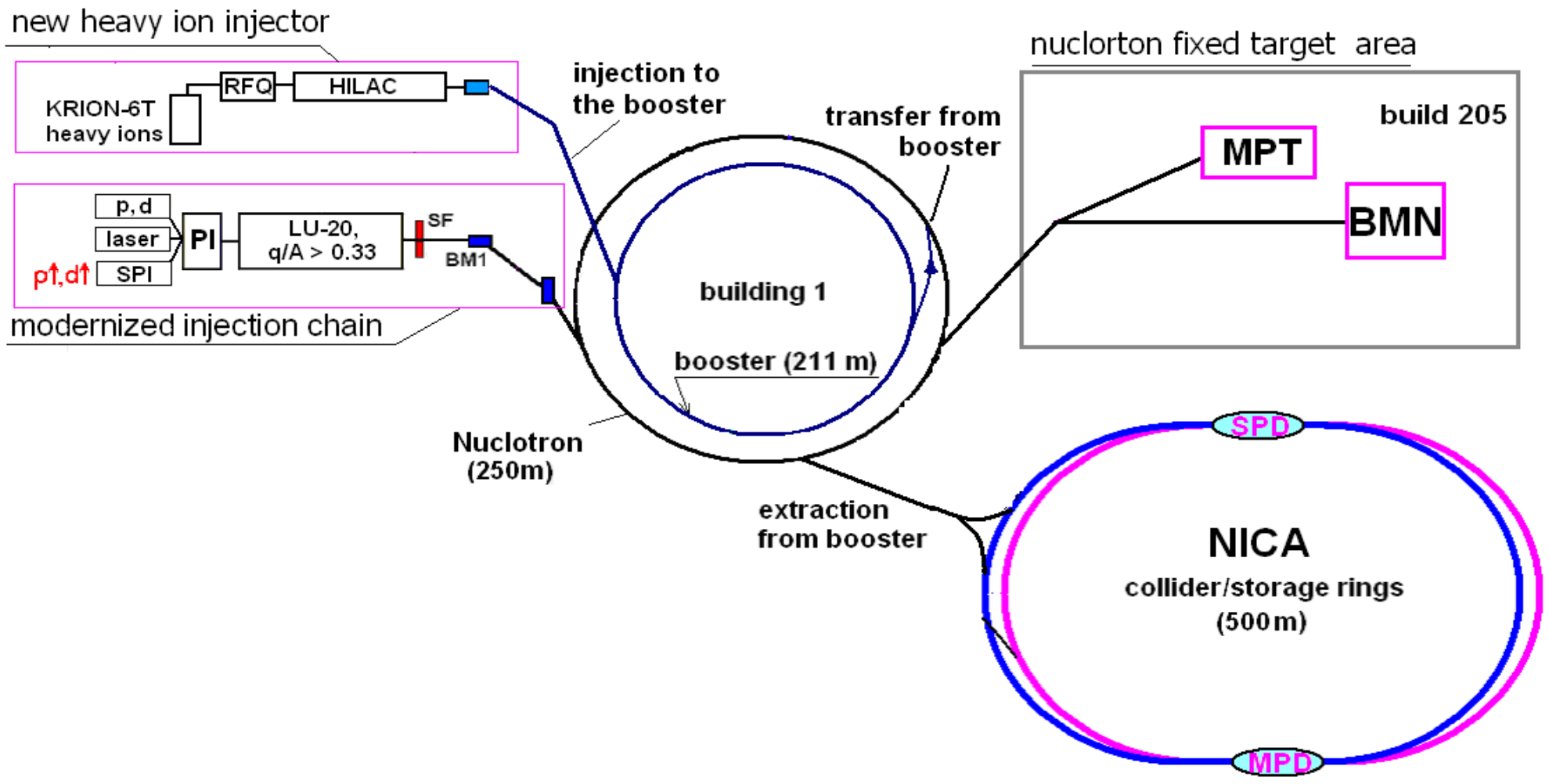

***Fig. 4.2:*** *The functional scheme of NICA complex.*

## 4.2. Source of polarized ions and injector.

The new polarized ion source is being commissioned now. It was designed and constructed as a universal pulsing high intensity source of polarized deuterons and protons based on a charge-exchange plasma ionizer. The output $D^+\uparrow$ $(H^+\uparrow)$ current of the source is expected to be at the level of 10 mA. The expected polarization is about 90% in the vector (±1) for $D^+\uparrow$ and $H^+\uparrow$ and tensor (+1,−2) for $D^+\uparrow$ modes. The project is carried out in cooperation with INR of RAS (Moscow). The equipment available from the CIPIOS ion source (IUCF, Bloomington, USA) is partially used for SPI. The source will deliver the 10 mks pulsed polarized proton or deuteron beam with intensity up to ~$2\cdot 10^{11}$ per pulse and repetition rate of 1 Hz [3].

Briefly, the SPI consists of several sections. The atomic beam section uses the permanent ($B$ = 1.4 T) and conventional electromagnet sextupoles ($B$ = 0.9 T) for beam focusing. The cryo-cooler section is used for cooling the atomic beam. In the radio-frequency transition section the atoms are polarized before they are focused into the ionizer. The resonant charge-exchange ionizer [4] produces pulses of positive ion plasma inside the solenoid. Nearly resonant charge-exchange reactions:

$$D^+ + H^0\uparrow \;\rightarrow H^+\uparrow + D^0 ,$$
$$H^+ + D^0\uparrow \;\rightarrow D^+\uparrow + H^0 , \qquad (4.2.1)$$

are used to produce polarized protons or deuterons. Spin orientation of $D^+\uparrow$ $(H^+\uparrow)$ at the exit of SPI is vertical. The polarized particles are focused through the extraction section into the injection Linac.

The Alvarez-type Linac LU-20 used as the Nuclotron injector was put into operation in 1974. It was originally designed as proton accelerator from 600 KeV to 20 MeV. Later it was modified to accelerate ions with charge-to-mass ratio $q/A$ > 0.33 up to 5 MeV/u at the $2\beta\lambda$ mode. The voltage transformer up to 700 kV is now used to feed the accelerating tube of the LU-20 pre-injector. The new pre-injector will be based on the RFQ section [5].

## 4.3. Acceleration of polarized protons and deuterons.

### *4.3.1. Polarized deuterons.*

Acceleration of polarized deuterons at the Synchrophasotron was achived for the first time in 1984 [6] and at Nuclotron in 2002 [7]. There are no dangerous spin resonances which could occur during the polarized deuterons acceleration in Nuclotron up to the kinetic energy of 5.6 GeV/u. This limit is practically very close to the maximum design energy of the Nuclotron (6 GeV/u for $q/A$ = ½). There are no doubts about the realization of the project in this case. The only problem in case of deuterons is changing the polarization directions from longitudinal to transversal or backward.

#### *4.3.2. Polarized protons.*

According to the initial NICA project, Nuclotron as the strong focusing synchrotron should accelerate polarized protons from the injection energy (20 MeV) up to the maximum design value of 12.6 GeV. The scheme considered below permits to accelerate polarized protons in Nuclotron up to 5 GeV and accelerate them further at NICA up to required energy.

Let us estimate first the expected proton beam intensity at the Nuclotron exit. The limitations and particle losses could come due to different reasons. Taking the SPI design current (10 mA) and estimated particle loss coefficient between the source and Nuclotron (0.5), RF capture (0.8), extraction efficiency (0.86) and other factors in the synchrotron (0.9), one can expect the output intensity up to $1.6 \cdot 10^{11}$ polarized protons per pulse.

For the successful crossing of numerous spin resonances in Nuclotron, the inserted devices like “siberian snakes” will be designed and installed into the accelerator lattice. Spin resonanses, occuring during the acceleration cycle at different combinations of the betatron ($\nu_x$ , $\nu_y$) and spin ($\nu$) oscillation frequencies, were analyzed in [8].Three cases were considered: $\nu = k$ , $\nu = k \pm \nu_y$ , $\nu = k \pm \nu_x$ , where $k$ = 0, 1, 2,.. . Dependence of the spin resonance frequency, $w_k$ (normalized to the value $w_d = 7.3 \cdot 10^{-4}$ corresponding to complete beam depolarization) on the proton energy for each of these cases is shown in Fig. 4.3. The “dangerous” resonances marked with black dots occure when the values of log($w_k$ / $w_d$) approch zero or -1. As one can see, there are four resonanses in the first case and two resonanses in the second and third cases.

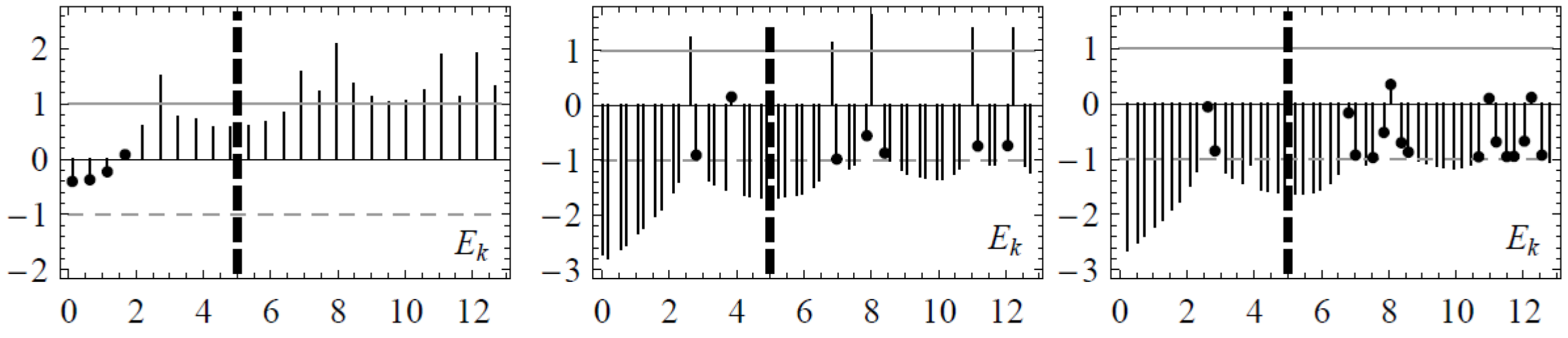

***Fig. 4.3****: Values of log ($w_k$ / $w_d$), caracterizing proton spin resonances in the Nuclotron, vs. the proton energy in GeV,calculated for: $\nu = k$ (left), $\nu = k \pm \nu_y$ (center), $\nu = k \pm \nu_x$ (right).*

To preserve polarization, we consider the siberian snake with solenoid magnetic field as an inserted device. Possible solution was found in the limited energy range up to 5 GeV indicated in Fig. 4.3 by vertical dashed lines. The proton spin dynamics along the Nuclotron ring is shown in Fig. 4.4 [9] assuming the snake (full or partial), operating synchronously with Nuclotron accelerating cycle, is placed in the second (after injection) straight section. The maximum magnetic field integral of the snake depends on the particle momentum and approximately equals 21 T·m at the Lorenz factor γ=6. It is not necessary to use a full snake to suppress the influence of spin resonances. One can use a partial snake as well. That reduces the longitudinal magnetic field integral by a factor 2. If the longitudinal magnetic field is introduced in the synchrotron straight section, the dependence of spin frequency $\nu$ on particle energy and spin angle $\varphi_z$ in the solenoid is defined by the relation: $\cos(\pi\nu) = \cos(\pi \gamma G)\cos\frac{\varphi_z}{2}$. Thus, even with a small longitudinal magnetic field, $\varphi_z/2\pi > |w_k|$, one can completely “exclude” the set of integer resonances, whereas suppressing of the intrinsic resonances is occurred if $\varphi_z/2\pi > |w_k|$.

The maximum longitudinal magnetic field integral at $\gamma = 6$ is reaching a value of 8.5 T·m, i.e. about twice as less than in the case of the full snake, $(\varphi_z = \pi)$.

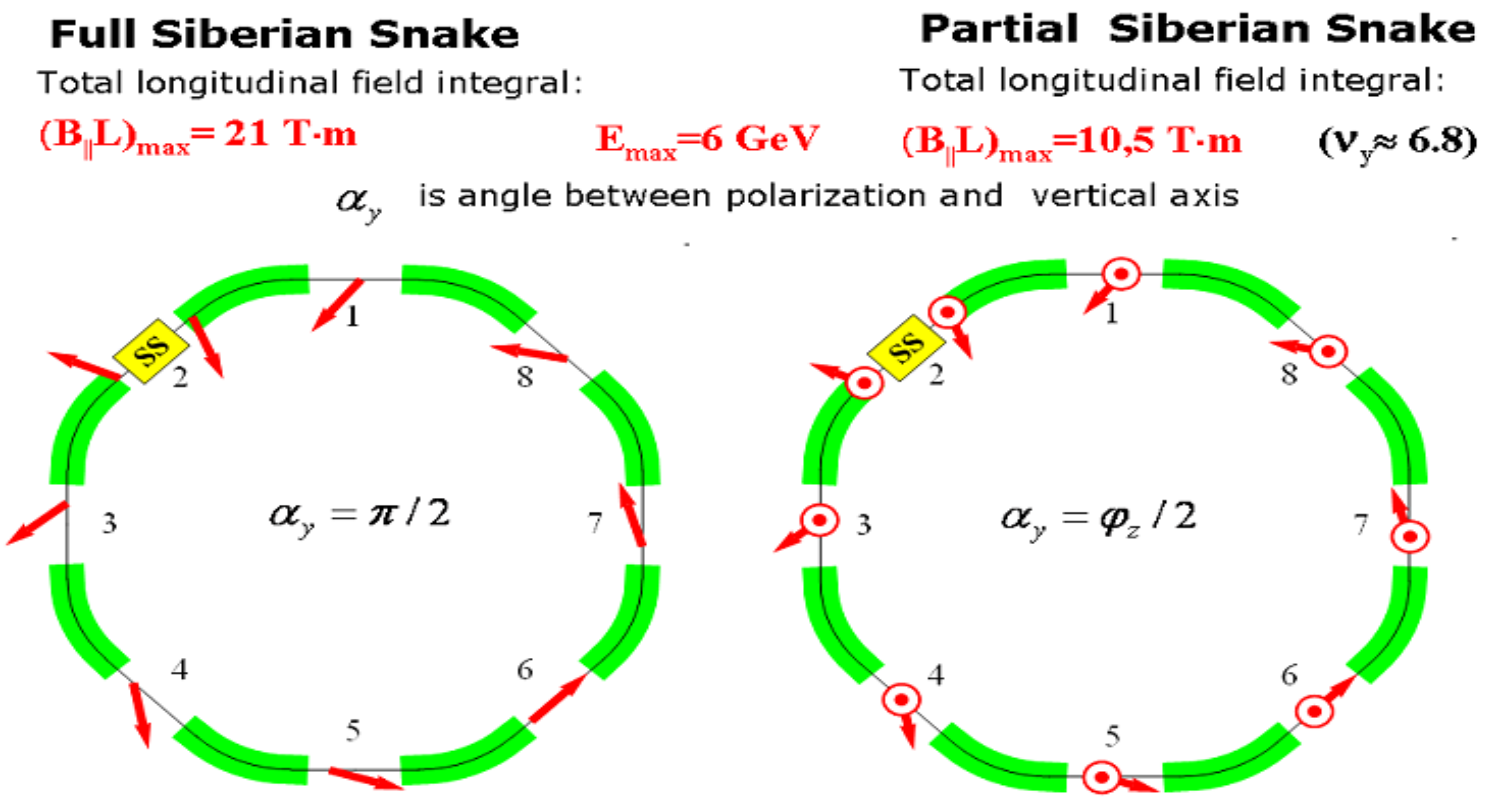

***Fig. 4.4:*** *Proton spin dynamics in the Nuclotron ring in the case of a full or partial snake.*

The snake structure – two solenoids and three quadrupoles (F – focusing and D - defocusing) and parameters of the insertion are shown in Fig. 4.5.

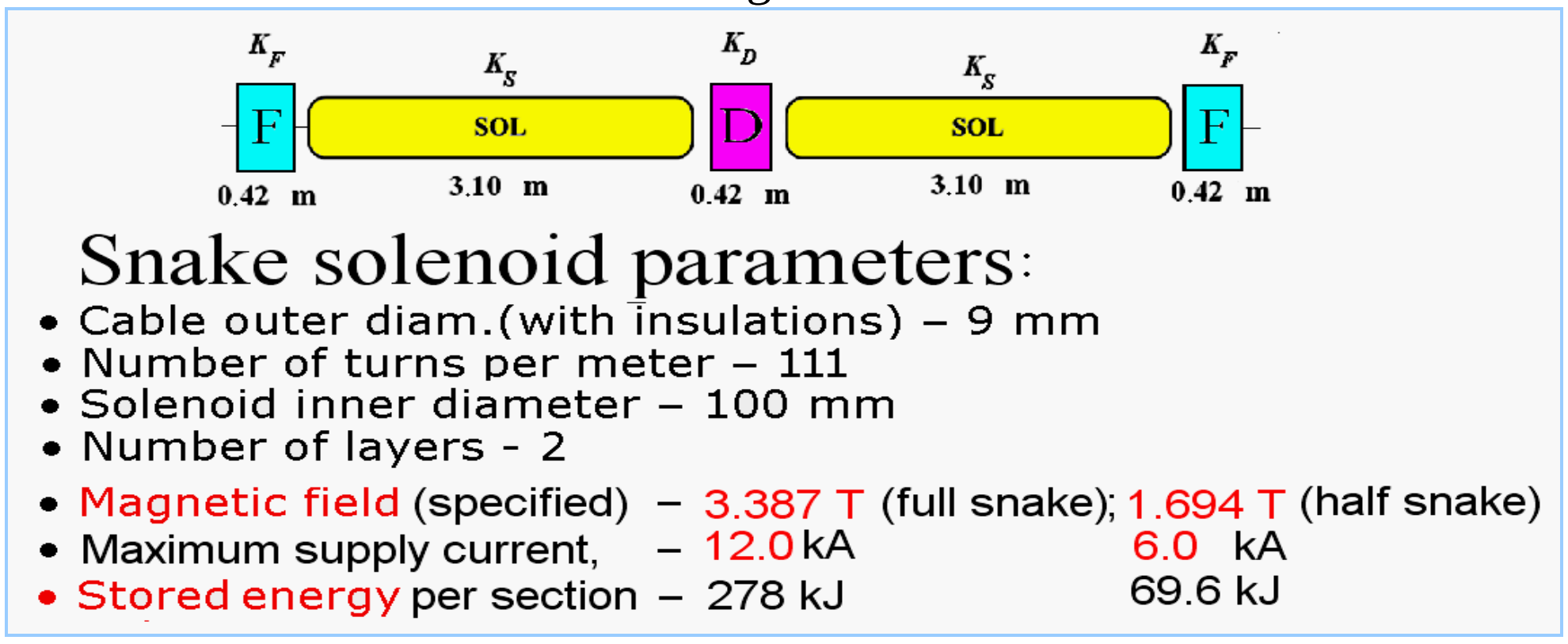

***Fig. 4.5:*** *Snake structure and parameters of insertion.*

### 4.4. NICA in the polarized proton and deuteron modes.

The novel scheme of the polarization control at NICA, suitable for protons and deuterons, is based on the idea of polarized beams manipulation in the vicinity of the zero spin tunes. This approach is actively developed at JLAB for the 8-shaped ring accelerator project. The zero spin tune is a natural regime for the just mentioned case. To provide zero spin tune regime at the collider of the racetrack symmetry, it is necessary to install two identical siberian snakes in the opposite straight sections (Fig. 4.6).

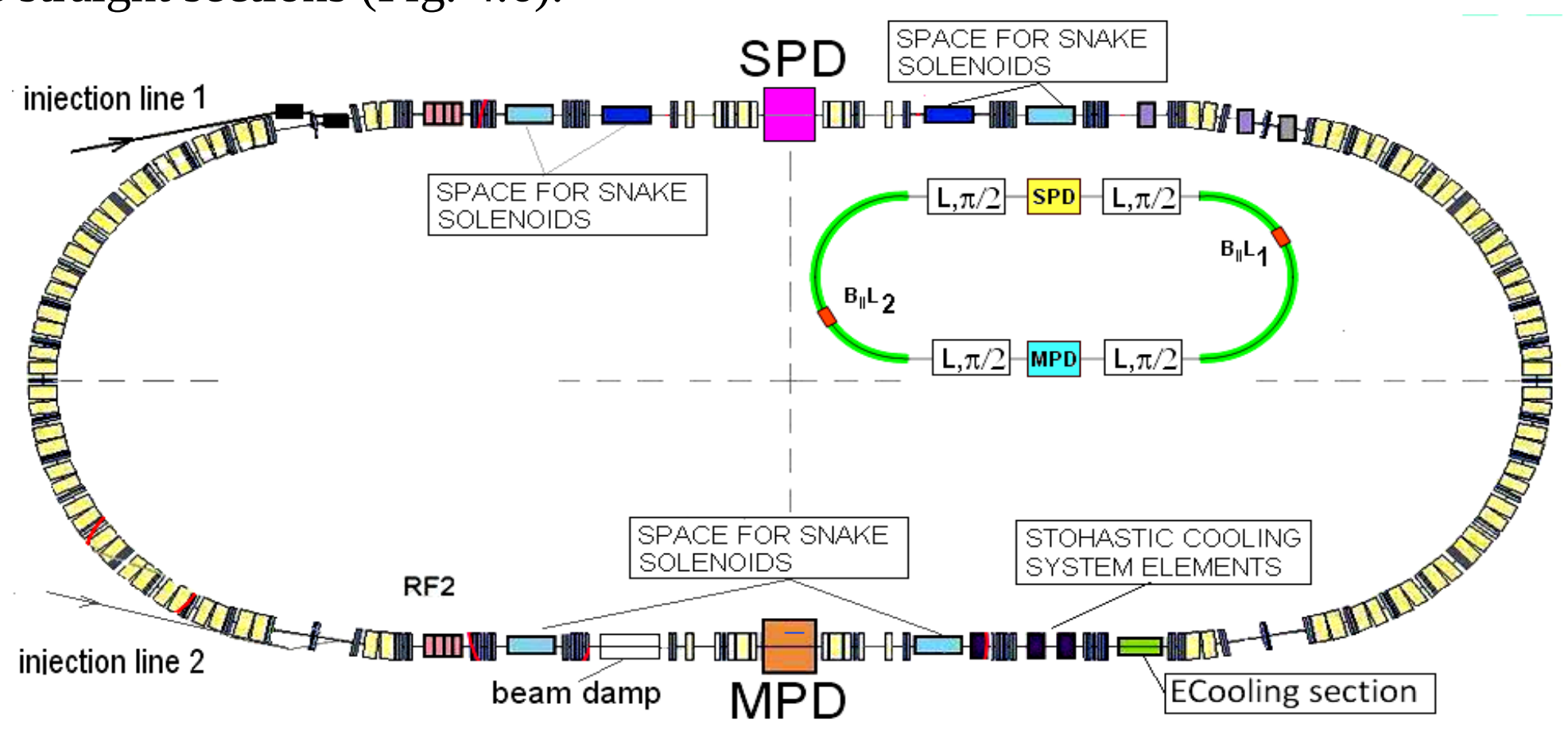

***Fig. 4.6:*** *Possible NICA structure for polarized proton and deuteron beams*

In this scheme any direction of the polarization is reproduced at any azimuth point after every turn. However, if one fixes the longitudinal (or vertical) polarization at SPD, the polarization vector at MPD will be rotated by some angle with respect to the direction of the particle velocity vector. This angle depends on the beam energy. If the direction of the polarization is fixed at MPD, some arbitrary polarization angle will occur at SPD. The control insertions $(\mathbf{B}_{||}\mathbf{L})_{1,2}$ can correct this angle [14].

***4.4.1. NICA luminosity.***

The NICA luminosity in the polarized proton mode is estimated for the proton kinetic energy region from 1 to 12.7 GeV [11], Fig. 4.7.

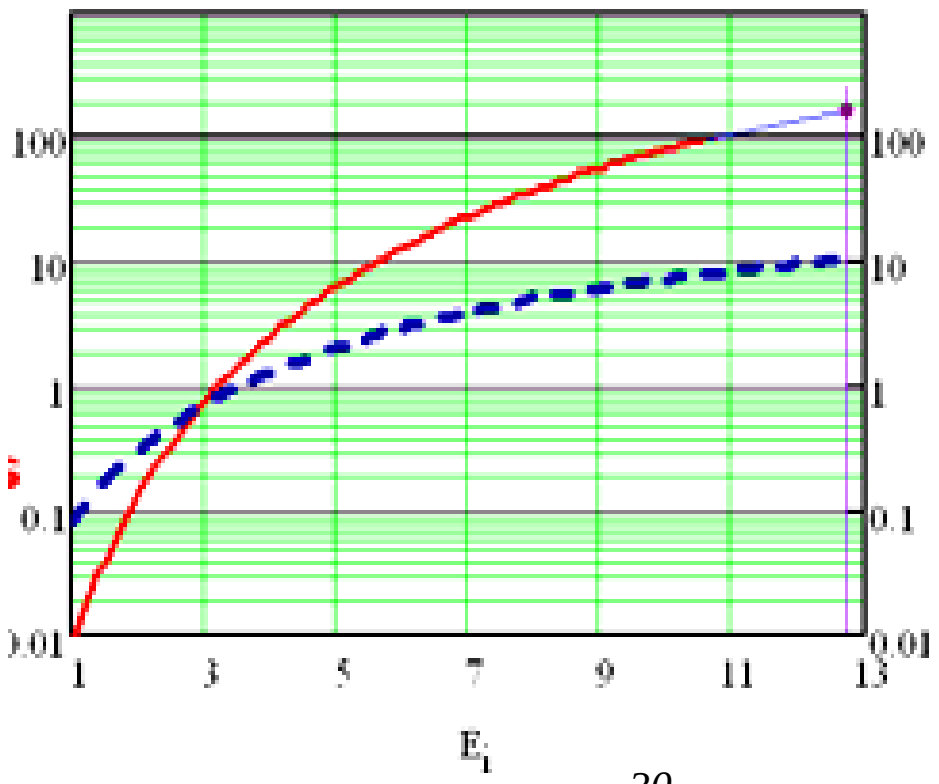

***Fig. 4.7:*** *NICA pp luminosity in units $10^{30}$ (left scale, solid line) and the number of particles per bunch in units $10^{11}$ cm $^{-2}$ s$^{-1}$ (right scale, dotted line) vs. the proton kinetic energy.*

The luminosity and total number of the stored particles has been calculated taking into account the beam space charge limits and other parameters listed below.

Parameters of NICA: circumference - 503 m,
number of intersection points (IP) - 2,
beta function $\beta_{min}$ in the IP - 0.35 m,
number of protons per bunch - $\sim 1\cdot 10^{12}$,
number of bunches - 22,
RMS bunch length - 0.5 m,
incoherent tune shift, $\Delta_{Lasslett}$ - 0.027,
beam-beam parameter, $\xi$ - 0.067,
beam emittance $\varepsilon_{nrm}$ , π mm mrad - 0.15 (normalized at 12.5 GeV).

The number of particles reaches a value about $2.2\cdot 10^{13}$ in each ring and the peak luminosity $L_{peack} = 2\cdot 10^{32}$ cm$^{-2}$s$^{-1}$ at 12.7 GeV. Assuming the cooling time $T_{cool}$ = 1500 s, the luminosity life time $T_{Llf}$ = 20000 s with the beam polarization not less than 70% and the machine reliability coefficient $k_r$ = 0.95, the average luminosity will be $L_{aver} = L_{peack}\cdot$ 0.86 or $1.7\cdot 10^{32}$ cm $^{-2}$ s$^{-1}$ [12] during the working time of the complex.

**So, feasible schemes of manipulations with polarized protons and deuterons are suggested [10, 14]. The final scheme of the polarized proton acceleration up to required energy and beam manipulations at NICA will be approved at the later stages of the NICA project.**

**4.5. Polarimetry at Linac, Nuclotron and NICA.**

Requirements to the polarization monitoring and measurements at NICA are the following:
- polarimeters should be installed at several key points of the NICA complex, namely: after Linac, inside the Nuclotron ring, at the beam transportation line to the collider and in both rings of the collider;
- evaluation of the polarization should be at the standard level for deuterons and protons;

- absolute calibration of the beam polarization should be possible;
- optimal use of the same experimental equipment at different places is desirable;
- permanent monitoring of the beam polarization is necessary.

***4.5.1. Deuteron and proton beam polarimetry in the energy range 200-2000 MeV.***

The *dp* elastic scattering at large angles (> 60º in CM system) is often used for the deuteron beam polarization measurements in the energy range 200-2000 MeV. Analyzing powers of this reaction have large values and were measured with high accuracy. The system of such measurements is designed at the LHEP (Fig. 4.9) in the framework of the project DSS (Deuteron Spin Structure). The polarimeter will use the Internal Target Station at Nuclotron. The setup is ready for operation. It will be tested further at the Nuclotron polarized deuteron beam first and then modified for measurements of the proton beam polarization in the energy range up to 5 GeV.

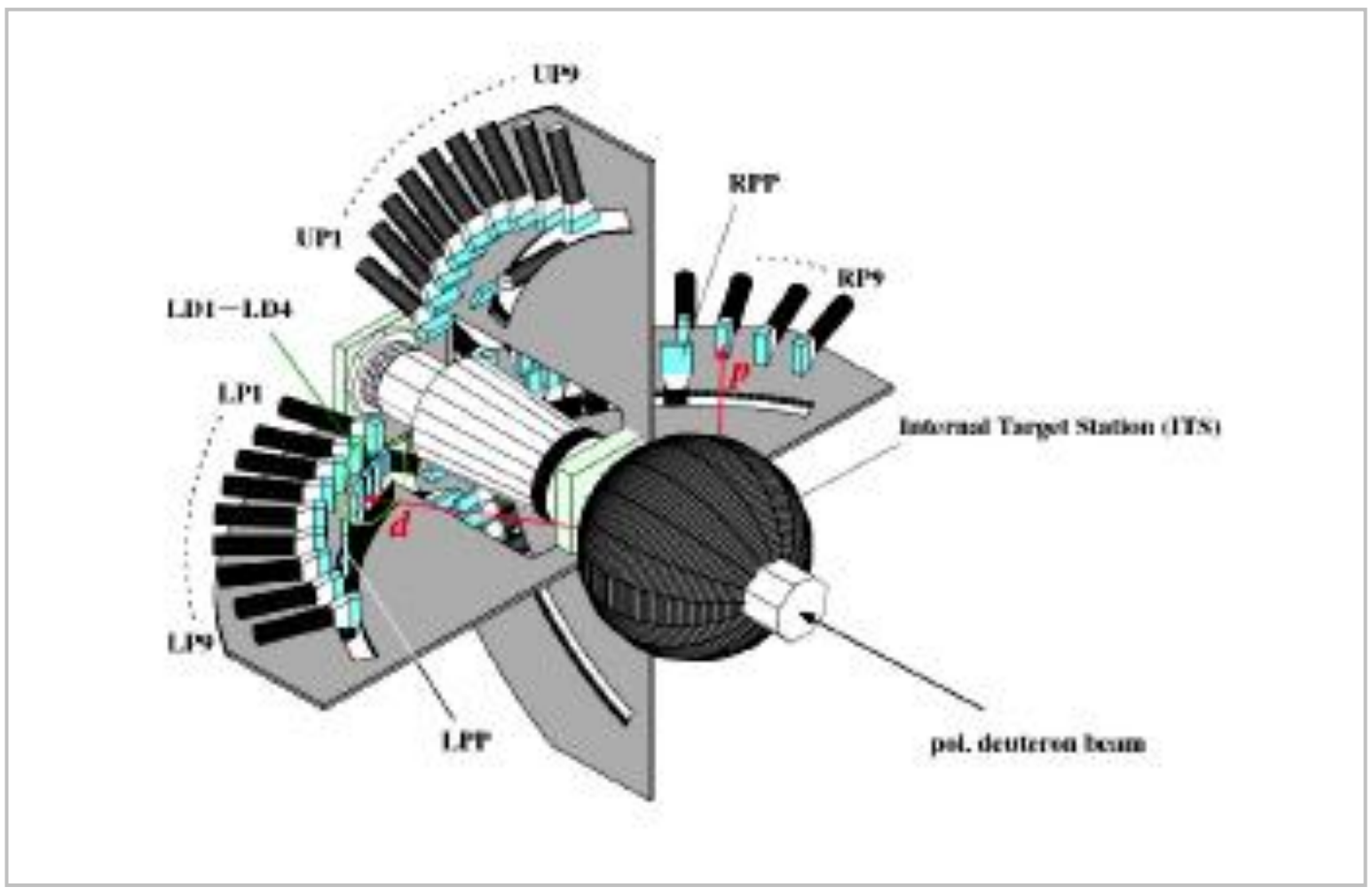

***Fig. 4.9:*** *The set-up to study dp elastic scattering at Nuclotron.*

***4.5.2. Proton beam polarimetry in the energy range above 2000 MeV.***

The proton beam polarization measurement in the energy range of NICA can be done using pC (proton-carbon) and CNI (Coulomb-Nuclear Interference) polarimeters. Since the hadronic spin-flip part of the amplitude at NICA energies is not negligible, CNI polarimeter is not an absolute one. To improve the systematic errors and to calibrate it, the polarimeter based on polarized *pp* elastic scattering will be designed. The place for polarized jet target is reserved at the collider.

Conceptual design of the NICA-SPIN polarization measurements is in progress.

## 5. Requirements to the spin physics detector (SPD).

Requirements for SPD are motivated by physics outlined in Section 2 and, first of all, by a topology of events and particles to be detected. The event topology for main processes to be studied is considered below.

SPD should operate at the highest possible luminosity. So, all the SPD sub-detectors should have high rate capabilities and preserve high efficiency during a long time.

It is useful to remember that in the energy range of NICA the total cross section of *pp* interactions is almost constant, about 40 mb, (Fig. 5.1, left), and expected event rates at the luminosity$10^{32}$ $cm^{-2}$ $s^{-1}$ will be $4{\cdot}10^6$ per second.

The average particle multiplicities estimated with PYTHIA at $\sqrt{s}$ = 24 GeV are the following: charged particles –13.5; neutral particles – 22.5. The typical invariant mass plot for di-lepton production is given in Fig. 5.1, right. The clean DY events can be detected in region of invariant mass 4 – 9 GeV.

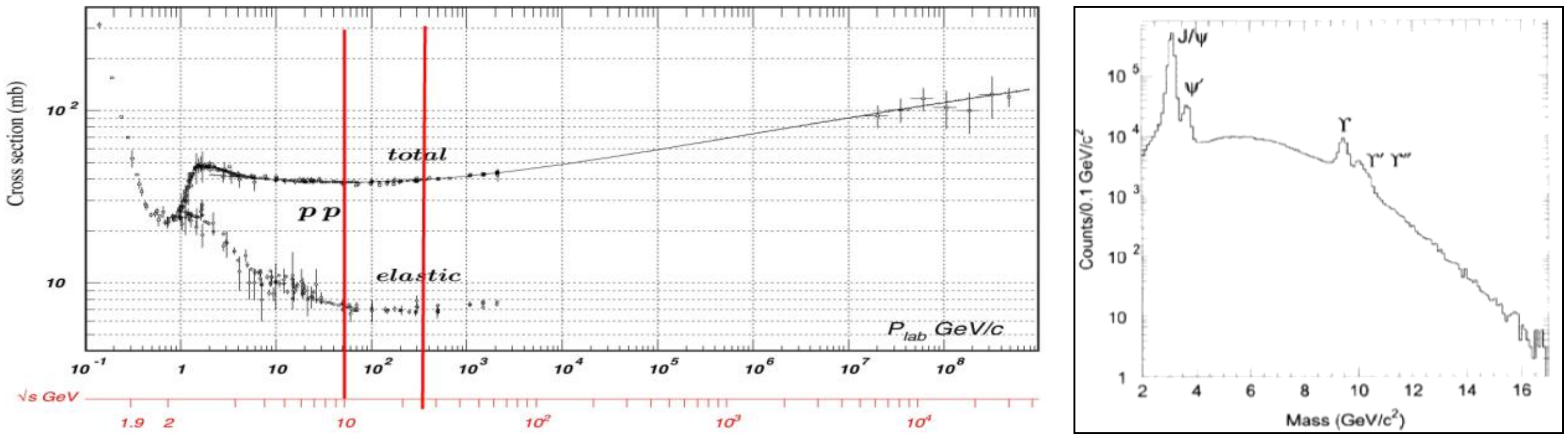

***Fig. 5.1:*** *Cross sections of pp interactions versus* $\sqrt{s}$ *(left) and the typical di-lepton invariant mass plot (right).*

## 5.1. Event topology.

### *5.1.1. Topology of DY events.*

The Feynman diagrams of the DY process and configuration of relevant vectors are given in Section 2. For physics purpose lepton pairs must be fully reconstructed using the sub-detectors of SPD. To determine a set and characteristics of the SPD sub-detectors, the DY ($\mu^-,\mu^+$) pairs to be detected were generated by MC method using the PYTHIA 6.4 code. The center of coordinates system was put at the beams intersection point (Z=0, the Z axis is along the beams).

The generated reaction is $pp \to (\mu^-,\mu^+) + X$ at $\sqrt{s}$ =24 GeV, which includes the leading order 2-2 quark level hard scattering sub-processes $q\bar{q} \to \gamma^* \to (\mu^-,\mu^+)$. The initial-state radiation (ISR) and final-state radiation (FSR) were switched on. The GRV 94L parameterization [1] of parton distributions was used. The di-muon invariant mass distribution is presented in Fig. 5.2. The cut $M_{\mu\mu}$ >2 GeV/c$^2$ was applied for other distributions.

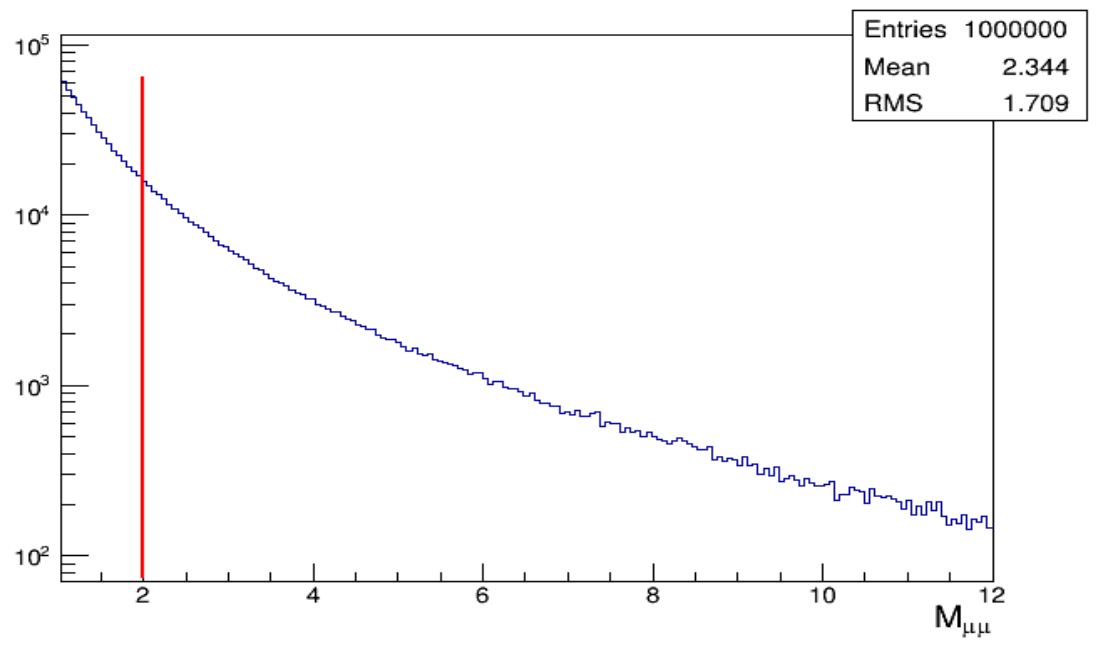

***Fig. 5.2****: Invariant mass distribution of di-muons.*

Momentum distributions of the single muon from the DY pairs with the invariant mass $M_{\mu\mu}$ >2 GeV/c$^2$ for different angular intervals volume looking from the beams intersection point (Z=0) are shown below (Fig. 5.3). The average muon momentum is equal to 2.5 GeV /c for all angles, 1.95 GeV/c – for the central angular region (35º ÷145º) and 3.5 GeV/c for the forward/backward region (3º ÷35º). The angular region 0º ÷3º, presumably, will be occupied by a beam pipe and events from this region will be lost. So, the momentum of particles to be measured in SPD is in the range ~ 0.5 - 12 GeV/c.

The distributions of the single muon polar angle and of the angle between muons in the pair are shown in Fig. 5.4. Most of the single muons are within the central angular region part of the volume. The minimal and maximal opening angles between muons are 20º and 180º, respectively. The maximal angle will be also limited by the beam pipe diameter the size of which should be minimal. These types of angular distributions require almost 4π geometry for the SPD.

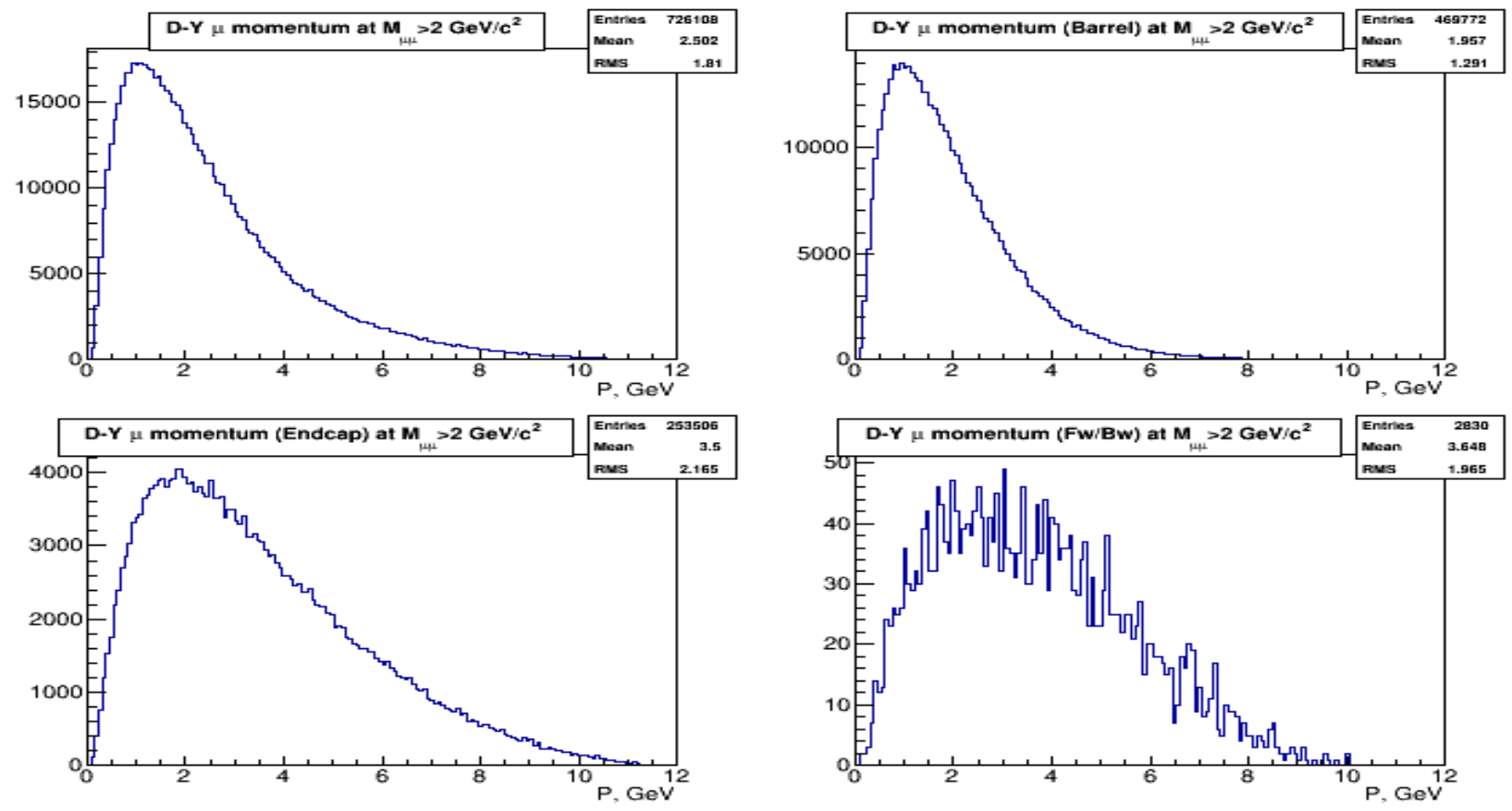

***Fig. 5.3:*** *Distributions of single muon momentum from the DY events for different angular intervals.* ***Upper:*** *left- all angles; right - 35°÷145°.* ***Bottom:*** *left- 3°÷35°, right - 0 °÷3°.*

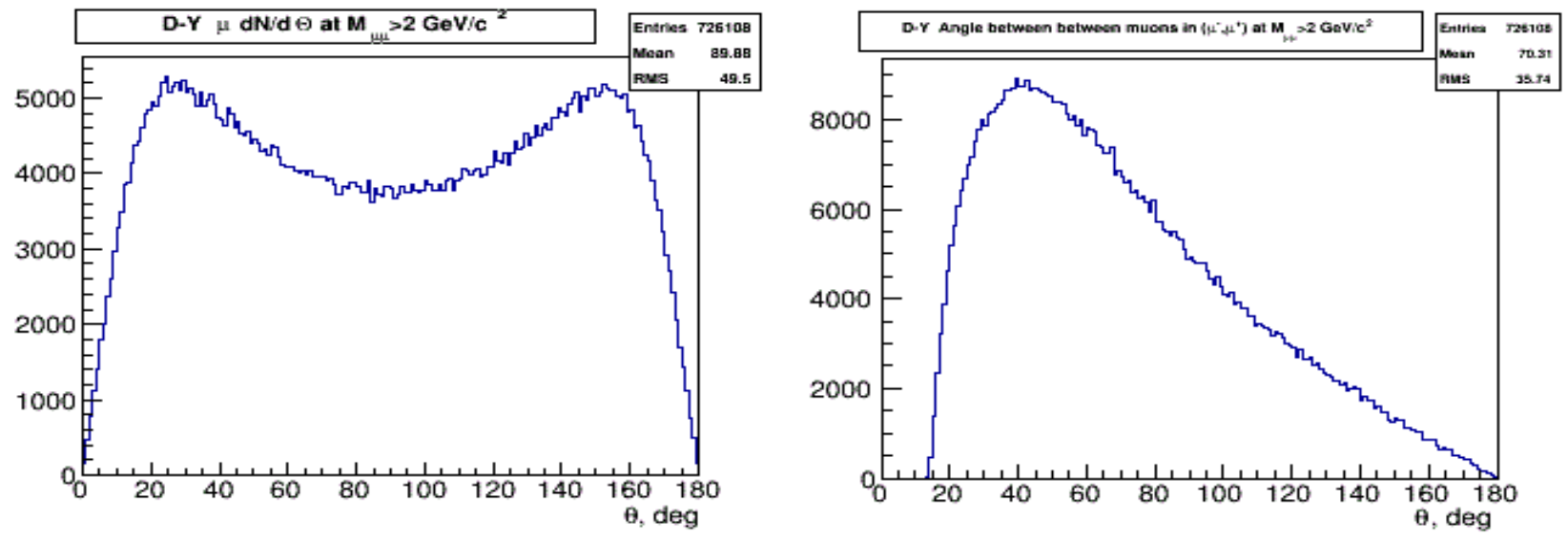

***Fig. 5.4:*** *Left – distribution of events as a function of the single muon polar angle. Right: the opening angle between two muons.*

The $e^+e^-$-pairs should have almost the same momentum and angular distributions as di-muon pairs.

The distributions of the muon transverse momentum are shown in Fig. 5.5.

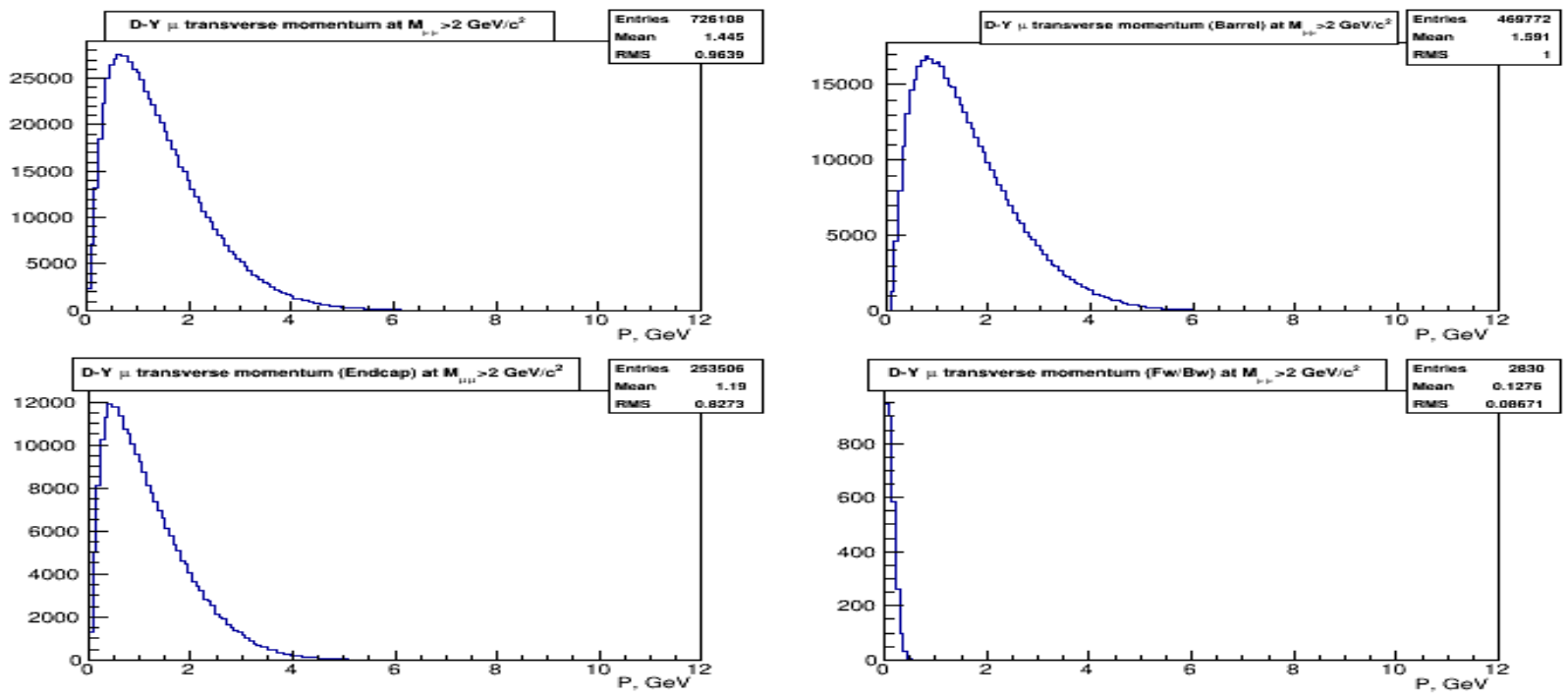

***Fig. 5.5:*** *Distributions of the muon transverse momentum from the DY events for different angular intervals. Upper: left- all angles; right - 35°÷145°. Bottom: left - 3°÷35°; right - 0 °÷3° .*

Taking into account the distributions shown above, the SPD, for the effective registration of the DY pairs, should have:

- almost 4π geometry;
- vertex detector;
- tracking system;
- momentum measurement;
- hadron, muon and electron identification systems in the energy range up to ~15 GeV.

### *5.1.2. Topology of J/Ψ events.*

The *J/Ψ* mesons produced in *pp* collisions at $\sqrt{s}$ =24 GeV and decayed into the charged lepton pairs have been simulated by MC with the PYTHIA 6.4 generator for the direct production mechanism (see Appendix). This mechanism includes the *J/Ψ* production via the processes of the gluon-gluon, gluon-quark and quark-quark fusions with production of intermediate states and its subsequent decays into the *J/Ψ*. The CTEQ 5L, LO parameterization [2] is used for the PDFs.

The momentum distributions of leptons from *J/Ψ* decays and of the opening angle of lepton pairs are shown in Fig. 5.6. The correlation between lepton polar angles is shown in Fig. 5.7. Most of the lepton pairs (61%) are within the 35º ÷145º angular interval; in 35% of pairs one lepton could be found in the 35º ÷145º angular interval whereas the other – in the 3º ÷35º interval. About 3% of leptons could be registered in the forward and backward 3º ÷35º angular intervals. A small part of the pairs will be lost due to the beam pipe. These types of angular distributions require almost 4π geometry for SPD.

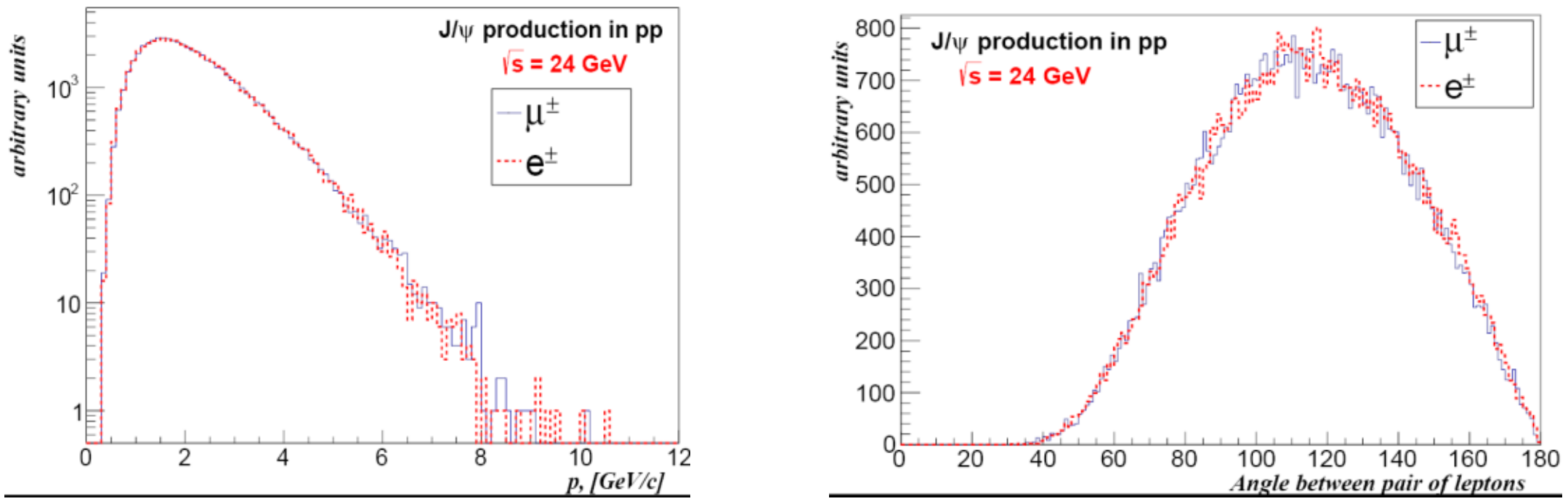

***Fig. 5.6:*** *Left - momentum distribution of leptons from J/Ψ decays; right – distribution of the lepton opening angles in the pair.*

The Feynman variable, $x_F$, and the transverse momentum, $p_T$, of directly produced *J/Ψ* mesons are shown in Fig. 5.8.

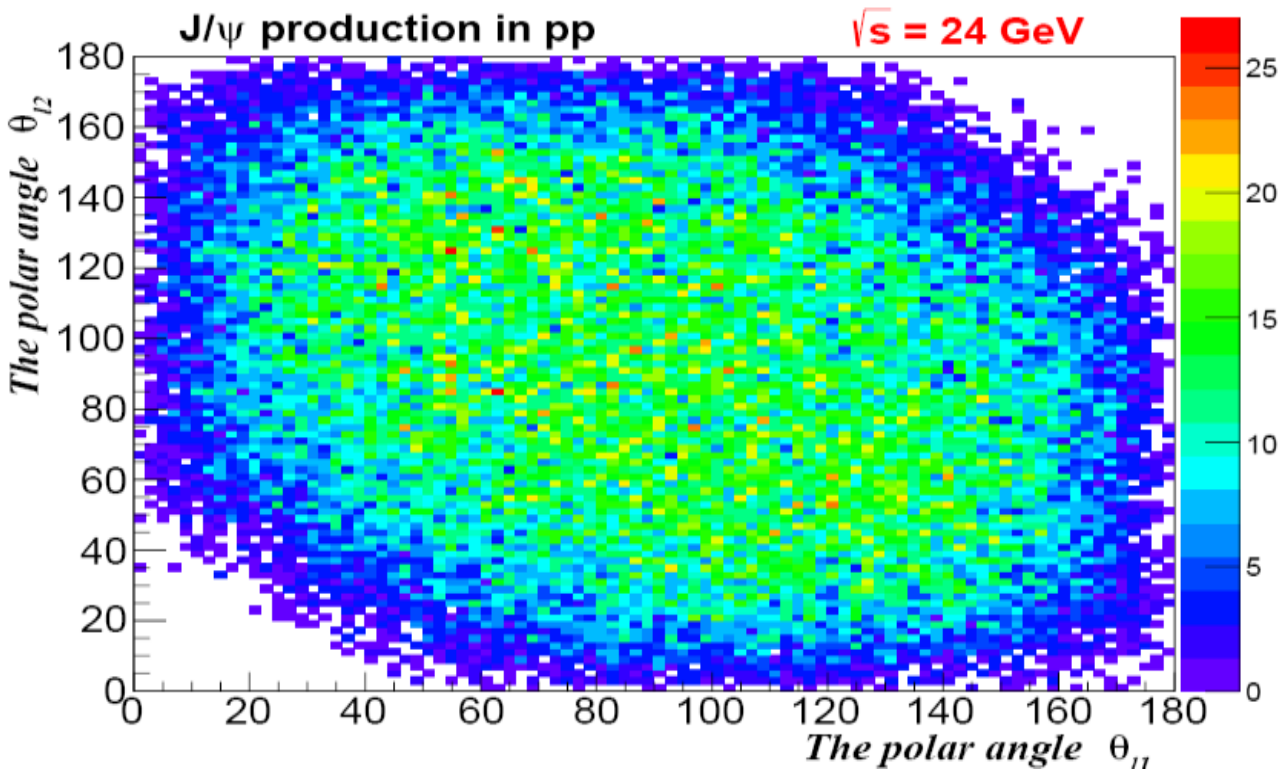

***Fig. 5.7:*** *Correlation between lepton polar angles in J/Ψ decays.*

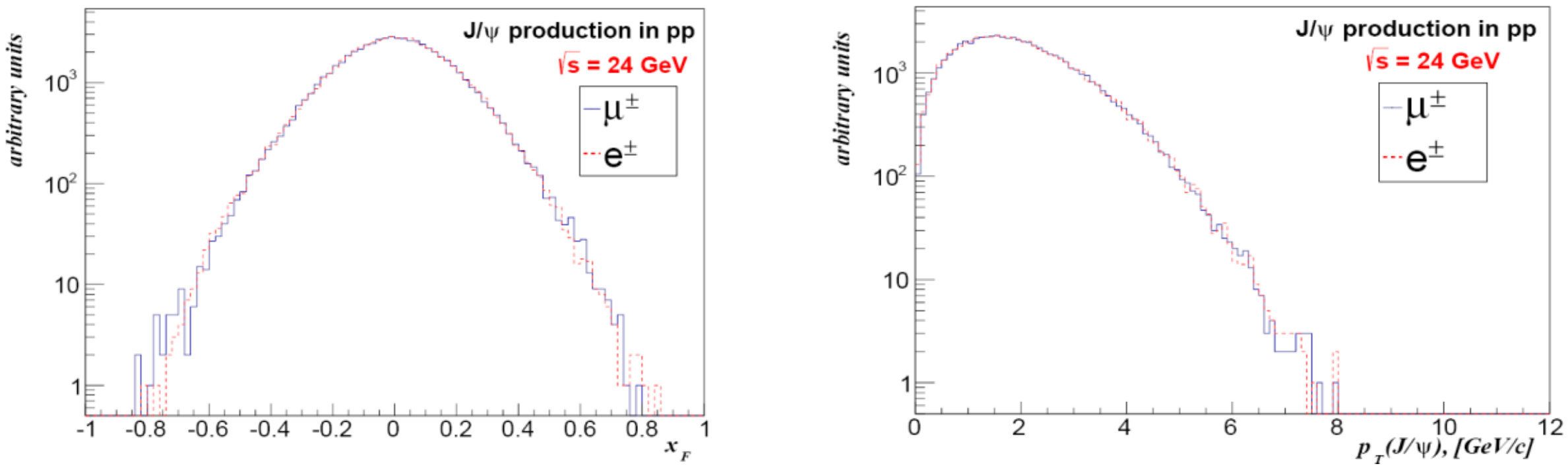

***Fig. 5.8:*** *Distributions of directly produced J/Ψ vs. $x_F$ (left) and vs. $p_T$ (right).*

### *5.1.3. Topology of the direct photon production.*

A sample of direct photons produced in *pp* collisions at $\sqrt{s}$=24 GeV has been generated by the MC method using the PYTHIA 6.4.2 code. The five hard processes with direct photons in the final state were used: $q+g \rightarrow q+\gamma$, $q+\bar{q} \rightarrow g+\gamma$, $g+g \rightarrow g+\gamma$, $q+\bar{q} \rightarrow \gamma+\gamma$ and $g+g \rightarrow \gamma+\gamma$. Relative probabilities of the first two processes are ~ 85% and 15%, respectively, while the contribution of all others is less than ~ 0.2%. CTEQ 5L is used for the set of PDFs. No special kinematic cuts are applied. The $p_T$ vs. $x_F$ distribution for direct photons is shown in Fig. 5.9.

The photon energy, $E_\gamma$, is plotted vs. the photon scattering angle, $\theta$, in Fig. 5.10 (left). The right part of this Figure shows the corresponding plot for minimum bias photons (mainly from $\pi^0$ decay). The MC simulations show that for $p_T > 4$ GeV signal-to-background ratio is about 5% that is in good agreement with the data of the UA6 experiment for non-polarized protons at $\sqrt{s}$ =24.3 GeV [3].

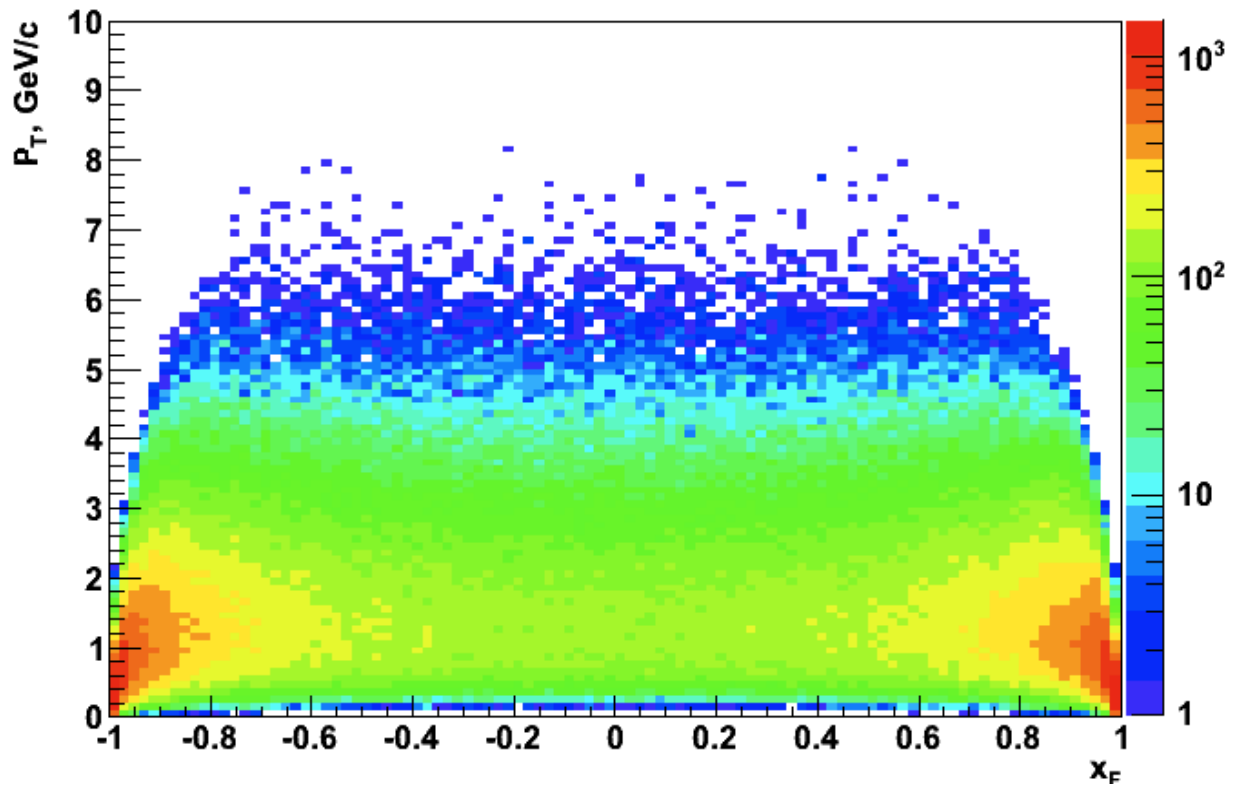

***Fig. 5.9:*** *The plot $p_T$ vs. $x_F$ for direct photons.*

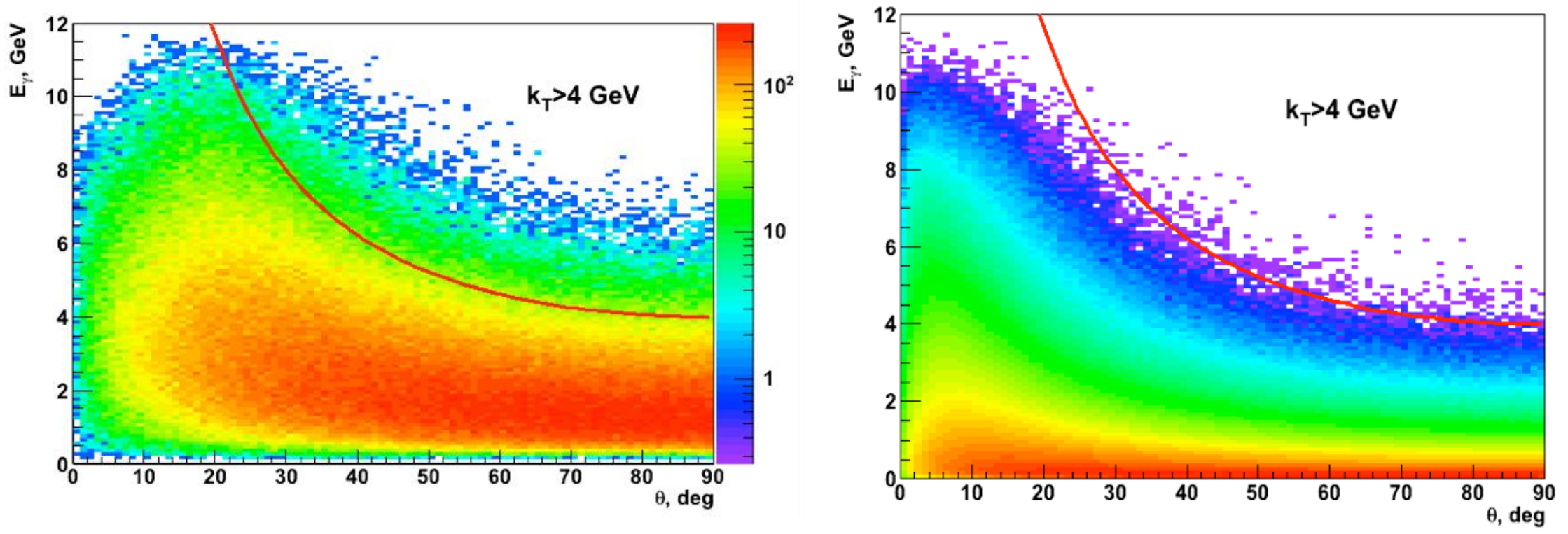

***Fig. 5.10:*** *Distribution of energy $E_\gamma$ as a function of scattering angle $\theta$:* **left** *- direct photons,* right *- minimum bias photons. Red lines correspond to the cut $p_T > 4$ GeV.*

For the effective registration and identification of direct photons, SPD should have:

- an electromagnetic calorimeter (ECAL)
- a tracking system capable to distinguish between clusters from neutral and charged particles in ECAL. It also should be capable to reconstruct the beam intersection point;
- a trigger system should include ECAL. Since for $A_N$ measurements quite energetic photons are needed only, for the main trigger one can require an energy of above 2-3 GeV deposited in any cell of ECAL;
- a DAQ system with a bandwidth up to 100 kHz;
- a luminosity monitor.

## 5.2. Possible layout of SPD.

### *5.2.1. Magnet: toroid vs. solenoid.*

Preliminary considerations of the event topologies (Sections 5.1.1 – 5.1.3) required SPD to be equipped with the following sub-detectors covering ~4π angular region around the beam intersection point: vertex detectors, tracking detectors, electromagnetic calorimeters, hadron and muon detectors. Some of them must be in the magnetic field for which there are two options: either toroid or solenoid type.

A toroid magnet provides a field free region around the beam pipe and, due to that, does not disturb the beam trajectories and polarizations. It can consist of 8 superconducting coils symmetrically placed around the beam axis (see Fig. 5.11). A support ring upstream (downstream) of the coils hosts the supply lines for electric power and for liquid helium. At the downstream end, a hexagonal plate compensates the magnetic forces to hold the coils in place. The field lines of ideal toroid magnet are always perpendicular to the particles originating from the beam intersection point. Since the field intensity increases inversely proportional to the radial distance, greater bending power is available for particles scattering at smaller angles and having higher momenta. These properties help to design a compact spectrometer that keeps the investment costs for the detector tolerable. The toroid magnet requires insertion of the coils into the tracking volume occupying a part of the azimuthal acceptance. Preliminary studies show that the use of superconducting coils, made by the *$Nb_3Sn$-Copper* core surrounded by a winding of aluminium for support and cooling, allows one to reach an azimuthal detector acceptance of about 85%.

For DY measurements there is a disadvantage of the toroid solution related to a high non-uniformity of acceptance as a function of azimuthal angle in the laboratory frame. The control of azimuthal acceptance is crucial for DY measurements, especially with transversely polarized beams. In theory such non-uniformity could be controlled by MC corrections, but in practice it may be rather difficult to provide reliable corrections when the acceptance variation vs. azimuthal angle is expected to be strong.

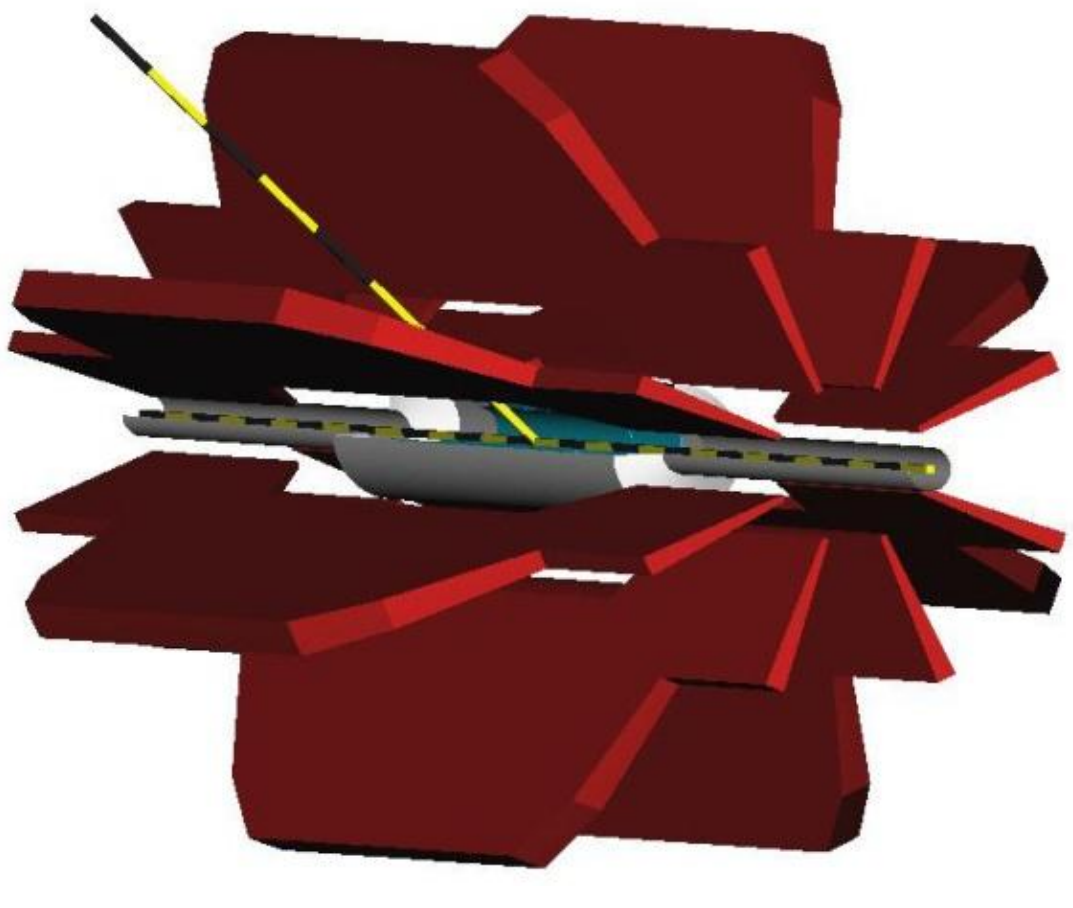

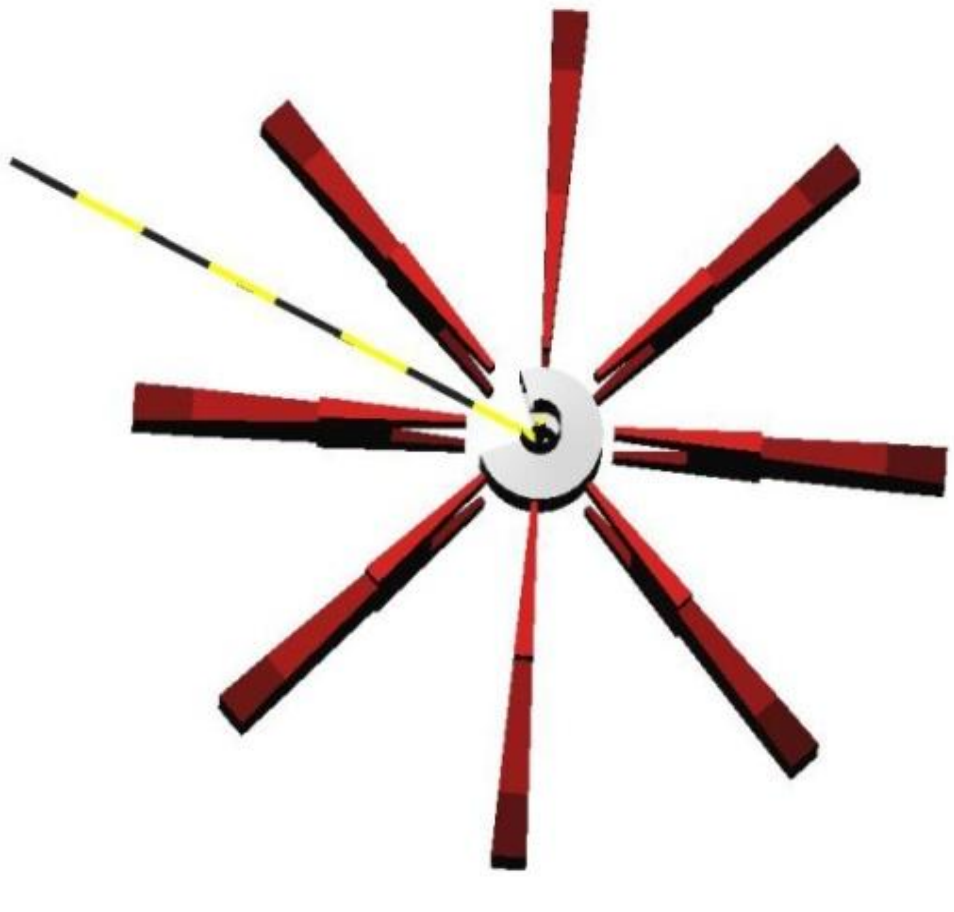

**Fig. 5.11:** Possible layout of SPD with the toroid magnet.

Possible SPD layout with the solenoid magnet is shown in Fig. 5.12. The magnet part of SPD, usually called "barrel", contains inside a vertex detector, tracking detectors and electromagnetic calorimeters (ECAL). Outside of the barrel one needs to have muon and hadron detectors. The end-cap part of SPD could contain a tracking, ECAL, and range systems. The solenoid SPD version could have almost 100% azimuthal acceptance, which is important, for example, for detection of some exclusive reactions. Disadvantage of the solenoid option is a presence of the magnetic field in the beam pipe region. This field can disturb beam particle trajectories and their polarization. Screening of this field should be studied.

The dimension of the SPD volume is still an open question. It should be optimized basing on compromise between the precisions and costs. The “almost 4π geometry” requested by DY and direct photons can be realized in the solenoid version of SPD if it has overall length and diameter of about 6 m.

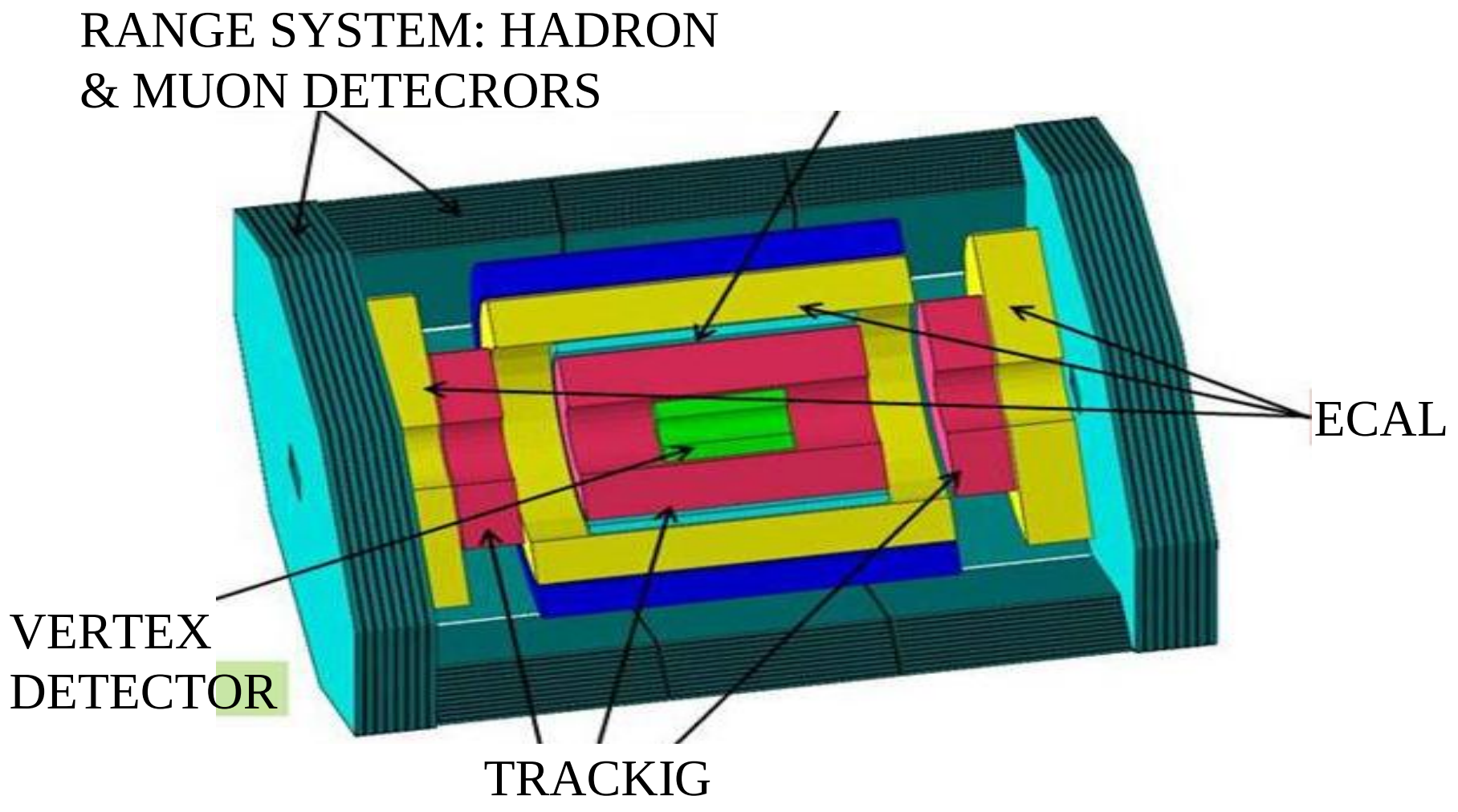

***Fig. 5.12:*** *Possible layout of SPD with the solenoid magnet.*

### *5.2.2. Vertex detector.*

The most obvious technology for the vertex detector (VD) is a silicon one (Fig. 5.13). It is approved for the MPD VD. Several layers of double sided silicon strips can provide a precise vertex reconstruction and tracking of the particles before they reach the general SPD tracking system. The design should use a small number of silicon layers to minimize the material budget. With a pitch of 50-100 μm it is possible to reach a spatial resolution of 20-30 μm. Such a spatial resolution would provide 50-80 μm for precision of the vertex reconstruction, good enough for reconstruction of the secondary decay vertices.

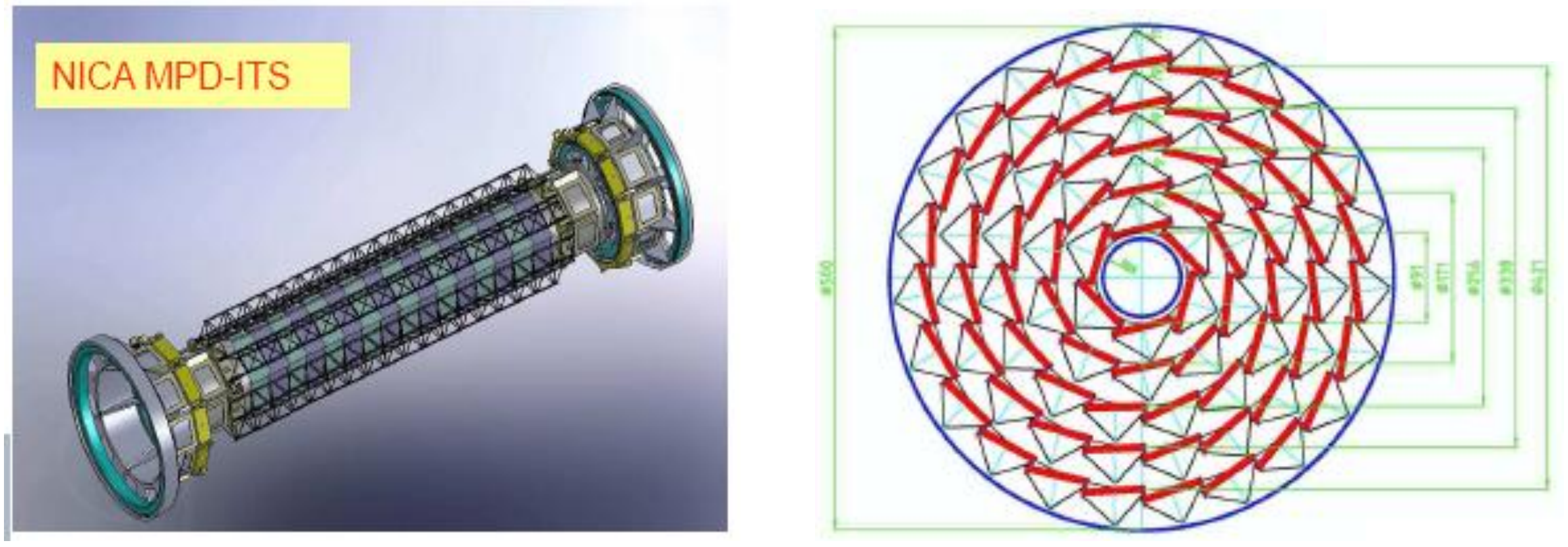

***Fig. 5.13****: Computer model of the MPD Silicon Tracking System (left) and front view of tracking layers (right).*

### *5.2.3. Tracking.*

There are several candidates for a tracking system: multiwire proportional chambers (MWPC), conventional drift chambers (DC) and their modification – thin wall drift tubes (straw chambers). The DCs are the good candidates for tracking detectors in the end-cap parts of SPD, while straw chambers are the best for the barrel part. Two groups have developed the technology of straw chamber production at JINR [6] with two-coordinate reed-out. The radial coordinate determination is organized via the electron drift time measurement while the measurement of the coordinate along the wire (z-coordinate) uses the cathode surface of the straw. Both technologies provide a radial coordinate resolution of 150-200 μm per plane. The chambers, assembled in modules consisting of several pairs of tracking planes, can have the radial coordinate resolution of about 50 μm. This can provide the momentum resolution at 0.3 – 0.5 T better than 5% over the NICA kinematic range. Straw tubes used by Baranov et al. Are made of the 30 μm nylon tape and have the coordinate resolution along the anode of about 1mm, while the Bazilev's et al. Tubes are made of double layers kapton of 25 μm thick (minimum) and have resolution along the anode of about 1 cm.

### *5.2.4. Electromagnetic calorimeters.*

The latest version of the electromagnetic calorimeter (ECAL) module, developed at JINR for the COMPASS-II experiment at CERN, Fig. 5.14 [7], can be a good candidate for ECAL in the barrel and end-cap parts of SPD. The module utilises new photon detector – Avalanche Multichannel Photon Detector (AMPD). AMPD can work in the strong magnetic fields. The modules have rectangular shape but can be produced also in the projective geometry which is better for SPD. The energy resolution of the module is about 10% at 1 GeV. The modules have a fast readout and can be used in the SPD trigger system. The module has 109 plates of the scintillator and absorber (*Pb*) of 12x12 cm in cross section and 0.8 (1.5) mm thick, respectively.

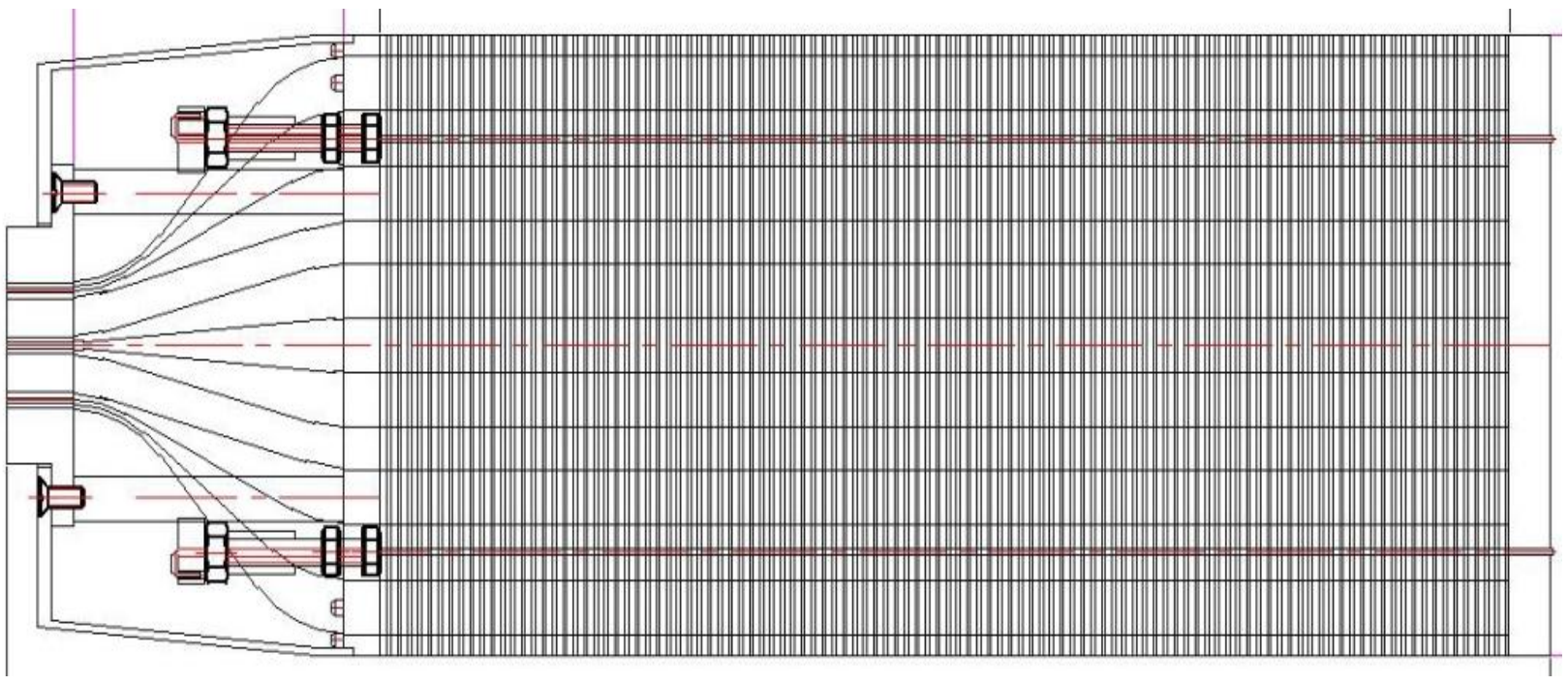

***Fig. 5.14:*** *ECAL module structure.*

The radiation length and Moliere radius is 1.64 and 3.5 cm, respectively. The light collection is performed with optical fibers dividing the module in nine logical sections (towers).

### *5.2.5. Hadron (muon) detectors.*

A system of mini-drift chambers interleaved with layers of iron and called the Range System (RS) is developed at JINR for FAIR/PANDA [8] (see Fig. 5.15). It can be used in the barrel part of SPD as a hadron and (or) muon detector for the Particle IDentification system (PID). RS can provide clean (> 99%) muon identification for muon energies greater than 1 GeV. The combination of responses from ECAL, RS and momentum reconstruction can be used for the identification of electrons, hadrons and muons in the energy range of the SPD.

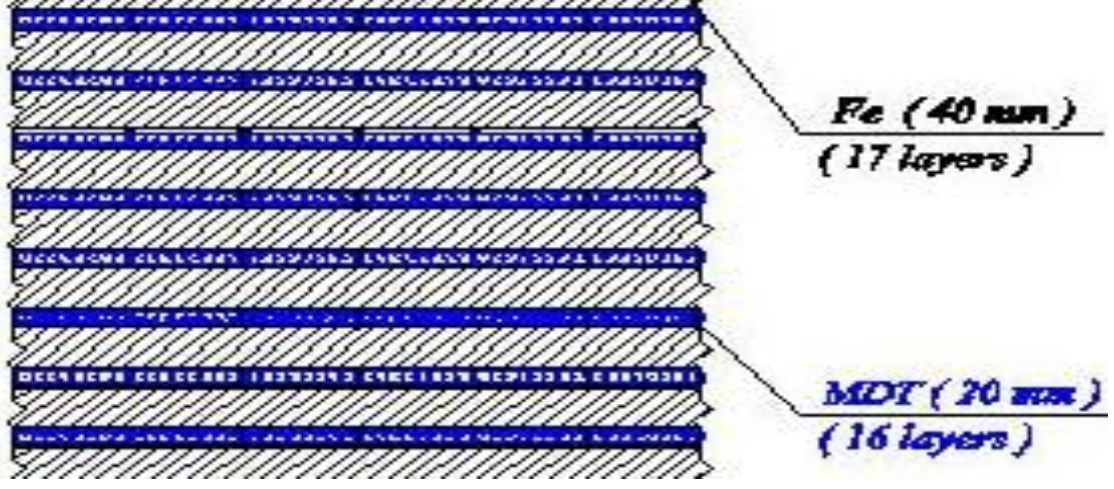

***Fig. 5.15:*** *Scheme of the RS. Dimension and thickness are subjects of optimization.*

The hadron and muon detectors in the end-cap parts of SPD are to be identified. As candidates for these detectors the COMPASS muon wall [9] can be considered. It consists of two layers of mini-drift chambers with a block of absorber between them.

The more elegant system for hadron and muon detectors of SPD can be constructed using calorimeters suggested for the future linear collider [11] (see Fig. 5.16). The prototype of the calorimeter module is under the tests. The module includes an electromagnetic and hadron parts. The hadron part consists of the 38 layers of iron (20 mm) and scintillator (5 mm) plates. The scintillator plate includes 216 tiles of 3·3, 6·6 and 12·12 cm. The light collection is performed with WLS fibers to the silicon PM with 1156 pixels and gain of ~$10^5$. This type of calorimeters can be used both in the barrel and end-cap parts of SPD, as well as in trigger system and as internal monitors of the beam polarization.

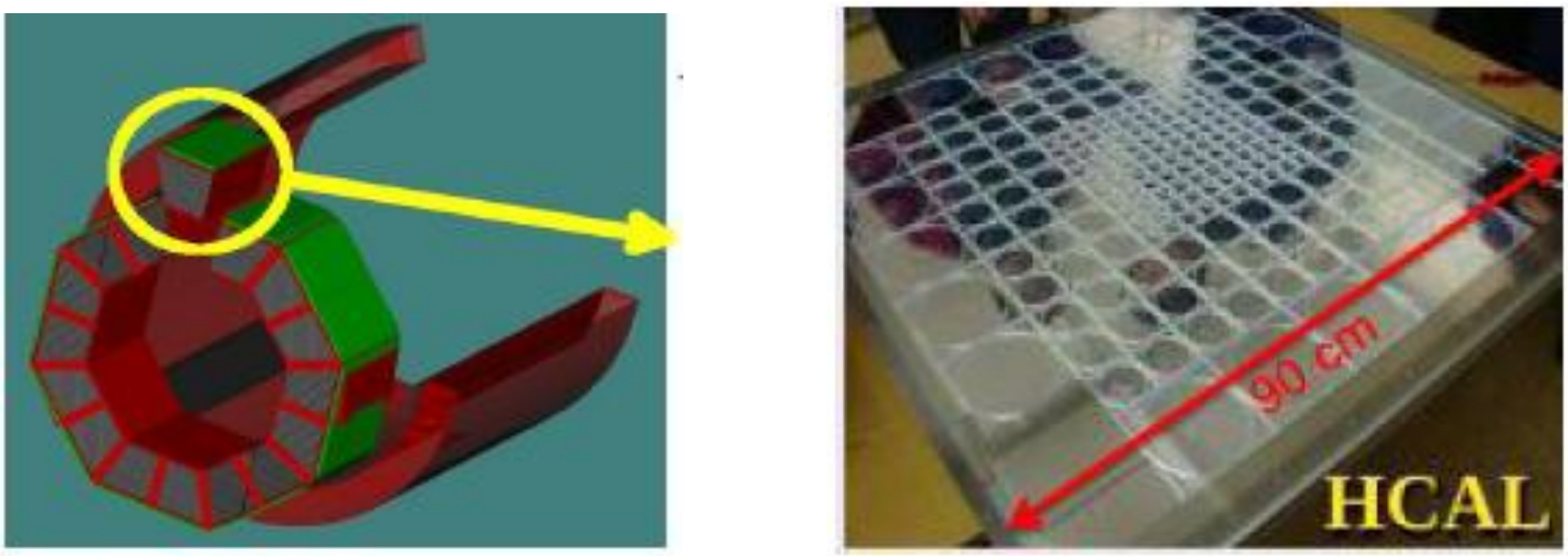

***Fig. 5.16:*** *The calorimeter modules for future linear collider (left) and structure of the hadron part of the calorimeter.*

### 5.3. Trigger system. (To be updated)

The main task of the trigger system is to provide separation of a particular reaction from all reactions occurred in collisions. Each of them will be pre-scaled with:

- two muons in the final state;
- electron/positron pairs in the final state;
- direct photons (decays of $\pi^0$, ω, η…);
- various types of charged hadrons in final states ( π+/-, K, p, …);
- other reactions.

Hodoscopes of scintillating counters and resistive plate chambers (RPC, Fig. 5.16 [10]) are proposed as option detectors for the SPD trigger system. The hodoscopes can be located before and after RS (or mounted in the last layers of RS). The ECAL modules will also be used in the trigger system.

The trigger system must accept an event rate of about 4 MHz, expected at the highest energy and luminosity, and reduce it to the DAQ level of about 100 KHz. For that it should be multilevel.

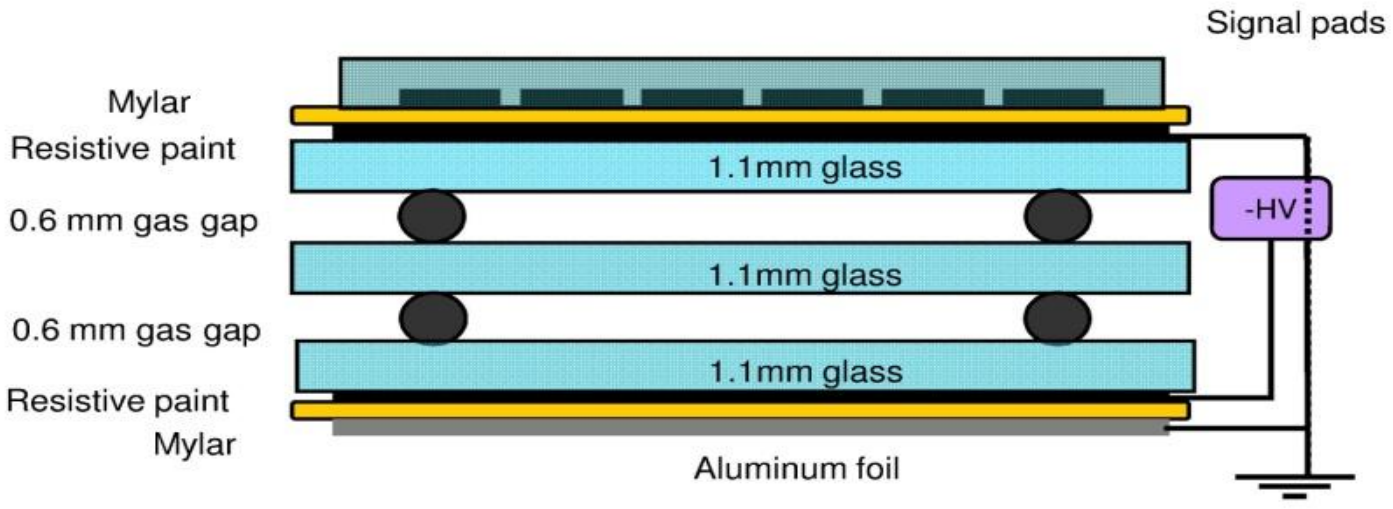

***Fig. 5.16**: Scheme of the RPC unit.*

## 5.4. Local polarimeters and luminosity monitors (to be updated)

### *5.4.1. Local polarimeters.*

Local polarimeters should provide information on the beam(s) polarization(s) at the beams intersection point. It means they should be incorporated in the SPD sub-detector system. Reaction, which can be used for this purpose, is an inclusive production of $\pi^0$ and $\pi^{\pm}$ mesons: *pp (pd, dd)* $\rightarrow \pi + X$. The single spin asymmetry, $A_N$, as a measure of the beam polarization, is rather large (see Fig. 5.17, left) and energy independent in the region of proposed experiments. The mesons could be detected with the ECAL and HCAL modules of the end-cap parts of SPD, as mentioned in Section 5.2.5. For studies of elastic (quasi-elastic) *pp*, *dp*, *dd* scattering reactions, SPD should be upgraded with a kind of specialized forward detector (FD) possessing tracking capabilities. The MicroMegas detectors will be good candidates for this purpose.

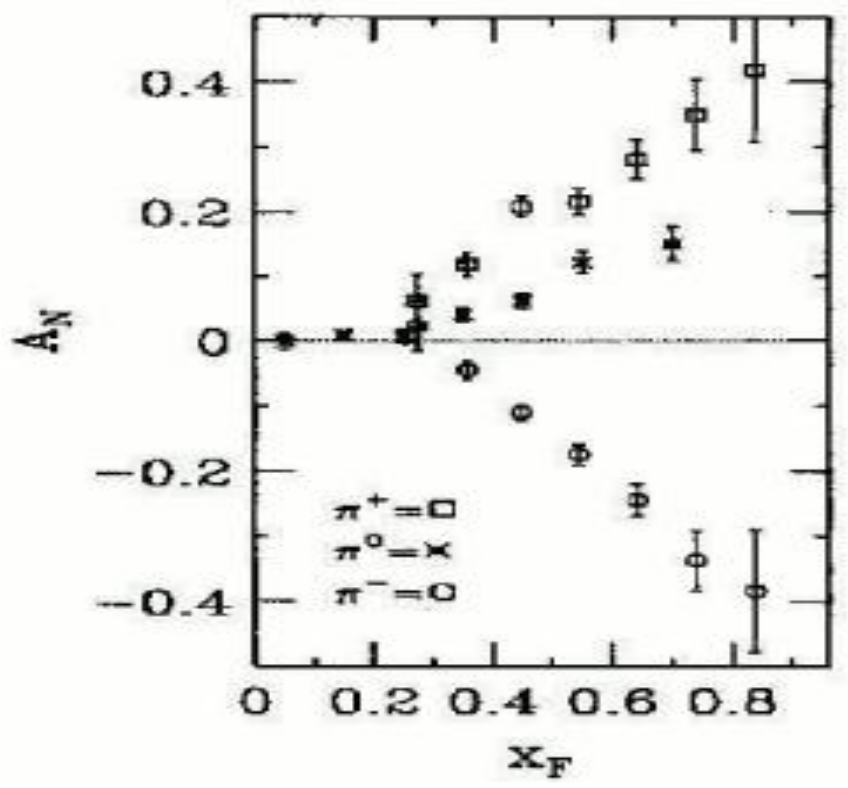

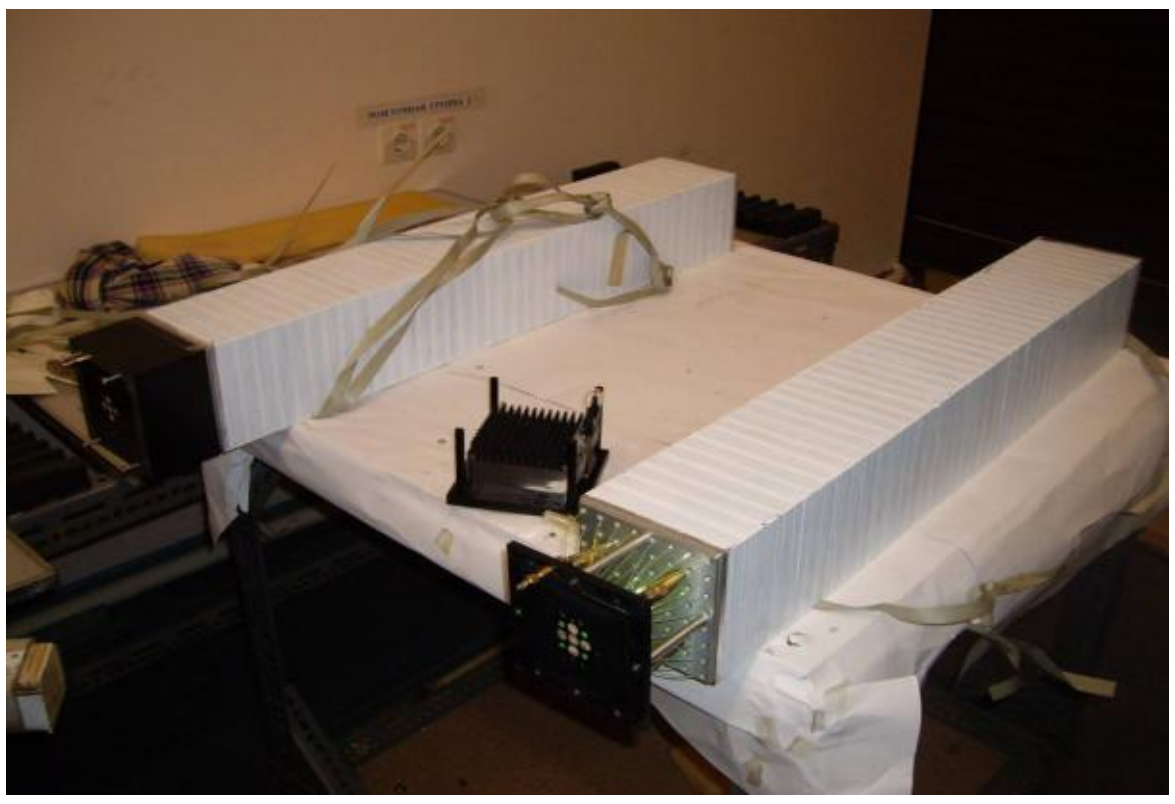

***Fig. 5.17: Left**: the single spin asymmetry for π mesons as a function of $x_F$.*
***Right**: the ZDC prototype modules developed at JINR*

### *5.4.2. Local luminosity monitors.*

The luminosity monitoring at SPD can be performed with the Zero Degree Calorimeters (ZDC) similar to those used at RHIC [10], or with ECAL modules of the SPD, or with ZDC similar to one developing for MPD at JINR. The last one is the "shashlyk"-type calorimeter consisting of electromagnetic (2 mm of *Pb* + 4 mm of scintillator, 40 layers) and hadron (16 mm of *Pb* + 4 mm of scintillator) parts. The total thickness of the calorimeter is 130 $X_0$ or 8.3 $\lambda_I$ (see Fig. 5.17, right). The design of local luminosity monitors will be proposed after finalizing the design of SPD.

## 5.5. Engineering infrastructure: the experimental area.

The plan view of the experimental area for SPD, extracted from the official NICA project documents (see preliminary drawing 318Б-063К-АР-АР, sheet 3), is shown in Fig. 5.18. There will be two rooms in the area – room numbers 128/1 and 128/2. The detector itself in the working position will be located in the room 128/1 (right side of the upper view in Fig. 5.18). The partial assembling/disassembling and maintenance of the detector can be performed in this room. The room 128/2 (left side of the upper view) is planned for general assembling of SPD. It is also a garage position for SPD between the working sessions.

Dimensions of the rooms (along/across the beams) are: for 128/1 – 22.5 m x 25 m= 562.5 m$^2$, for 128/2 – 24 m x 42 m = 1008 m$^2$. Both rooms have a height 19.85 m from the floor level to the roof. The floor is reinforced to keep the uniformly distributed weight 2 t/m$^2$ in the room 128/1 and 16 t/m$^2$ in 128/2. The whole area (128/1 and 128/2) is located in a hollow, depth 3.49 m below the median plane of the Collider (1.99 m below clean floor level of the Collider).

SPD, assembled on a rolling cart platform in the room 128/2, will be transported to 128/1 by rails. The total weight of assembled SPD should be less than 1200 tons. The assembly room 128/2 is equipped with a bridge crane of 50 tons lifting capacity. Crane provides the movement of the SPD components from the unloading space to the assembly space. The height from the floor to the bottom of the crane hook is 18.5 meters. The crane service zone is 22 m long in transverse direction. The crane has additional hook with lifting capacity of 10 t.

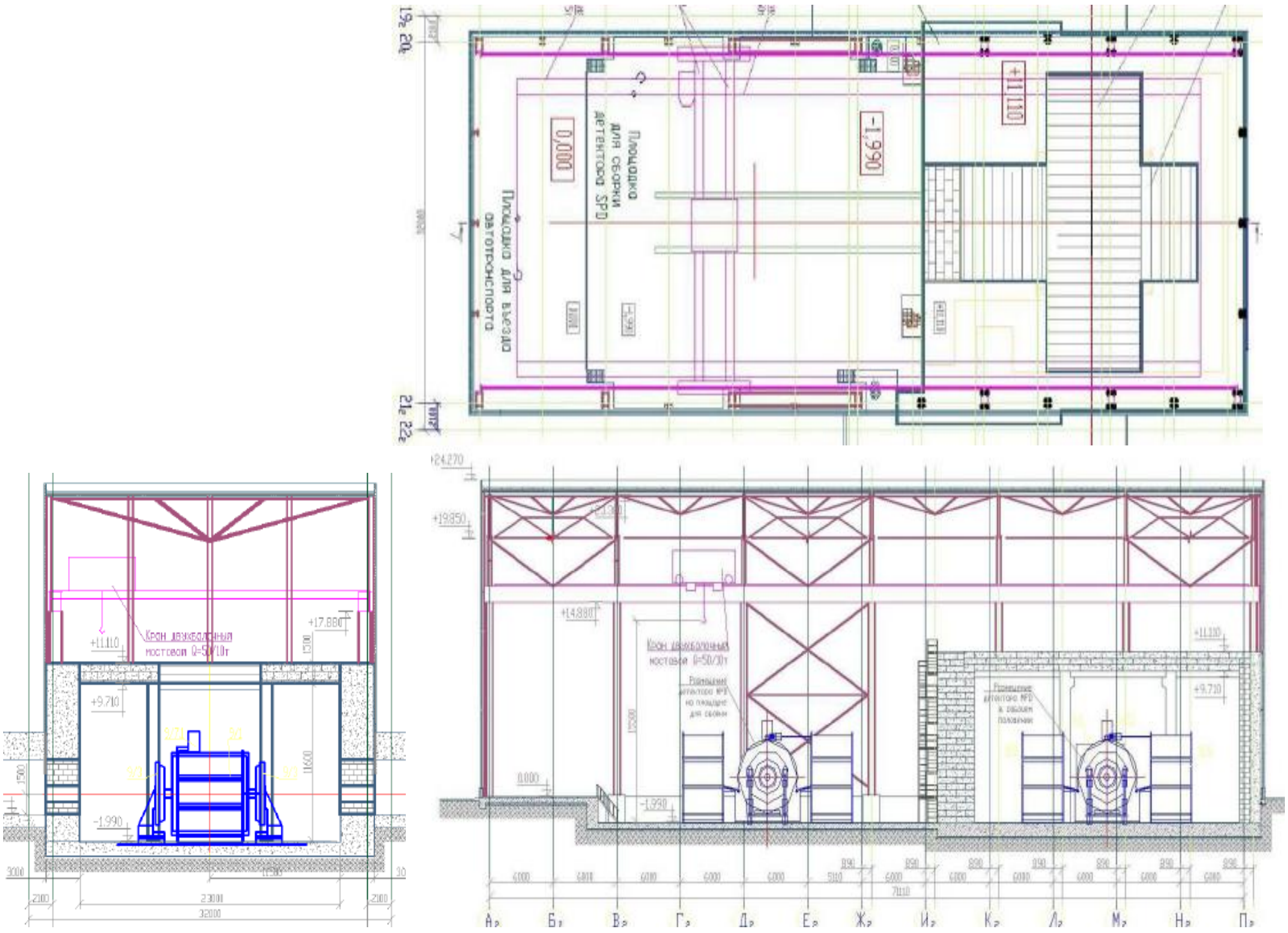

***Fig. 5.18:*** *Drawings of the SPD experimental area.*
***Upper****: a view from the top.* ***Down****: views from the sides.*

## 5.6. Data acquisition (DAQ) and data base (DB) systems. (to be updated)

### *5.6.1. SPD data acquisition system.*

As an example of possible SPD DAQ, one can consider the modernizing DAQ of the COMPASS experiment [12].

### *5.6.2. Data base of SPD.*

The data base (DB) of the experiment is a backbone of its software system. DB should contain information on the experiment configuration, data on sub-detectors, information on physics and calibrations data, software logistics, user’s information, etc. The SPD DB design can be similar to some of the high energy physics experiment [13].

Sections 5.7 – 5.10 to be written at the Proposal stage of the project.

# 6. Proposed measurements with SPD.

We propose to perform measurements of asymmetries of the DY pair's production in collisions of polarized protons and deuterons (Eqs. 2.1.0) which provide an access to all collinear and TMD PDFs of quarks and anti-quarks in nucleons. The measurements of asymmetries in production of *J/Ψ* and direct photons will be performed simultaneously with DY using dedicated triggers. The set of these measurements will supply complete information for tests of the quark-parton model of nucleons at the twist-two level with minimal systematic errors.

## 6.1. Estimations of DY pairs and *J/Ψ* production rates.

### *6.1.1. Estimations of the DY production rates and precisions of asymmetry measurements.*

Estimation of the DY pair's production rate at SPD was performed using the expression [1] for the differential and total cross sections of the *pp* interactions:

$$\frac{d^2\sigma}{dQ^2 dx_1} = \frac{1}{sx_1}\frac{4\pi\alpha^2}{9Q^2}\sum_{f,\bar{f}} e_f^2[f(x_1,Q^2)\bar{f}(x_2,Q^2)]_{x_2=Q^2/sx_1}$$

$$\sigma_{tot} = \int_{Q^2_{min}}^{Q^2_{max}} dQ^2 \int_{x_{min}}^{1} dx_1 \frac{d^2\sigma}{dQ^2 dx_1}, \qquad (6.1.1)$$

where $Q$ is the invariant mass of the lepton pair, $M_{l-l+}$, $x_1$ ($x_2$)≡ $x_a$ ($x_b$) is the Bjorken variable of colliding hadron, $s$ is the *pp* center of mass energy squared. The Table 2 shows values of the cross-sections and expected statistics for DY events per 7000 hours of data taking and 100% acceptance of SPD at two energies.

Table 2: Estimation of the cross-section and number of DY events.

| Lower cut on $M_{l+l-}$ , GeV | 2.0 | 3.0 | 3.5 | 4.0 |
|---|---|---|---|---|
| $\sqrt{s}$=24 GeV ($L$ = 1.0·$10^{32}$ cm$^{-2}$ s$^{-1}$) | | | | |
| $\sigma_{DY}$ total, nb | 1.15 | 0.20 | 0.12 | 0.06 |
| events per 7000h, $10^3$ | 1800 | 313 | 179 | 92 |
| $\sqrt{s}$=26 GeV ($L$ = 1.2·$10^{32}$ cm$^{-2}$ s$^{-1}$) | | | | |
| $\sigma_{DY}$ total, nb | 1.30 | 0.24 | 0.14 | 0.07 |
| events per 7000h, $10^3$ | 2490 | 460 | 269 | 142 |

The dependence of the total cross section and of number of DY events per year versus the cut on the minimal $M_{l-l+}$ is shown in Fig. 6.1.

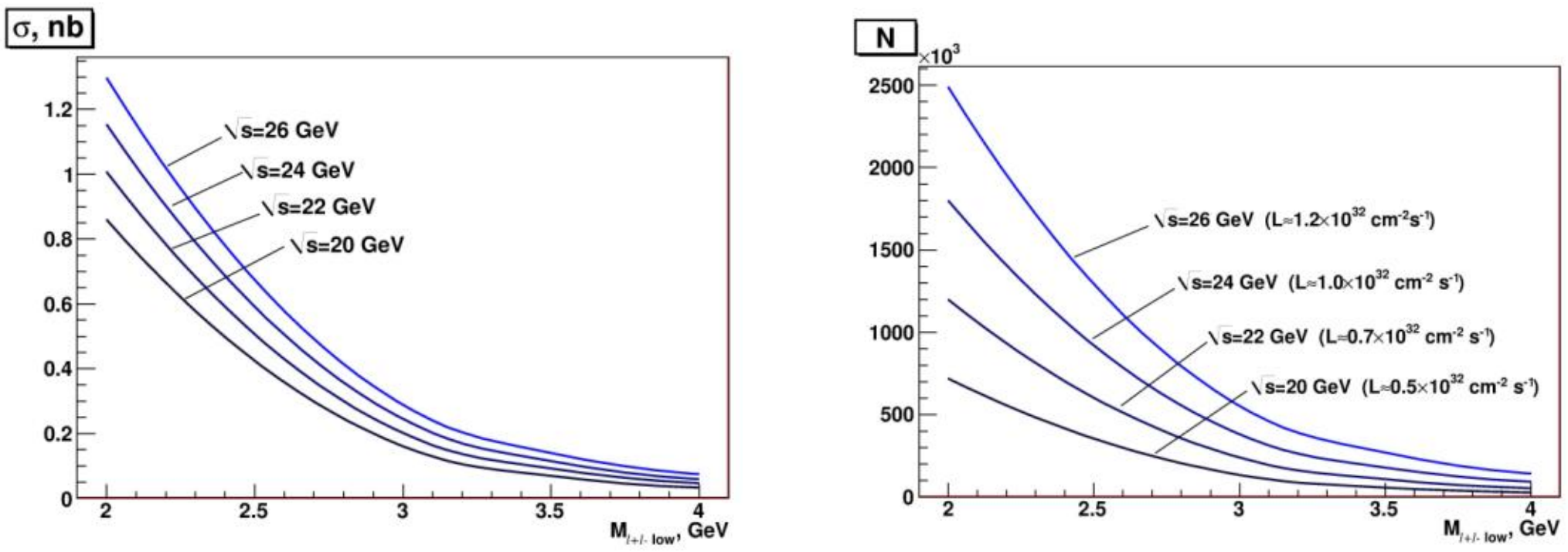

***Fig. 6.1****: Cross section (left) and number of DY events (right) versus the minimal invariant mass of lepton pair for various proton beam energies.*

To estimate the precision of measurements, the set of original software packages for MC simulations, including generators for Sivers, Boer-Mulders and Transversity PDFs, were developed [2]. With these packages a sample of 100K DY events was generated in the region of $Q^2 > 11$ GeV$^2$ for comparison with expected asymmetries.

Let us first estimate the $q_T$- weighted integrated asymmetry (Sivers) $A_{UT}^{w\left[\sin(\phi-\phi_S)\frac{q_T}{M_N}\right]}\Big|_{pp\uparrow\to l^+l^-X}$ given by Eq. (2.1.12). For this purpose three different fits for the Sivers function have been used: Fit I: $xf_{1uT}^{\perp(1)} = -xf_{1dT}^{\perp(1)} = 0.4x(1-x)^5$; Fit II: $xf_{1uT}^{\perp(1)} = -xf_{1dT}^{\perp(1)} = 0.1x^{0.3}(1-x)^5$ of Ref.[3] and Fit III: $xf_{1uT}^{\perp(1)} = -xf_{1dT}^{\perp(1)} = (0.17...0.18)x^{0.66}(1-x)^5$ of Ref. [4]. For the first moment of the Sivers PDF entering Eq. (2.1.12) the model (with the positive sign) proposed in Ref. [4] is used:

$$\frac{\bar{f}_{1qT}^{\perp(1)}}{f_{1qT}^{\perp(1)}} = \frac{\bar{f}_{1u}(x) + \bar{f}_{1d}(x)}{f_{1u}(x) + f_{1d}(x)}. \qquad (6.1.2)$$

The estimated asymmetry as a function of $x_p - x_{p\uparrow}$ is shown in Fig. 6.2.

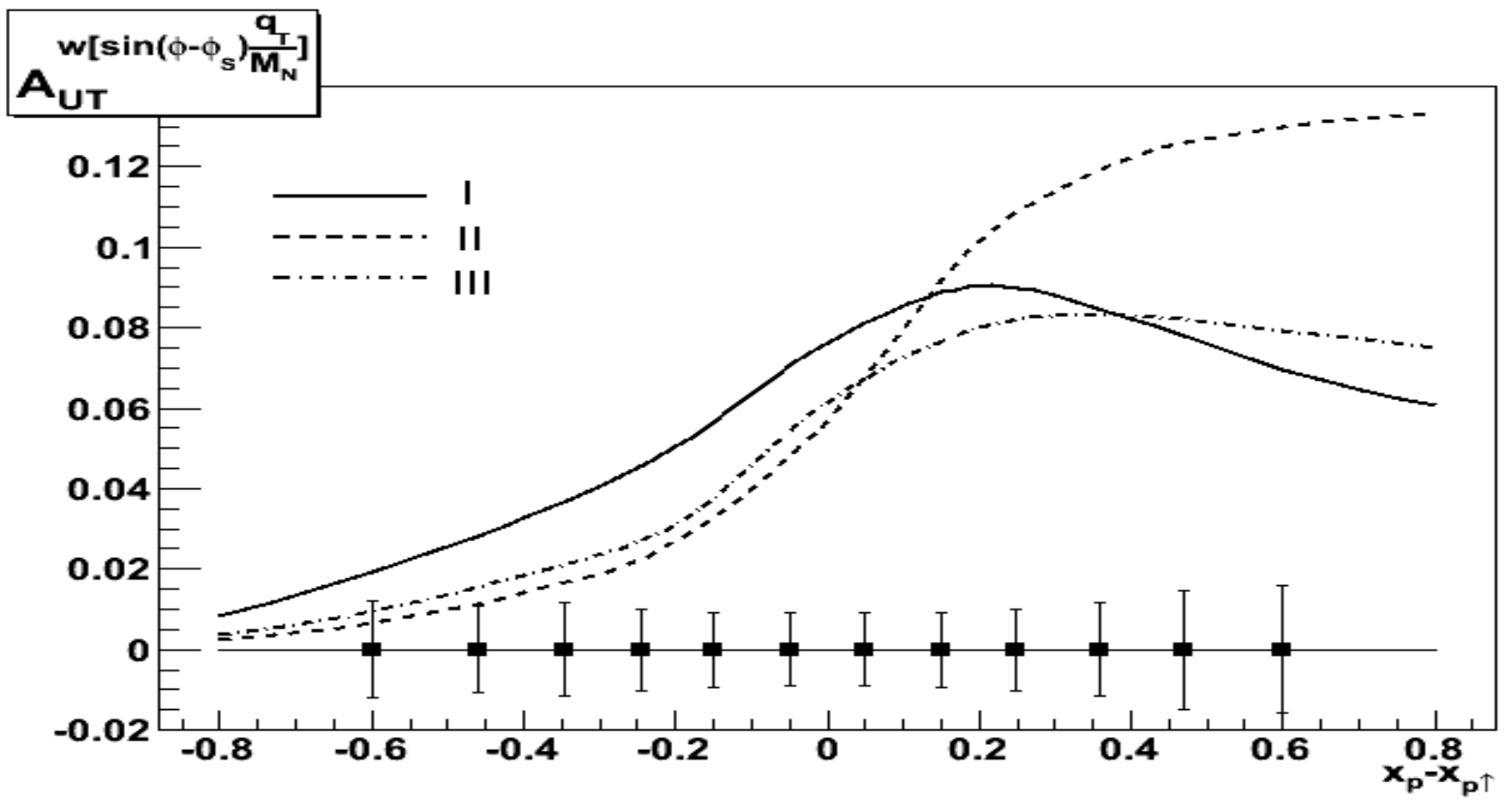

***Fig. 6.2:*** *Estimated Sivers asymmetry* $A_{UT}^{w\left[\sin(\phi-\phi_S)\frac{q_T}{M_N}\right]}$ *at* $\sqrt{s}$ *=26 GeV with* $Q^2$ *= 15 GeV*$^2$*. Numbers I, II, III denote corresponding fits. Points with the expected statistical errors correspond to 100K of generated events.*

As one can see from this Figure, the expected integrated Sivers asymmetries depend on the PDF parameterization and vary in the whole region of $x_p - x_{p\uparrow}$ from about 1 to 12%. Statistics of 100K is marginally enough to distinguish between the fits.

Now one can estimate the $q_T$- weighted integrated asymmetry (Boer-Mulders) $A_{UT}^{w\left[\sin(\phi+\phi_S)\frac{q_T}{M_N}\right]}\Big|_{pp\uparrow}$ given by Eq. (2.1.13). Since the Boer-Mulders PDF and its first moment are still poorly known, the Boer's model (Eq. (50) in Ref.[5]) has been used which provides the good fit for the NA10 [6] and E615 [7] data on the anomalously large cos*(2φ)* dependence of DY cross sections. This model gives for the first moment (2.1.15) entering Eq. (2.1.13) the value $h_{1q}^{\perp(1)}(x) = 0.163 f_1(x)$. For the first moment of the Boer-Mulders sea part PDF, the following relation is used:

$$\frac{\bar{h}_{1qT}^{\perp(1)}(x)}{h_{1T}^{\perp(1)q}(x)} = \frac{\bar{f}_{1q}(x)}{f_{1q}(x)}. \qquad (6.1.3)$$

The Transversity PDF $h_1$ was extracted recently from the combined data of HERMES, COMPASS and BELLE collaborations. However, due to large experimental uncertainties, in a course of extraction a number of approximations were used. Particularly the sea part of Transversity PDF was assumed to be zero. But, in the case of *pp* collisions, the sea PDFs play the important role. That is why two versions of the evolution model for the transversity are considered here. In the first version of the model the transversity for quarks and anti-quarks

$$h_{1q}(x,Q_0^2)=\frac{1}{2}\left[q(x,Q_0^2)+\Delta q(x,Q_0^2)\right],\ \ \bar{h}_{1q}(x,Q_0^2)=\frac{1}{2}\left[\bar{q}(x,Q_0^2)+\Delta\bar{q}(x,Q_0^2)\right] \quad (6.1.4)$$

are assumed to be equal to the helicity PDF $\Delta q$ ($h_{1q}=\Delta q,\ \bar{h}_{1q}=\Delta\bar{q}$) at the low initial $Q_0^2=0.23\ GeV^2$, and then they are evolved with DGLAP equations. In the second model [8, 9] the transversity PDFs are assumed to be equal to $h_{1q}=(\Delta q+q)/2\ and\ \bar{h}_{1q}=(\Delta\bar{q}+\bar{q})/2$ at the same initial scale, and then $h_{1q}$ and $\bar{h}_{1q}$ are again evolved with DGLAP. This model is considered as more realistic one. The results of estimations for the NICA energy are presented in Fig. 6.3.

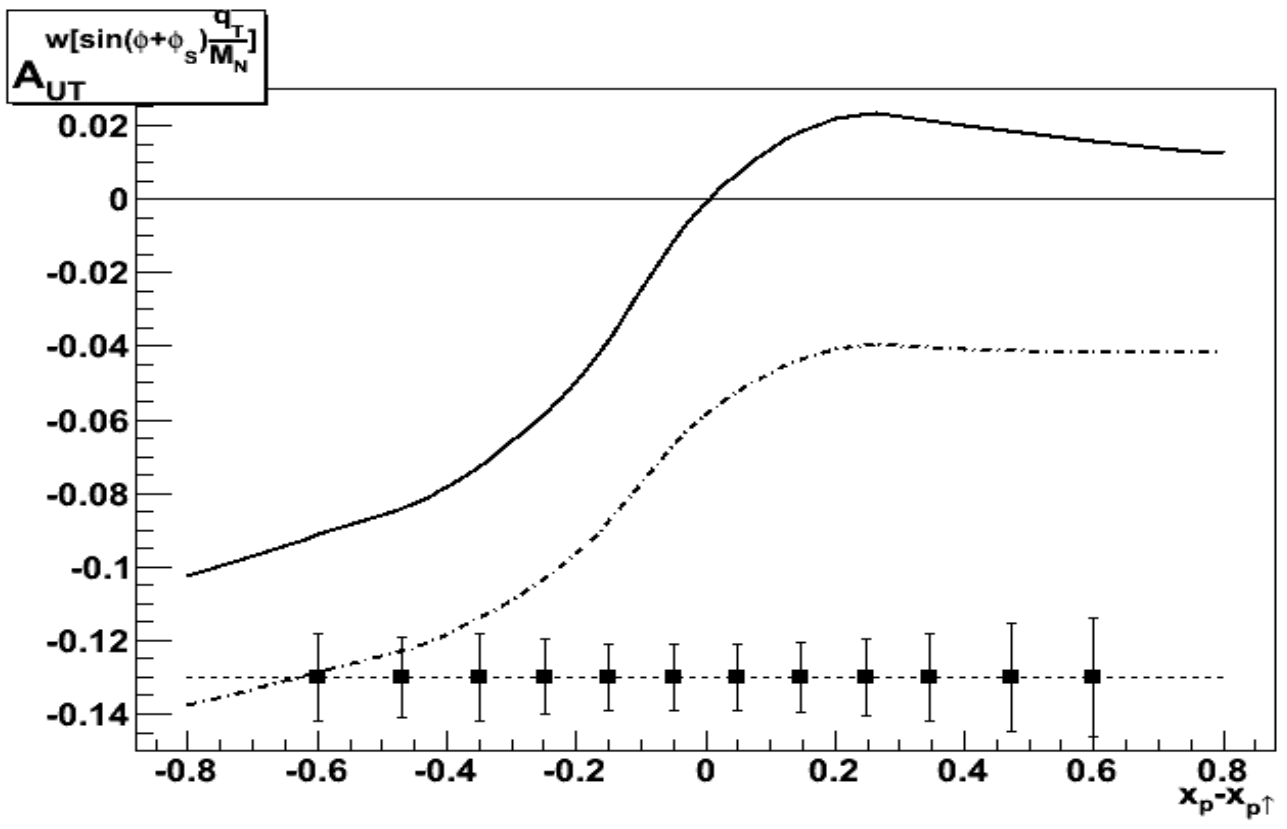

***Fig. 6.3:*** *Estimations of Boer-Mulders asymmetry* $A_{UT}^{w\left[\sin(\phi+\phi_S)\frac{q_T}{M_N}\right]}$ *at* $\sqrt{s}$ *=26 GeV and* $Q^2$ *= 15* *GeV*$^2$*. The solid and dotted curves correspond to the first and second versions of the evolution model, respectively. Points with bars show the expected statistical errors obtained with 100K of events.*

***6.1.2. Estimations of the J/Ψ production rates and precisions of asymmetry measurements.***

Statistics of the *J/Ψ* and DY events (with cut on $M_{l\text{-}l+,}$= 4 GeV) expected to be recorded ("per year") in 7000 hours of data taking with 100% efficiency of SPD are given in Table 3.

Table 3: Comparison of the *J/Ψ* and DY statistics

| $\sqrt{s}$ , GeV | 24 | 26 | $\sqrt{s}$ , GeV | 24 | 26 |
|---|---|---|---|---|---|
| $\sigma_{J/\Psi}\cdot B_{e+e-}$ , *nb* | 12 | 16 | $\sigma_{DY}$ , *nb* | 0.06 | 0.07 |
| Events "per year" | $18\cdot10^6$ | $23\cdot10^6$ | Events "per year" | $92\cdot10^3$ | $142\cdot10^3$ |

## 6.2. Estimations of direct photon production rates.

Estimation of the direct photon production rates based on PYTHIA6 Monte-Carlo simulation is presented in Table 4 for two values of colliding proton energies. Event rates are given for all and for leading processes of direct photon production considered in PYTHIA assuming 7000 hours of operation at maximal luminosity. The last column gives the rates corresponding to the cut on transverse momentum of photons suggested in Section 5.1.3. Statistical accuracies of $A_N$ and $A_{LL}$ measurements at NICA, suggested in Section 2.3, have been estimated assuming the beam polarizations (both transversal and longitudinal) equal to $P = \pm\ 0.8$ and overall detector efficiency (acceptance, efficiency of event reconstruction and selection criteria) of about 50%. Under such assumption, after 7000h of data taking the $A_N$ and $A_{LL}$ could be measured with statistical accuracy of ~0.11% and ~0.18%, respectively, in each of 18 $x_F$ bins (-0.9< $x_F$ < 0.9). Large statistics of events provide opportunities to measure the asymmetries as a function of $x_F$ and $p_T$.

Table 4: Estimated rates of the direct photon production.

| $\sqrt{s}$=24 GeV $L = 1.0 \times 10^{32}$, $cm^{-1}s^{-1}$ | $\sigma_{tot}$, nbarn | $\sigma_{P_T>4\ GeV/c}$, nbarn | Events/year, $10^6$ | Events/year, $10^6$ ($P_T > 4\ GeV/c$) |
|---|---|---|---|---|
| All processes | 1290 | 42 | 3260 | 105 |
| $qg \to q\gamma$ | 1080 | 33 | 2730 | 84 |
| $q\bar{q} \to g\gamma$ | 210 | 9 | 530 | 21 |
| $\sqrt{s}$=26 GeV $L = 1.2 \times 10^{32}$, $cm^{-1}s^{-1}$ | $\sigma_{tot}$, nbarn | $\sigma_{P_T>4\ GeV/c}$, nbarn | Events/year, $10^6$ | Events/year, $10^6$ ($P_T > 4\ GeV/c$) |
| All processes | 1440 | 48 | 4340 | 144 |
| $qg \to q\gamma$ | 1220 | 38 | 3680 | 116 |
| $q\bar{q} \to g\gamma$ | 240 | 10 | 660 | 28 |

To minimize systematic uncertainties, precision of luminosity and beam polarization should be under control, as well as accuracy of $\pi^0$, $\eta$ and other background rejection.

The expected statistics for "one year" of data taking, equivalent to 7000 hours (80 % of time) with maximal luminosity are given in Section 6 without taking into account the experiment overall efficiency. For more realistic expectations usually one uses the GEANT package which takes into account the detection and identification possibilities including the geometry and detection efficiencies. But at this stage of the project such estimations cannot be done because the overall structure of the SPD is not defined. Basing on the previous experience, one can expect that this efficiency will be not less than 50%.

6.3 – 6.5. TO BE WRITTEN AT THE PROPOSAL STAGE OF THE PROJECT

## 7. Time lines of experiments.

### 7.1. Possible data taking scenario.

At the first step of the project it is reasonable to start measurements with non-polarized protons (*pp*) and with non-polarized deuterons (*dd*), (*pd*). These data would provide a cross checks of our results with very precise world data on $f_1$ PDFs. At the same time new data on the Boer-Mulders PDF will be obtained.

At the second step the measurements should be performed with longitudinally polarized protons and deuterons in *pp*, *pd* and *dd* collisions with the beam polarizations *UL*, *LU*, *LL* to obtain asymmetries $A_{LU}$, $A_{UL}$ and $A_{LL}$ (Eqs.2.1.10) in each case. These data will be cross checked by existing data on $g_1$ PDF and provide new information on the Worm-gear-L PDF in proton and neutron (*u* & *d* quarks).

At the third step (the most important) measurements should be performed with transverse beam polarization in *pp*, *pd* and *dd* collisions (*UT*, *TU* and *TT*) to obtain asymmetries $A_{UT}$, $A_{TU}$ and $A_{TT}$ in each case. These data will be cross checked by existing data on Transversity PDF and provide new information on the Sivers, Worm-gear-T and Pretzelosity PDFs in proton and neutron (*u* & *d* quarks).

Finally, at the fourth step (the most difficult) measurements should be performed with *pp*, *pd* and *dd* beams when one beam polarized longitudinally while other – transversally in order to measure asymmetries $A_{LT}$ and $A_{TL}$ in each case. These data will provide new information and cross checks of our results on Transversity, Worm-gear-L, Pretzelosity and Worm-gear-T PDFs.

Following recommendations of the PAC, the corresponding Proposal (including the time lines of experiments) could be prepared by the end of 2015.

## 8. References.

### 8.1. References to Section 1.

### 8.2. References to Section 2.

### 8.4. References for Section 4.

### 8.5. References for Section 5.

### 8.6. References for Section 6.

## APPENDIX 1

Table 1: Cross sections of $J/\psi$ production in $pp$ colisions at $\sqrt{s} = 24$ GeV

| ISUB | process | cross section, [mb] | Comments |
|---|---|---|---|
| 'colour singlet' approach | | | |
| 86 | $gg \to J/\psi g$ | $1.429 \cdot 10^{-6}$ | |
| 87 | $gg \to \chi_{0c} g \to J/\psi\gamma$ | $3.348 \cdot 10^{-6}$ | |
| 88 | $gg \to \chi_{1c} g \to J/\psi$ | $3.954 \cdot 10^{-7}$ | |
| 89 | $gg \to \chi_{2c} g \to J/\psi$ | $2.736 \cdot 10^{-6}$ | |
| 106 | $gg \to J/\psi\gamma$ | $3.894 \cdot 10^{-8}$ | |
| 'colour octet' mechanism | | | |
| 421 | $gg \to c\bar{c}[^3S_1^{(1)}]g$ | $1.653 \cdot 10^{-6}$ | |
| 422 | $gg \to c\bar{c}[^3S_1^{(8)}]g$ | $5.762 \cdot 10^{-7}$ | |
| 423 | $gg \to c\bar{c}[^1S_0^{(8)}]g$ | $1.742 \cdot 10^{-6}$ | |
| 424 | $gg \to c\bar{c}[^3P_J^{(8)}]g$ | $3.609 \cdot 10^{-6}$ | |
| 425 | $gq \to qc\bar{c}[^3S_1^{(8)}]$ | $1.510 \cdot 10^{-6}$ | |
| 426 | $gq \to qc\bar{c}[^1S_0^{(8)}]$ | $1.817 \cdot 10^{-6}$ | |
| 427 | $gq \to qc\bar{c}[^3P_J^{(8)}]$ | $4.154 \cdot 10^{-6}$ | |
| 428 | $q\bar{q} \to gc\bar{c}[^3S_1^{(8)}]$ | $2.686 \cdot 10^{-7}$ | |
| 429 | $q\bar{q} \to gc\bar{c}[^1S_0^{(8)}]$ | $1.072 \cdot 10^{-8}$ | |
| 430 | $q\bar{q} \to gc\bar{c}[^3P_J^{(8)}]$ | $7.200 \cdot 10^{-8}$ | |
| 431 | $gg \to c\bar{c}[^3P_0^{(1)}]$ | $1.948 \cdot 10^{-5}$ | $\chi_{0c}$ |
| 432 | $gg \to c\bar{c}[^3P_1^{(1)}]$ | $2.300 \cdot 10^{-6}$ | $\chi_{1c}$ |
| 433 | $gg \to c\bar{c}[^3P_2^{(1)}]$ | $1.592 \cdot 10^{-5}$ | $\chi_{2c}$ |
| 434 | $g\bar{q} \to qc\bar{c}[^3P_0^{(1)}]$ | $1.844 \cdot 10^{-5}$ | $\chi_{0c}$ |
| 435 | $g\bar{q} \to qc\bar{c}[^3P_1^{(1)}]$ | $4.802 \cdot 10^{-6}$ | $\chi_{1c}$ |
| 436 | $g\bar{q} \to qc\bar{c}[^3P_2^{(1)}]$ | $1.836 \cdot 10^{-5}$ | $\chi_{2c}$ |
| 437 | $q\bar{q} \to gc\bar{c}[^3P_0^{(1)}]$ | $8.471 \cdot 10^{-9}$ | $\chi_{0c}$ |
| 438 | $q\bar{q} \to gc\bar{c}[^3P_1^{(1)}]$ | $4.703 \cdot 10^{-7}$ | $\chi_{1c}$ |
| 439 | $q\bar{q} \to gc\bar{c}[^3P_2^{(1)}]$ | $3.571 \cdot 10^{-7}$ | $\chi_{2c}$ |